\documentclass[journal=jctcce,manuscript=article,keywords=true]{achemso}
\setkeys{acs}{articletitle=true,chaptertitle=true}
\usepackage[T1]{fontenc}
\usepackage{amsmath,amssymb,graphicx,booktabs,longtable,array,ragged2e}
\usepackage{caption,needspace,xcolor,xurl,placeins,float}
\usepackage[normalem]{ulem}
\usepackage[hidelinks]{hyperref}

\AtBeginDocument{\singlespacing}
\SectionNumbersOn
\newif\ifShowRevisions
\newcommand{\added}[1]{#1}
\newcommand{\deleted}[1]{}
\newcommand{\sourceonly}[1]{}
\newcommand{\sourceattachment}[1]{}

\newcommand{\code}[1]{{\UrlFont\nolinkurl{#1}}}
\ifdefined\ShowTerminologyRevisions
  \newcommand{\termremoved}[1]{\textcolor{red}{\sout{#1}}}
  \newcommand{\termadded}[1]{\textbf{#1}}
\else
  \newcommand{\termremoved}[1]{\ignorespaces}
  \newcommand{\termadded}[1]{#1}
\fi

\newcommand{\USTCSchool}{School of Artificial Intelligence and Data Science,
University of Science and Technology of China, Hefei 230026, P.R.~China}
\newcommand{\USTCSuzhou}{Suzhou Institute for Advanced Research,
University of Science and Technology of China, Suzhou 215123, P.R.~China}
\newcommand{\ShanghaiLab}{SII-ECNU Joint Autonomous Chemistry Laboratory,
Shanghai Innovation Institute, Shanghai 200003, P.R.~China}
\newcommand{\SuzhouAI}{Suzhou Big Data \& AI Research and Engineering Center,
Suzhou 215123, P.R.~China}

\author{Zitian Lu}
\affiliation{\USTCSchool}
\alsoaffiliation{\USTCSuzhou}

\author{Shengrui Xu}
\affiliation{\USTCSchool}
\alsoaffiliation{\ShanghaiLab}
\alsoaffiliation{\USTCSuzhou}

\author{Tong Zhu}
\affiliation{Shanghai Engineering Research Center of Molecular Therapeutics and
New Drug Development, School of Chemistry and Molecular Engineering,
East China Normal University, Shanghai 200062, P.R.~China}
\alsoaffiliation{\ShanghaiLab}
\email{tzhu@lps.ecnu.edu.cn}

\author{Qiangqiang Gu}
\affiliation{\USTCSchool}
\alsoaffiliation{\USTCSuzhou}
\alsoaffiliation{\SuzhouAI}
\email{guqq@ustc.edu.cn}

\author{Jinzhe Zeng}
\affiliation{\USTCSchool}
\alsoaffiliation{State Key Laboratory of Precision and Intelligent Chemistry,
University of Science and Technology of China, Hefei 230026, P.R.~China}
\alsoaffiliation{\USTCSuzhou}
\alsoaffiliation{\SuzhouAI}
\email{jinzhe.zeng@ustc.edu.cn}

\title{xTBloom: GPU Batching for High-Throughput GFN-xTB Calculations}


\begin{document}
\begin{abstract}
\hypertarget{fs-doxcnat0PETxfpe9b6sPXqLgFAe}{}
Conformational searches and multiwindow free-energy sampling require \termremoved{repeated} \termadded{large numbers of} GFN-xTB energy and force evaluations \termremoved{for large numbers of conformations or independent sampling windows} \termadded{across molecular conformations and sampling windows}. Processing each system separately on a GPU makes it difficult to amortize memory management, \termadded{host--device} data transfer, and kernel-launch overhead. \termremoved{We} \termadded{Here, we} present xTBloom, a GPU batching library for GFN-xTB energy and force evaluations. xTBloom sorts evaluation queues by atomic orbital (AO) count before forming batches, accommodates size differences through a native ragged layout, maintains independent self-consistent charge (SCC) and failure states for each system, and selects execution paths according to data residency. We focus the quantitative evaluation on GFN2-xTB through cross-program comparisons on QM9 and OMol25, CPU\textendash{}CUDA comparisons, \termremoved{finite differences} \termadded{finite-difference force checks}, and controlled failure tests. On a near-size OMol25 workload, \termremoved{GPU evaluation was approximately 2.9 times faster than xTBloom's own single-core CPU backend} \termadded{GPU execution achieved a 2.9-fold speedup over xTBloom's single-core CPU backend}. A companion LAMMPS plugin integrates cross-window batching with periodic quantum mechanical/molecular mechanical (QM/MM) simulation. In a short synchronized test with 48 windows on one GPU, the complete coupled \termremoved{scheme} \termadded{workflow} achieved an aggregate speedup of approximately 14 \termremoved{relative to sequential window execution} \termadded{over sequential execution of the same windows}. These results demonstrate the throughput benefits of batched GFN-xTB evaluation for repeated electronic-structure calculations and multiwindow sampling.

\end{abstract}

\FloatBarrier
\Needspace{9\baselineskip}
\hypertarget{fs-doxcnNyF0TQiB9WaTHHTJsHDp6c}{}
\section{Introduction}

\hypertarget{fs-doxcnuD3kzBGzClJBMJMcO6r45b}{}
GFN1-xTB~\citep{Grimme_JChemTheoryComput_2017_v13_p1989} and GFN2-xTB~\citep{Bannwarth_JChemTheoryComput_2019_v15_p1652,Bannwarth_WiresComputMolSci_2021_v11_pe1493} are semiempirical tight-binding quantum-chemical methods applicable to a broad range of elements, molecular structures, and noncovalent interactions. Their relatively low computational cost has supported widespread use in conformational searches, geometry optimization, reaction exploration, and multiscale simulation~\citep{Bannwarth_WiresComputMolSci_2021_v11_pe1493,Katbashev_JPhysChemA_2025_v129_p2667}. These applications repeatedly evaluate energies and forces for many molecules, conformations, or sampled frames. Conformational search tools such as CREST use GFN2-xTB to evaluate large numbers of conformations and explore chemical space~\citep{Grimme_JChemTheoryComput_2019_v15_p2847,Pracht_PhysChemChemPhys_2020_v22_p7169,Pracht_JChemPhys_2024_v160_p114110}. QMugs~\citep{Isert_SciData_2022_v9_p273} uses GFN2-xTB to optimize drug-like molecular conformations and records the resulting energies, structures, and electronic properties. Studies of path sampling, geometry optimization, and reaction screening likewise use members of the GFN-xTB family for repeated energy and force evaluations~\citep{Li_JacsAu_2024_v4_p4757,Yang_Research_2025_v8_p0837,Li_JChemInfModel_2025_v65_p2297,Li_CleanEnergy_2026_v10_p157}. In quantum mechanical/molecular mechanical (QM/MM) simulation combined with machine-learning correction potentials, GFN-xTB also provides a low-level baseline for dynamics~\citep{Giese_JPhysChemB_2024_v128_p6257,Pan_JChemTheoryComput_2021_v17_p5745} and higher-level energy corrections~\citep{Li_ChemRxiv_2025_pv2}. These multimolecule, multiconformation, and multiwindow tasks create a demand for batched energy and force evaluation.

\hypertarget{fs-doxcnWg6XkNPOy66JYTCiJFGhTc}{}
We target batches of similarly sized molecules, multiple conformations of one molecule, and independent sampling windows. These tasks provide many independent calculations that can be aggregated, while retaining differences in system size and self-consistent charge (SCC) convergence within a batch. Efficient execution therefore requires control of data-organization and scheduling costs. Per-system calls limit the amortization of memory management, data transfer, and GPU kernel launches, and padding all inputs to the largest dimension in a batch adds redundant storage. Successful results also remain useful when a small number of systems fail to converge. We therefore organize batches by size to aggregate parallel work while accommodating within-batch differences and preserving per-system results.

\hypertarget{fs-doxcnNHFOhu3T0jzXFgy6qkG4Kd}{}
GPU implementations of semiempirical methods~\citep{Steinbach_JChemTheoryComput_2025_v21_p7335,Wu_JChemTheoryComput_2012_v8_p2272,Zhou_JChemTheoryComput_2020_v16_p4951,Athavale_JChemTheoryComput_2025_v21_p9498} and other electronic-structure approaches~\citep{Ufimtsev_JChemTheoryComput_2008_v4_p222,Seritan_WiresComputMolSci_2021_v11_pe1494,Barca_JChemTheoryComput_2021_v17_p7486,Andrade_JChemTheoryComput_2021_v17_p7447,Ju_JChemTheoryComput_2024_v20_p10838} offer several routes to accelerated solution. Steinbach and Bannwarth demonstrated asynchronous execution of multiple GFN2-xTB SCF calculations on one GPU and applied it to conformational sampling~\citep{Steinbach_JChemTheoryComput_2025_v21_p7335}. Integral optimizations for xTB reduce redundant work by reusing recurrence intermediates and screening shell pairs~\citep{Sevi_xtb_2026_pr1465}. The embeddable interfaces of xTB and tblite operate on individual systems~\citep{Bannwarth_WiresComputMolSci_2021_v11_pe1493,Katbashev_JPhysChemA_2025_v129_p2667}, with the caller organizing multisystem workloads. dxtb supports CUDA and batched SCC with padded tensors for mixed-size inputs~\citep{Friede_JChemPhys_2024_v161_p062501}. TBMaLT also uses padded batches and removes converged systems during SCC~\citep{McSloy_JChemPhys_2023_v158_p034801}. GPU acceleration of DFTB for large systems addresses Hamiltonian diagonalization and density-matrix construction~\citep{Vuong_JChemPhys_2023_v158_p084802}, and parallel capabilities have also been developed systematically in DFTB+~\citep{Hourahine_JPhysChemA_2025_v129_p5373}. Here we examine queue organization and natural-size storage in the target small-matrix regime and connect batched evaluation to cross-window execution in periodic QM/MM.

\hypertarget{fs-doxcn6DgDoKOTqKvMPEydyUtdnf}{}
We developed xTBloom as a GPU batching library for GFN-xTB energy and force evaluation, with GFN2-xTB as the main quantitative target. xTBloom sorts work by atomic orbital (AO) count, accommodates within-batch size differences through native ragged inputs, and maintains each system's SCC state and results independently. Its CPU and CUDA backends share a C application binary interface (ABI). Python interfaces for ASE~\citep{HjorthLarsen_JPhysCondensMatter_2017_v29_p273002} and dpdata~\citep{Zeng_JChemInfModel_2025_v65_p11497} connect batched electronic-structure evaluation to molecular simulation and data-generation workflows.

\hypertarget{fs-doxcnlVXVSlVlc2Z6CNAhIbNIFE}{}
To assess integration and acceleration in molecular simulation, we use a companion plugin for LAMMPS~\citep{Thompson_ComputPhysCommun_2022_v271_p108171} to perform cross-window QM/MM umbrella sampling on a single GPU and incorporate a Deep Potential\textendash{}Range Correction (DPRc) model. Numerical consistency tests and GPU benchmarks evaluate xTBloom itself. Free-energy calculations with the uncorrected QM/MM model and the DPRc-corrected model demonstrate its application, while matched timings against sequential window execution quantify the throughput benefit of cross-window batching.

\FloatBarrier
\Needspace{9\baselineskip}
\hypertarget{fs-doxcnqov26iPSrIZrW0NOf9Y5dg}{}
\section{Theoretical background}

\FloatBarrier
\Needspace{5\baselineskip}
\hypertarget{fs-doxcn6iXOHTqekiQAAX6cABA8wf}{}
\subsection{GFN-xTB energies and self-consistent electronic structure}

\hypertarget{fs-doxcnxH0U1uwrv4vySxZ8L6sr2f}{}
GFN1-xTB~\citep{Grimme_JChemTheoryComput_2017_v13_p1989} and GFN2-xTB~\citep{Bannwarth_JChemTheoryComput_2019_v15_p1652} combine parameterized electronic, repulsive, dispersion, and related energy contributions. GFN2-xTB additionally includes multipolar electrostatics and density-dependent dispersion~\citep{Bannwarth_WiresComputMolSci_2021_v11_pe1493}. For model \(m\) and nuclear coordinates \(\mathbf R\), we denote the finite-electronic-temperature energy scale by \ensuremath{\tau} and the dimensionless electronic entropy by \(\widetilde S_{\mathrm e}\).

\hypertarget{fs-doxcnN05bRHMYNe5nvNDbni5kUr}{}
\begin{equation}
m\in\{\mathrm{GFN1\text{-}xTB},\mathrm{GFN2\text{-}xTB}\},\qquad \tau=k_{\mathrm B}T_{\mathrm e},\qquad \widetilde S_{\mathrm e}=S_{\mathrm e}/k_{\mathrm B}.
\end{equation}

\Needspace{5\baselineskip}
\hypertarget{fs-doxcnntZ6RoxrIBogV5d8Hi5yGb}{}
The variational objective of the self-consistent electronic problem can then be written as

\hypertarget{fs-doxcnsFs4llvUnwy20OYLuZEcOd}{}
\begin{equation}
\mathcal F_m(\mathbf P,\mathbf x;\mathbf R,\tau)=E_m(\mathbf P,\mathbf x;\mathbf R)-\tau\widetilde S_{\mathrm e}.
\end{equation}

\hypertarget{fs-doxcneXd5w65sBomXFdIa86rT3X}{}
Here, \(\mathbf P\) is the total electronic density matrix, obtained by summing the density matrices of the spin channels. The vector \(\mathbf x\) collectively denotes the SCC variables determined from those spin-channel density matrices. For GFN2-xTB, these variables include shell charges and atomic multipoles, together with the corresponding spin layout when the spin-polarization extension is enabled. At finite electronic temperature, the xTBloom public interface reports the electronic Helmholtz free energy.

\hypertarget{fs-doxcnGlYhBOm8AQRcKO8nia4qif}{}
The schematic energy decomposition for GFN2-xTB and its optional spin-polarization extension separates geometric and electronic-state dependencies~\citep{Grimme_JChemTheoryComput_2017_v13_p1989,Bannwarth_JChemTheoryComput_2019_v15_p1652,Bannwarth_WiresComputMolSci_2021_v11_pe1493,Neugebauer_JComputChem_2023_v44_p2120}

\hypertarget{fs-doxcnVaBly6d8oNgHcAdXHqIdmc}{}
\begin{equation}
\begin{aligned}E_{\mathrm{GFN2}}={}&\operatorname{Tr}[\mathbf P\mathbf H^0(\mathbf R)]+E_{\mathrm{iso}}(\mathbf q;\mathbf R)+E_{\mathrm{aniso}}(\mathbf q,\boldsymbol\mu,\boldsymbol\Theta;\mathbf R)\\&+E_{\mathrm{D4}}^{(2)}(\mathbf q;\mathbf R)+E_{\mathrm{rep}}(\mathbf R)+E_{\mathrm{D4}}^{(3)}(\mathbf R)+E_{\mathrm{spin}}.\end{aligned}
\end{equation}

\hypertarget{fs-doxcngUngaNgWHBcAWFNq373e7N}{}
Here, \(\mathbf q\) denotes shell and atomic charges, and \ensuremath{\mu} and \ensuremath{\Theta} denote atomic dipole and quadrupole moments. The isotropic and anisotropic terms each contain their corresponding electrostatic and exchange-correlation contributions. The spGFN2-xTB extension introduces the spin-dependent energy \(\boldsymbol{E_{\mathrm{spin}}}\)~\citep{Neugebauer_JComputChem_2023_v44_p2120}. The original GFN2-xTB model has \(\boldsymbol{E_{\mathrm{spin}}=0}\). At fixed geometry, the overlap matrix, zeroth-order Hamiltonian, and geometric interaction factors can be constructed in advance. The repulsive term and the charge-independent D4 three-body term are also constant. Charges, multipoles, and their associated potentials change with the electronic state, and the D4 two-body term participates in the charge-dependent self-consistent calculation. Geometry-dependent data can therefore be reused across SCC iterations, while density-dependent energies and potentials are updated at each iteration.

\hypertarget{fs-doxcnXubPhINiCA5uxaym3B83oe}{}
A GFN-xTB evaluation first prepares the structure and parameters and constructs geometry-dependent matrices and interaction factors. It then assembles the Hamiltonian from the current SCC state and enters the self-consistent iteration. For each spin channel \ensuremath{\sigma}, it solves the generalized eigenproblem

\hypertarget{fs-doxcn3at5mKW0NQHXh5rn0quMhc}{}
\begin{equation}
\mathbf H^\sigma(\mathbf x;\mathbf R)\mathbf C^\sigma=\mathbf S(\mathbf R)\mathbf C^\sigma\boldsymbol\varepsilon^\sigma,
\end{equation}

\hypertarget{fs-doxcnr9Vd35X8BQYvkZlSVrhWDb}{}
Here, \(\mathbf C\) is the orbital-coefficient matrix and \ensuremath{\varepsilon} is the diagonal matrix of orbital energies. The orbital coefficients and the diagonal occupation matrix \(\mathbf f\) determine the density matrix

\hypertarget{fs-doxcnAiNchWWAEIzO4F02npbioc}{}
\begin{equation}
\mathbf P^\sigma=\mathbf C^\sigma\mathbf f^\sigma(\mathbf C^\sigma)^{\mathsf T},\qquad\mathbf P=\sum_\sigma\mathbf P^\sigma.
\end{equation}

\hypertarget{fs-doxcnfc869RidFtYYCTP8P00iKf}{}
Each spin channel satisfies orbital normalization in the nonorthogonal basis and an electron-number constraint,

\hypertarget{fs-doxcnScarTqhNjpPOVBSXkKqTwg}{}
\begin{equation}
(\mathbf C^\sigma)^{\mathsf T}\mathbf S\mathbf C^\sigma=\mathbf I,\qquad\operatorname{Tr}(\mathbf P^\sigma\mathbf S)=\sum_i f_i^\sigma=N_\sigma.
\end{equation}

\hypertarget{fs-doxcn1LdNkxzyTZ05QqRgzKTsVh}{}
The model's reference valence-electron counts and the molecular charge determine the total number of valence electrons. The spin constraint further specifies the electron number in each spin channel, \(N_\sigma\). Restricted calculations share spatial orbitals while determining the occupations and electronic-entropy contributions of the two physical spin channels separately. Population analysis in a nonorthogonal atomic orbital basis incorporates the overlap matrix. The resulting shell charges and atomic multipoles determine the corresponding potentials for the next Hamiltonian construction.

\hypertarget{fs-doxcnCoEEGIOcrDx81WoS3iAqad}{}
Self-consistency requires agreement between the input SCC state and the output state obtained from the updated density. A mixing algorithm uses their residual and its history to generate the next input and suppress oscillations. In the common-orbital GFN2-xTB calculations considered here, convergence requires both the root-mean-square (RMS) residual of the full SCC state and the absolute change in electronic free energy to be strictly smaller than their respective tolerances. The residual contains shell-charge, atomic-dipole, and quadrupole components. SI S1.2 defines their units and normalization. Each iteration evaluates the free energy from the updated density and its associated electronic variables. The final energy and forces are evaluated after the convergence conditions are met.

\hypertarget{fs-doxcnoiUEaVkB7lGafDfuQuOkte}{}
The cost of the eigensolver depends primarily on the AO dimension, while short-range interaction costs depend on atom count and spatial distribution. The number of SCC iterations determines how often electronic-structure operations are repeated. Differences in orbital spectra, initial electronic states, and charge response can produce different iteration counts even at the same AO dimension. Organizing work by atom count alone therefore gives an incomplete account of these costs.

\FloatBarrier
\Needspace{5\baselineskip}
\hypertarget{fs-doxcnFJY9CxJmy2lA2IrdAWKkpc}{}
\subsection{Energies and analytic forces in QM/MM coupling}

\hypertarget{fs-doxcnupiYmaKYyDSwDc3DAeVAEc}{}
At finite electronic temperature, analytic forces are derivatives of the Helmholtz free energy, including its electronic-entropy contribution. Coordinate derivatives also involve the overlap, Hamiltonian, and other geometry-dependent terms. Energy definitions, occupation conventions, and force definitions must therefore be consistent in geometry optimization and dynamics.

\hypertarget{fs-doxcn3oji26HAeG97RembGHIfaf}{}
Analytic forces connect electronic-structure evaluation to geometry optimization and molecular dynamics. In QM/MM models~\citep{Lin_TheorChemAcc_2007_v117_p185,Senn_AngewChemIntEd_2009_v48_p1198}, the electronic state of the quantum region responds to the classical environment, which also contributes through classical interactions and electrostatic embedding. Interoperable free-energy software frameworks that combine QM/MM and machine-learning potentials provide related workflow context~\citep{Tao_JChemPhys_2024_v160_p224104}.

\hypertarget{fs-doxcnZZVFoSgPIoQzxSjRajPvPg}{}
The total QM/MM energy includes the retained classical bonded terms, MM\textendash{}MM and QM\textendash{}MM Lennard\textendash{}Jones interactions, MM\textendash{}MM electrostatics, and the GFN2-xTB QM/MM electronic free energy. Biased sampling adds the corresponding bias potential. Near-field MM point charges enter Hamiltonian construction during SCC. The response of the periodic environment to QM atomic charges and its energy contribution are written as

\hypertarget{fs-doxcnxGsFFm6DvYq8U2OQ9BfKwf}{}
\begin{equation}
\boldsymbol\phi_{\mathrm{per}}=\mathbf b+\mathbf A\mathbf q, \qquad E_{\mathrm{per}} =\mathbf q^{\mathsf T}\mathbf b+ \frac12\mathbf q^{\mathsf T}\mathbf A\mathbf q .
\end{equation}

\hypertarget{fs-doxcn7VF8i1xe6iuHAn9kVvXtIe}{}
Here, \(\mathbf q\) is the vector of net QM atomic charges, \(\mathbf b\) is the potential shift independent of the current QM charges, and \(\mathbf A\) is a symmetric response matrix. Consequently,

\hypertarget{fs-doxcnRQV0Owlg4yXlKwDS5g3XVe}{}
\begin{equation}
\boldsymbol\phi_{\mathrm{per}} =\frac{\partial E_{\mathrm{per}}}{\partial\mathbf q}
\end{equation}

\hypertarget{fs-doxcnGzIL4RcsG8dlZPJ1oAt2Ne}{}
Periodic electrostatics includes complementary Ewald terms~\citep{Darden_JChemPhys_1993_v98_p10089,Essmann_JChemPhys_1995_v103_p8577}, self-interaction corrections, and subtraction of interactions already included explicitly~\citep{Nam_JChemTheoryComput_2005_v1_p2,Giese_JChemTheoryComput_2016_v12_p2611}. It also requires the coordinate derivatives of the external operators.

\hypertarget{fs-doxcnbwJ7x4HjL3WS92dRdBw3tf}{}
Explicit MM point charges enter the electronic-structure problem through screened shell-monopole interactions. For fixed geometry and point-charge inputs, this external potential can be precomputed and added to the shell potentials in each SCC iteration. Changes in QM charges still modify the self-consistent energy and electronic state. The periodic response includes feedback from the current QM charges and is updated with the SCC state. Both explicit point charges and the periodic response therefore participate in the embedded self-consistent solution.

\Needspace{7\baselineskip}
\hypertarget{fs-doxcnnUqu9GEswhBTma2WrGEBIb}{}
For a given geometry and set of external inputs, SCC convergence yields the embedded-system free energy, including the model energy terms, explicit point charges, and periodic response. Denoting this free energy by \(\overline{\mathcal F}_m^{\mathrm{emb}}(\mathbf R;\mathbf b,\mathbf A,\tau)\), xTBloom returns the QM atomic force

\hypertarget{fs-doxcnJsCqoNGyFHz2ykS3GktLWf}{}
\begin{equation}
\mathbf F_I^{\mathrm{xTBloom}}=-\left.\frac{\partial\overline{\mathcal F}_m^{\mathrm{emb}}}{\partial\mathbf R_I}\right|_{\mathbf b,\mathbf A,\tau}.
\end{equation}

\hypertarget{fs-doxcnCiDILCoIeWz7LAPH04EkEf}{}
This derivative includes the geometric dependencies represented within the library and holds \(\mathbf b\), \(\mathbf A\), and the electronic-temperature energy scale \ensuremath{\tau} fixed. xTBloom evaluates the analytic derivatives of explicit point-charge interactions with respect to both QM-atom and point-charge coordinates and returns point-charge forces when requested. For a periodic response that depends on coordinates or the cell, the coupling driver supplies the explicit derivatives of \(\mathbf b\) and \(\mathbf A\) and combines them with classical-interaction and bias forces to obtain the complete QM/MM force.

\hypertarget{fs-doxcnxmwLErlkLrSSNRARvxyOse}{}
Deep Potential\textendash{}Range Correction (DPRc)~\citep{Zeng_JChemTheoryComput_2021_v17_p6993} learns the energy and force differences between higher-level and lower-level QM/MM descriptions. Related range-corrected \ensuremath{\Delta}MLP work examines the transferability of MACE graph neural networks in QM/MM~\citep{Giese_JPhysChemB_2025_v129_p5477}. The difference potential and its negative gradient are added to the lower-level QM/MM baseline~\citep{Ramakrishnan_JChemTheoryComput_2015_v11_p2087}.

\hypertarget{fs-doxcniB7NPcWtD2Q2jSFRp4VJEf}{}
Let \(\boldsymbol{U_{\mathrm{high}}}\) and \(\boldsymbol{U_{\mathrm{low}}}\) denote the higher-level and lower-level effective QM/MM potential energies. The difference potential and corrected energy are then

\hypertarget{fs-doxcnz7mEaS2x0GnOufHf61tPZd}{}
\begin{equation}
\begin{aligned}\boldsymbol{\Delta U_\theta(\mathbf R)}&\boldsymbol{\simeq U_{\mathrm{high}}(\mathbf R)-U_{\mathrm{low}}(\mathbf R)},\\\boldsymbol{U_{\mathrm{corr}}(\mathbf R)}&\boldsymbol{=U_{\mathrm{low}}(\mathbf R)+\Delta U_\theta(\mathbf R).}\end{aligned}
\end{equation}

\hypertarget{fs-doxcnaSak5Tq87lxIM5bJV1XZAD}{}
Here, \ensuremath{\theta} denotes the model parameters, and \(\mathbf R\) contains the coordinates of the QM and MM atoms.

\hypertarget{fs-doxcnYWKjujbHt4LTWab4pV8Kuf}{}
The difference labels and production simulations use the same lower-level energy definition. At finite electronic temperature, the electronic contribution is the free energy consistent with the analytic forces.

\hypertarget{fs-doxcnBjprKxVzbkE1sodU3ZfpRc}{}
The corrected energy and forces retain the lower-level QM/MM baseline, so advancing the dynamics continues to require repeated GFN-xTB electronic-structure solutions.

\FloatBarrier
\Needspace{9\baselineskip}
\hypertarget{fs-doxcnXwRYPg8Ipe35riX04tRxne}{}
\section{Methods and implementation}

\hypertarget{fs-doxcnjTP5FMpdXDoYgc2kHxeonb}{}
xTBloom solves individual molecules independently and executes batches on the GPU. Systems share computational resources while retaining their own electronic states, convergence histories, and outputs. The complete execution scheme organizes batches through AO sorting and adjusts work according to the active SCC state. Successive evaluations reuse topology plans, workspaces, and compatible electronic states to reduce repeated preparation and data transfer.

\FloatBarrier
\Needspace{5\baselineskip}
\hypertarget{fs-doxcnXJ9Sy4O1qbXPU53F3rcdTd}{}
\subsection{Batch organization and data layout for batched GPU evaluation}

\hypertarget{fs-doxcnPD4fQs7cUnz6Cp50R8HPPb}{}
Even size-grouped molecular batches can differ in atom count, orbital dimension, and interaction count. These operations favor different data organizations. Atom-based layouts provide only an indirect measure of eigenproblem size, while uniform padding to the largest system in a batch introduces redundant storage. xTBloom separates the external molecular representation from the execution layout of each computational stage. Callers provide inputs at each system's natural size, and the execution layer organizes GPU tasks by the actual computational dimensions.

\hypertarget{fs-doxcnv3yrBmnxbYeIdjK8hiE5Lg}{}
The public interface represents ragged batches using flat atom arrays, offsets, and system-level charge and spin information. Preparation establishes mappings among atoms, shells, orbitals, and matrix storage. \mbox{Original} system indices link internal tasks to output locations. Internal matrices may be packed or padded to meet solver requirements, with system identity and result order preserved throughout these layout transformations.

\hypertarget{fs-doxcn6pUbOMpMwKycfhF7VYeP92}{}
Before partitioning a queue into batches, xTBloom sorts the entire queue by AO count to concentrate systems of the same dimension within a call. Matrix solves within each batch are grouped by exact AO dimension, and the solver path is selected from the matrix size and group population. System mappings restore results to the original input order. Short-range geometric processing selects dense enumeration or spatial-grid atom-pair lists according to atom count. Each short-range interaction uses its own cutoff definition. Long-range electrostatics is exempt from a uniform short-range cutoff.

\FloatBarrier
\Needspace{5\baselineskip}
\hypertarget{fs-doxcnFXtGl3L9GxIroyjAd7oDwc}{}
\subsection{Batched SCC execution for nonuniform convergence}

\hypertarget{fs-doxcnPa7SPaWusPw5FH0o7bL0rb}{}
Systems within one size group may still require different numbers of SCC iterations. xTBloom maintains a separate electronic state, mixing history, iteration count, and termination state for each system and advances it according to its own convergence conditions. After each update, unconverged systems without numerical failures form the active set for the next iteration. Converged systems retain their final electronic states. Systems with numerical failures enter the corresponding failure state. Electronic variables are frozen for both classes of terminated systems.

\hypertarget{fs-doxcnwuXvpc1NweDARAQJipq6Zd}{}
The execution layer handles the shrinking active set through compaction or masking. For eligible closed-shell batches with small AO dimensions, it compacts active members and performs subsequent eigensolves and orbital back-transformation on those members. Other eligible batches retain their allocated computational capacity and mask state updates for terminated members. Compaction reduces some redundant computation, while masking avoids repeated batch reorganization. Supplementary scans show that the benefit of active-capacity scheduling depends jointly on AO dimension, solver-bucket capacity, and active fraction. Reorganization and dispatch costs dominate most measurements with larger AO dimensions and bucket capacities, and substantial component-level gains occur mainly for very small active sets (SI S5.3). Supported GPU execution paths perform convergence checks and schedule subsequent iterations on the device, reducing per-iteration host interaction. The SI specifies the implementation conditions.

\hypertarget{fs-doxcncLGnyYczePSES08VTnzr0b}{}
SCC nonconvergence and eigensolver failures attributable to an individual system are returned as per-system statuses. When the call returns successfully, all requested floating-point results for a failed member are filled with NaNs, while healthy members retain their successful results. Input-validation, resource-allocation, and device or stream runtime failures are handled as call-level errors. SI S2.2 defines output-commit semantics.

\FloatBarrier
\Needspace{5\baselineskip}
\hypertarget{fs-doxcnBEAYOey6YhutQtbPRsdXOe}{}
\subsection{Data reuse and residency for successive evaluations}

\hypertarget{fs-doxcnocxxdxKh8alu36UzcQzNNb}{}
Geometry optimization and molecular dynamics repeatedly evaluate one topology, and each geometry requires multiple SCC iterations. xTBloom organizes data reuse at the topology, geometry, and electronic-state levels. Each class of data is reused over its applicable computational stages to reduce repeated preparation and memory-management costs. Figure 1c summarizes these reuse levels.

\hypertarget{fs-doxcn2jy6UmBqeAiTzleYDrjjOg}{}
A fixed-topology plan stores elements, system boundaries, charges, spin information, and computational policy and owns reusable workspace. The plan and storage remain reusable as coordinates change, with geometry-dependent numerical quantities rebuilt for the new geometry. Within the SCC iterations of one geometry, the overlap matrix and its Cholesky factor remain constant and can be reused to transform each generalized eigenproblem. Electronic-state contributions to the Hamiltonian are updated at each iteration. Topology and workspace reuse thus spans successive geometries. Overlap-matrix and Cholesky-factor reuse applies within a fixed geometry.

\hypertarget{fs-doxcnA3g9lJBwWtO1Wj95ohWvUe}{}
Electronic-state warm starts exploit continuity between successive geometries. For the most recent fully converged batch in the same context, the runtime can reuse its electronic state as the SCC initial guess for a new geometry when the policy remains compatible. The runtime rejects warm starts when the topology or policy is incompatible. Initial-state reuse controls iteration initialization, while workspace reuse controls storage management. Each evaluation assesses convergence for its current geometry.

\hypertarget{fs-doxcnCP6vsO1DkPP99vuTu93G7f}{}
On GPU execution paths, transfers follow the actual locations of the input and output buffers, supporting host, device, and mixed memory modes. Computational workspace remains on the device, and results are delivered to their specified destinations. GPU-resident data can therefore pass directly to subsequent computations. The same solver interface supports host-driven independent batches, repeated calls, and device-side workflows.

\begin{figure}[!htbp]
\centering
\hypertarget{fs-doxcn9Djd0EW3NDvd3czWaiiqSc}{}
\includegraphics[width=\linewidth,height=.66\textheight,keepaspectratio]{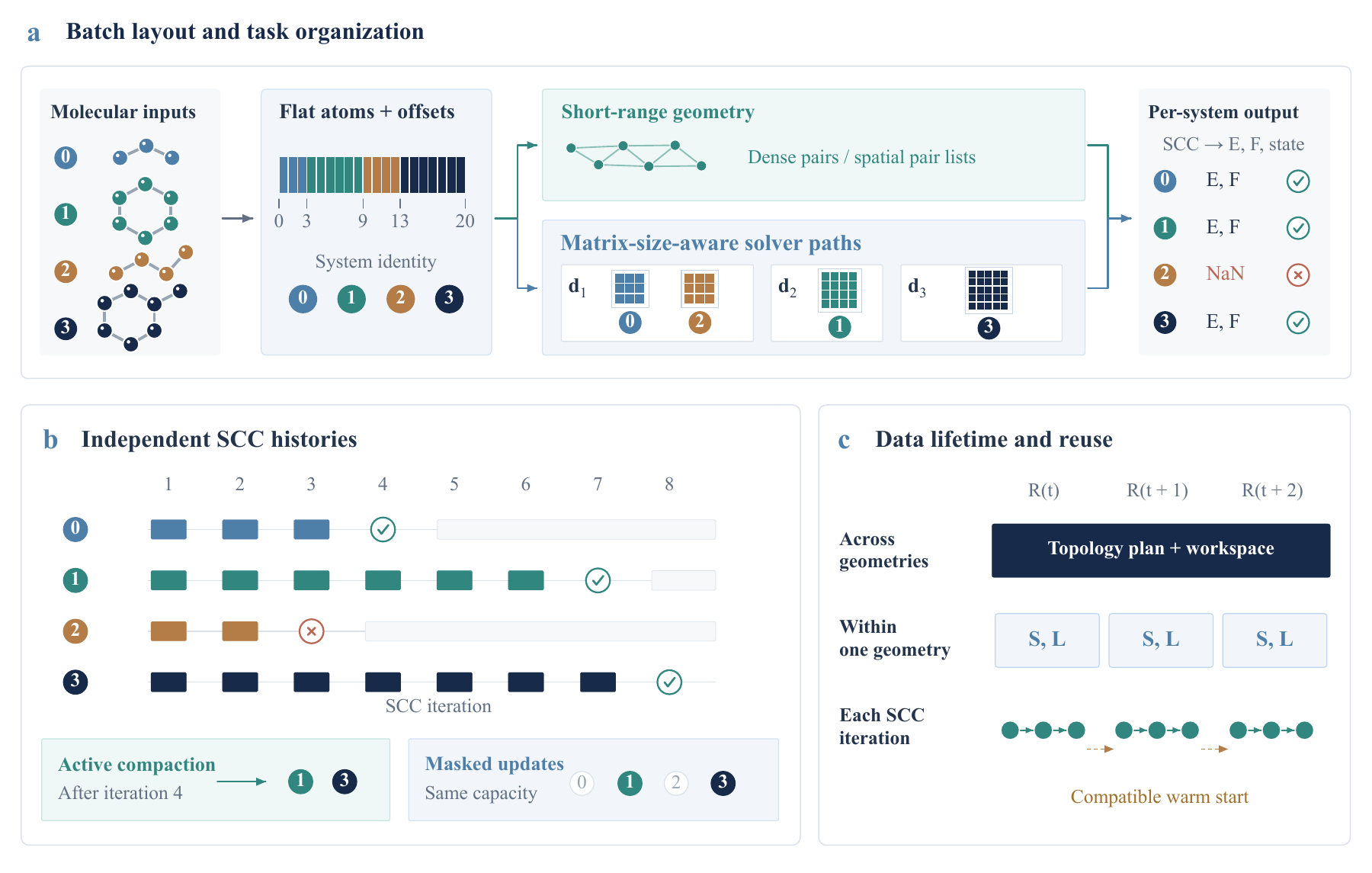}
\caption*{\hypertarget{fs-doxcnOq2ykXQqe75aUlry5fGl2b}{}Figure 1. GPU batch organization and data reuse. (a) Flat atom arrays and offsets preserve system identity; solver paths depend on matrix size and group population. (b) Each system has an independent SCC history and termination state. Active members are processed by compaction or masking; lanes and iteration counts are schematic. (c) Topology plans and workspaces are reused across compatible geometries, whereas the overlap matrix \(S\) and its Cholesky factor \(L\) are reused within a fixed geometry. Dashed arrows indicate electronic-state warm starts after full convergence of the previous geometry under a compatible policy.}
\end{figure}

\FloatBarrier
\Needspace{5\baselineskip}
\hypertarget{fs-doxcnMdFgExIxaELO8j8hsneqbf}{}
\subsection{LAMMPS QM/MM plugin and cross-window batching on one GPU}

\hypertarget{fs-doxcnykLX6KaGpybrepUOyh0tBe}{}
We developed a QM/MM plugin in the companion LAMMPS project to apply cross-window GPU batching to molecular dynamics. The plugin uses synchronized multipartition execution (Figure 2) to batch QM/MM coupling evaluations across windows at the same integration step, allowing independent windows to share one GPU. Batching covers GFN2-xTB electronic-structure evaluation, classical nonbonded interactions, periodic electrostatics using the particle--particle particle--mesh (PPPM) method with fast Fourier transforms (FFTs), and DeePMD/DPRc inference when enabled. Each window retains its own coordinates, integrator, and electronic-state history.

\begin{figure}[!htbp]
\centering
\hypertarget{fs-doxcnbPJxU90a41W04HN0QT6VOh}{}
\includegraphics[width=\linewidth,height=.66\textheight,keepaspectratio]{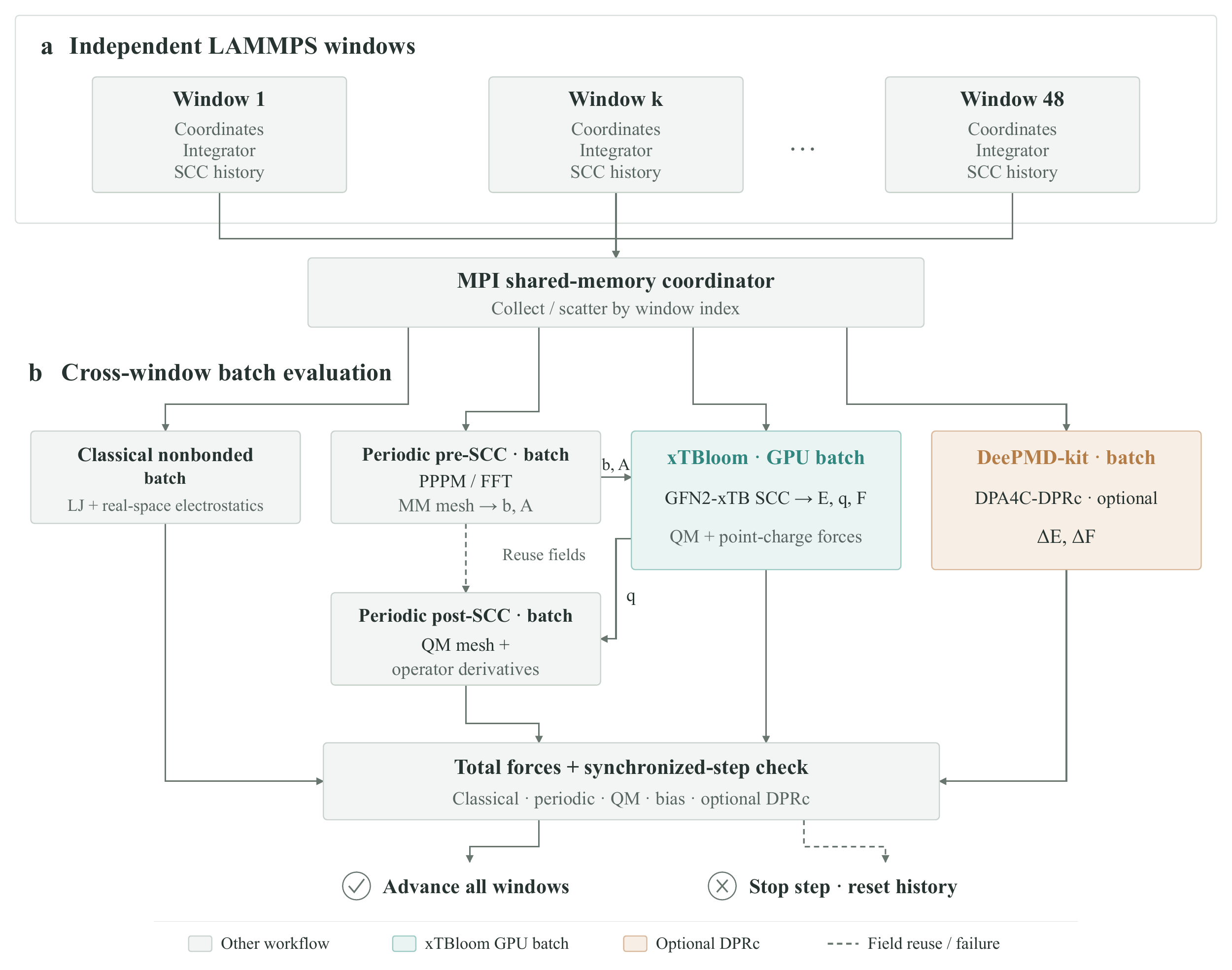}
\caption*{\hypertarget{fs-doxcnAmkIyJePMECRQDBeObepvh}{}Figure 2. Cross-window QM/MM batching on one GPU. (a) Independent LAMMPS windows retain their own coordinates and SCC histories; an MPI shared-memory coordinator collects and distributes data by window index. (b) Classical, periodic, GFN2-xTB and optional DPRc computations are batched across windows, followed by force assembly and a synchronized-step check. Periodic preprocessing supplies \(\mathbf b\) and \(\mathbf A\) to xTBloom, and the converged QM charges \(\mathbf q\) enter postprocessing. Gray marks the windows and coordination, classical and periodic calculations, and force assembly; teal marks xTBloom GPU batching, and ochre marks optional DPRc. Dashed arrows indicate field reuse or, at the cross, stopping the step and resetting electronic-state history after a failure.}
\end{figure}

\hypertarget{fs-doxcnyEF7gLX4FKYYUzIvzYWNrg}{}
Each LAMMPS partition advances one window trajectory and handles classical interactions, periodic boundaries, the thermostat, and bias potentials. Partition root processes coordinate through intranode MPI shared memory to collect QM coordinates, MM point charges, and periodic-response inputs at synchronized integration steps. A coordinator calls the xTBloom batch executor and distributes energies, forces, charges, and SCC statuses using fixed window indices. This mapping preserves the electronic-state history of each window across successive geometries. After all windows complete their current-step evaluations successfully, the dynamics driver assembles the total forces and advances integration. A solver failure in any window stops the current synchronized step.

\hypertarget{fs-doxcnxDxd0dOX2mIDayEXYxOsmh}{}
The plugin evaluates periodic electrostatics in two stages. Before SCC, it constructs the periodic environmental potential \(\mathbf b\) and charge-response matrix \(\mathbf A\) from the current geometry and MM charges. During SCC, the periodic feedback is updated from the current QM charges through \(\mathbf b+\mathbf A\mathbf q\). After convergence, the final QM charges complete the QM mesh evaluation. The two stages reuse field and spectral data. The LAMMPS coupling layer supplies external-operator derivatives and combines them with explicit point-charge reaction forces and classical forces. Periodic electrostatics, electronic structure, and the classical force field therefore use a consistent charge state within each integration step.

\hypertarget{fs-doxcnpZs3dIfFUY4b9d05j5ZlXe}{}
In DPRc-corrected simulations, each window evaluates the lower-level GFN2-xTB QM/MM baseline using the QM coordinates, MM environment, and constraint state from the same time step. DPA4C~\citep{Li_arXiv_2026_p2608.19041} represents the local environmental dependence, and DeePMD-kit~\citep{Wang_ComputPhysCommun_2018_v228_p178,Zeng_JChemTheoryComput_2025_v21_p4375} evaluates the difference potential and its coordinate derivatives. The correction omits MM\textendash{}MM interaction terms. The dynamics workflow adds the difference potential and its negative gradient to the lower-level QM/MM energy and forces, together with the bias potential and its forces. Section 2.2 defines the corresponding energies.

\FloatBarrier
\Needspace{9\baselineskip}
\hypertarget{fs-doxcnFadByrTtdP6KiQAsE7FSEb}{}
\section{Computational details}

\hypertarget{fs-doxcnLt7ApR7svQwrpxLvz0V1UU}{}
We used four groups of experiments to evaluate numerical agreement, full-evaluation performance, execution design and application integration. Cross-program comparisons, finite differences and controlled mixed batches tested numerical behavior and failure handling. Full evaluations compared the GPU implementation with CPU and external-program baselines. AO-sorting and active-capacity experiments examined the effects of batch composition and SCC scheduling in Section 5.2.2. Cross-window QM/MM umbrella sampling tested integration into molecular simulation, with matched sequential-window and batched timings used to measure application throughput.

\FloatBarrier
\Needspace{5\baselineskip}
\hypertarget{fs-doxcn0OxxN2gn4htIp84ZgY4cJg}{}
\subsection{Numerical comparisons and per-system failure isolation}

\hypertarget{fs-doxcn2dec2FOf7Xk8dXnbNm3aUd}{}
The numerical tests used the standard conformance suite, QM9 and OMol25. QM9~\citep{Ramakrishnan_SciData_2014_v1_p140022} provided a baseline of small, neutral CHONF molecules. OMol25~\citep{Levine_arXiv_2025_p2505.08762} covered a wider range of elements, charges, spins and nonequilibrium geometries. Its numerical stress subset contributed difficult configurations to the numerical statistics. A separate operational stress set was used to assess per-system completion in Section 5.3. Table 1 summarizes the samples and evaluation tasks.

\hypertarget{fs-doxcnCQNOOAA9GMCxbQtHJWL9cf}{}
\termremoved{Native} \termadded{The} xTB \termadded{program} and tblite provided the GFN2-xTB numerical references. We also compared the energies, forces and execution statuses returned by the xTBloom CPU and CUDA backends. The optional spin-polarization extension was disabled in every numerical comparison in this section. Open-shell inputs retained their supplied unpaired-electron counts under a restricted, common-orbital treatment. Results were aligned by system identifier. Program-specific model and numerical settings are given in SI S1.2, and the performance protocol is described in Section 4.2.

\hypertarget{fs-doxcnSaq5SOKg319xlg0tuTolMe}{}
For each system, we recorded input support, execution status and numerical-test outcomes. SCC nonconvergence, GPU memory exhaustion, timeouts, reference-calculation failures and implementation errors were recorded separately. Analytic forces were tested by finite differences. Controlled mixed batches tested the return of per-system results when one member failed to converge (SI S2.2).

\hypertarget{fs-doxcnulZmFocJ5wVAHogHqbpQ5b}{}
Each of the 20,000 numerical-test systems in Table 1 was evaluated with xTBloom CPU, \termremoved{native} xTB CPU, tblite CPU and xTBloom CUDA. For the CPU-program comparisons, the energy-difference threshold was \(5\times10^{-7}\) Eh. The force metric was the maximum absolute difference over all Cartesian components of each system, with a threshold of \(5\times10^{-6}\) Eh/bohr. The corresponding CPU--CUDA thresholds were \(1\times10^{-6}\) Eh and \(1\times10^{-6}\) Eh/bohr. CPU-program error distributions and joint pass rates used the common-success set of all three CPU programs. CPU--CUDA statistics used the intersection of successful results from the two backends.

\hypertarget{fs-doxcnNPxNYU6AjW9Ux3kkfjIqbe}{}
Operational coverage was tested separately on 2,000 difficult OMol25 configurations spanning open-shell, charged, metal-containing, large, environmental and reactive inputs (Table 1). Each engine evaluated these inputs individually. The execution environment and resource allocation are described in Section 4.2.

\begin{table}[H]
\begingroup
\fontsize{9.5}{12.8}\selectfont
\setlength{\tabcolsep}{3pt}
\renewcommand{\arraystretch}{1.18}
\setlength{\LTcapwidth}{\linewidth}
\hypertarget{fs-doxcn2SVdlKGuypCEEcuiSNdOKh}{}
\caption*{\hypertarget{fs-doxcnAuCY4rh7NINFt6EE0gTMge}{}Table 1. Datasets and numerical-validation tasks.}
\begin{tabular}{@{}>{\RaggedRight\arraybackslash}p{\dimexpr 0.203252\linewidth-1.219512\tabcolsep\relax}>{\RaggedRight\arraybackslash}p{\dimexpr 0.121951\linewidth-0.731707\tabcolsep\relax}>{\RaggedRight\arraybackslash}p{\dimexpr 0.337398\linewidth-2.024390\tabcolsep\relax}>{\RaggedRight\arraybackslash}p{\dimexpr 0.337398\linewidth-2.024390\tabcolsep\relax}@{}}
\toprule
\bfseries Dataset & \bfseries Systems or configu\-rations & \bfseries Coverage & \bfseries Evaluation metrics \\
\midrule
QM9 numerical set & 10,000 & Small neutral molecules & Three-program CPU differences, CPU--CUDA differences and execution status \\
OMol25 main numerical set & 8,000 & Multiple elements, charges, spins and nonequilibrium geometries & As above \\
OMol25 numerical stress subset & 2,000 & Difficult electronic states and geometries & As above \\
OMol25 performance set & 2,048 & Similar sizes, neutral singlets and metal-free light main-group elements & Full-queue time, throughput and output differences \\
OMol25 operational stress set & 2,000 & Open-shell, charged, metal-containing, large, environmental and reactive inputs & Per-system success rate and failure categories \\
\bottomrule
\end{tabular}
\endgroup
\end{table}

\FloatBarrier
\Needspace{5\baselineskip}
\hypertarget{fs-doxcnHHNeKmmr21ikLTHjCEB1Gd}{}
\subsection{Performance workloads and timing protocol}

\hypertarget{fs-doxcnN3bYlSvKo9QvA5ZTB37d4c}{}
The CPU backend uses the same model, parameters and result definitions as the CUDA backend and serves as a numerical reference and single-core performance baseline. The primary performance comparison evaluates the full implementation. Sorting and batch-size comparisons are described in Section 4.3.

\hypertarget{fs-doxcnyJK4chzGICCTX6qkzF5phe}{}
The performance workload comprised a fixed queue of 2,048 similarly sized OMol25 configurations. The inputs were neutral, singlet, metal-free light main-group systems with 100--119 GFN2 AOs. xTBloom sorted the queue by AO count and evaluated CUDA batches of \(B=256\). The CPU baselines used single-core, sequential per-system calls to xTBloom and \termremoved{native} xTB. We tested dxtb at \(B=1\), 64, 128 and 256. Its fastest configuration that completed the full queue, \(B=64\), was used for the primary comparison. The remaining configurations and complete results are reported in the SI.

\hypertarget{fs-doxcnUtPQmc7mp3tRiYuIJpuFGc}{}
All configurations evaluated double-precision GFN2-xTB energies and forces at an electronic temperature of 300 K, with an SCC limit of 500 iterations and a FRESH initial state. Each CPU configuration used one physical core, and each GPU configuration used one RTX 5090. xTBloom was built in Release mode. Software versions, build options, reference-program settings and sample composition are given in the SI.

\hypertarget{fs-doxcnM6i5aTtWT4lFtdVcDCgdaf}{}
The timing protocol specified one warmup traversal and three measured traversals of the full input queue for each configuration. We report the median primary evaluation time and its minimum--maximum range. Primary time includes per-batch adapter and descriptor preparation, required transfers, SCC and force calculation, synchronization and complete publication of host results. xTBloom preparation includes context creation and reuse handling. Input reading, AO sorting, Executor construction, destruction of the final retained context and compressed result storage fall outside the primary timing interval. Full-input throughput is the number of configurations divided by the primary time for the complete queue. Execution status, finite outputs and CPU--CUDA agreement were checked alongside timing. Configuration-specific timing scopes are defined in SI S4.2.\par\Needspace{5\baselineskip} The same-topology workload used an ideal geometry and locally perturbed configurations. For \(B=512\), each batch contained the ideal geometry and 511 configurations with independent Gaussian perturbations (0.020 bohr standard deviation per Cartesian coordinate). xTBloom CPU and CUDA were compared with sequential xTB/tblite calls under a one-physical-core host constraint. The primary configuration was \(B=512\) with a cold electronic state. The \(B=128\) auto-warm configuration examined consecutive evaluations. Configuration generation, hardware, core affinity and initialization protocols are given in SI S6.

\FloatBarrier
\Needspace{5\baselineskip}
\hypertarget{fs-doxcnDJoyFTDItDG0WZYnBAVHib}{}
\subsection{Ablation and configuration-sensitivity protocols}

\hypertarget{fs-doxcnWeSqFGNtVvjYWmZj3ygnBd}{}
The AO-sorting ablation compared the full implementation with original-order batching on the same fixed queue at \(B=256\). Both configurations retained grouping by exact AO dimension within each batch. All other computational and timing settings followed Section 4.2. Per-system results were aligned by their original identifiers to compare energies, forces, execution statuses and SCC iteration counts. A separate original-order scan at \(B=1\), 64 and 256 measured batch-size sensitivity.

\hypertarget{fs-doxcnSUsxdWcXu7Pvh9e5dI2Nae}{}
The primary AO-matrix layout comparison used the same inputs and software settings to test natural-size storage against storage padded to the largest AO dimension in each batch. Batch membership, actual-dimension eigensolves, precision and SCC settings were held fixed. Full energy-and-force evaluation times and independently sampled GPU memory were measured. The complete protocol is given in SI S5.4.

\hypertarget{fs-doxcnx6bbkGyrZe5Y41HItSFY3f}{}
The SCC-scheduling ablation compared an active-capacity dispatch chain with a fixed-capacity CUDA Graph using the same batch inputs, initial electronic states and SCC-loop execution counts. The AO and SCC-scheduling experiments used their respective workloads and settings, detailed in SI S5.1 and Table S5a. The scheduling results are reported in Section 5.2.2. An additional single-iteration SCC component comparison held the initial state and termination mask fixed while varying AO count, solver-bucket capacity and active fraction. Its protocol and results are given in SI S5.3.

\FloatBarrier
\Needspace{5\baselineskip}
\hypertarget{fs-doxcn8tqtGV0igDRP9OPjkrRN1e}{}
\subsection{QM/MM and DPRc free-energy simulations}

\hypertarget{fs-doxcnRwvciHof4ryaoHasT7Ptee}{}
The free-energy application used the ethylene phosphate model reaction with a methoxy leaving group studied by Giese et al~\citep{Giese_JPhysChemA_2022_v126_p8519}. The periodic system contained a 16-atom QM region comprising two complete molecules and 2,974 TIP4P-Ew water molecules~\citep{Horn_JChemPhys_2004_v120_p9665}. The QM region had a total charge of \(-2e\) and used closed-shell GFN2-xTB. We completed separate sampling calculations with uncorrected GFN2-xTB QM/MM and GFN2-xTB QM/MM+DPA4C-DPRc, retaining the same classical environment and constraints.

\hypertarget{fs-doxcnm5QElrVIp9tOI13vNDLNCh}{}
The reaction coordinate was \(\xi=r(\mathrm{P},\mathrm{O5}')-r(\mathrm{P},\mathrm{O2}')\), where O2\ensuremath{\prime} is the attacking oxygen and O5\ensuremath{\prime} is the leaving oxygen~\citep{Giese_JPhysChemA_2022_v126_p8519}. We placed 48 umbrella windows at 0.1 Å intervals from \(-3.1\) to 1.6 Å~\citep{Torrie_JComputPhys_1977_v23_p187,Kastner_WiresComputMolSci_2011_v1_p932}. Each window was equilibrated for 100 ps and then sampled in three 250 ps production trials with distinct random-number streams, giving 36 ns per Hamiltonian. LAMMPS~\citep{Thompson_ComputPhysCommun_2022_v271_p108171} propagated the dynamics with a 1 fs time step and a Langevin thermostat~\citep{Schneider_PhysRevB_1978_v17_p1302} at 298 K. Colvars~\citep{Fiorin_MolPhys_2013_v111_p3345} applied the biases, and SHAKE~\citep{Ryckaert_JComputPhys_1977_v23_p327} constrained the water geometry. A coordinator allowed all 48 windows to share one GPU for batched QM/MM evaluation.

\hypertarget{fs-doxcnHkI02daf4cPQh5FmQmUryc}{}
The DPRc dataset contained 6,560 periodic configurations labeled with the energy and force differences between PBE0/6-31G* QM/MM~\citep{Adamo_JChemPhys_1999_v110_p6158,Hariharan_TheorChimActa_1973_v28_p213} and GFN2-xTB QM/MM. The high-level calculations used AmberTools with QUICK~\citep{Case_JChemInfModel_2023_v63_p6183,Cruzeiro_JChemInfModel_2021_v61_p2109}. A DPA4C correction model was used for the DPRc production trajectories. Label construction, training, window-level data splits and held-out errors are described in SI S8.

\hypertarget{fs-doxcnkNnuhf0CB5RkuMawqqHtqd}{}
The weighted histogram analysis method (WHAM) combined the window samples to obtain the potential of mean force (PMF) for each Hamiltonian~\citep{Kumar_JComputChem_1992_v13_p1011}. The resulting profiles were compared with the digitized gwTP-PBE0 reference of Giese et al~\citep{Giese_JPhysChemA_2022_v126_p8519}. Statistical analysis and the comparison protocol are described in SI S9.

\hypertarget{fs-doxcn74zp80BO4hbCxbB8WTwxwc}{}
Application performance was measured in short synchronized runs on one RTX PRO 6000, comparing cross-window batches with sequential execution of the same windows. The short-run timings used the DPRc model available before retraining. Production sampling used the model retrained on 6,560 configurations. Training and data splits are described in SI S8, and the full timing protocol is given in SI S7.1.

\FloatBarrier
\Needspace{9\baselineskip}
\hypertarget{fs-doxcn4TFrHm8ok2XVlk1vO0sSlg}{}
\section{Results and discussion}

\FloatBarrier
\Needspace{5\baselineskip}
\hypertarget{fs-doxcniThIGbJQSGf6AkCUcNVzoc}{}
\subsection{Numerical agreement and analytic-force tests}

\hypertarget{fs-doxcnzMxabLwXQJPPtBdfWnWQuh}{}
Among the systems successfully evaluated by all three CPU programs, xTBloom and tblite jointly passed the energy and force criteria for 99.985\% of configurations. xTBloom's GFN2 parameters were generated from tblite data, parts of its numerical implementation used tblite as a reference, and the relevant numerical settings were aligned for this comparison. The xTBloom--xTB and tblite--xTB differences were similar in magnitude (Table 2a). Both comparisons involving \termremoved{native} xTB had lower joint pass rates for restricted open-shell inputs than for closed-shell inputs (Table S1f). Full error distributions, statistics stratified by closed-shell and restricted open-shell inputs, outlier configurations and detailed settings are reported in SI S1.

\begin{table}[H]
\begingroup
\fontsize{10}{13.5}\selectfont
\setlength{\tabcolsep}{3pt}
\renewcommand{\arraystretch}{1.18}
\setlength{\LTcapwidth}{\linewidth}
\hypertarget{fs-doxcn9BZK5GtMZm4OWeW6K659oc}{}
\caption*{\hypertarget{fs-doxcnedXDKzwXuggFra4a8CX0yb}{}Table 2a. The 95th percentiles of absolute differences over 19,897 systems successfully evaluated by all three CPU programs.}
\begin{tabular}{@{}>{\RaggedRight\arraybackslash}p{\dimexpr 0.359584\linewidth-1.438336\tabcolsep\relax}>{\RaggedRight\arraybackslash}p{\dimexpr 0.267459\linewidth-1.069837\tabcolsep\relax}>{\RaggedRight\arraybackslash}p{\dimexpr 0.372957\linewidth-1.491828\tabcolsep\relax}@{}}
\toprule
\bfseries Program pair & \bfseries Energy difference / Eh & \bfseries Maximum force-component difference / (Eh/\allowbreak{}bohr) \\
\midrule
xTBloom\textendash{}tblite & 2.26\ensuremath{\times}10\textsuperscript{-10} & 1.48\ensuremath{\times}10\textsuperscript{-8} \\
xTBloom\textendash{}xTB & 2.94\ensuremath{\times}10\textsuperscript{-5} & 6.39\ensuremath{\times}10\textsuperscript{-6} \\
tblite\textendash{}xTB & 2.94\ensuremath{\times}10\textsuperscript{-5} & 6.39\ensuremath{\times}10\textsuperscript{-6} \\
\bottomrule
\end{tabular}
\endgroup
\end{table}

\hypertarget{fs-doxcndvsB2Y60onVSjo39XPxdRL}{}
The CPU and CUDA backends jointly passed the energy and force criteria for 99.995\% of their common-success systems. Table 2b summarizes the numerical differences and execution statuses.

\begin{table}[H]
\begingroup
\fontsize{10}{13.5}\selectfont
\setlength{\tabcolsep}{3pt}
\renewcommand{\arraystretch}{1.18}
\setlength{\LTcapwidth}{\linewidth}
\hypertarget{fs-doxcny1OXsuNuo89hOWxHb3VKpe}{}
\caption*{\hypertarget{fs-doxcnuyrQnvak9aWiI1PZHd0Zfb}{}Table 2b. Numerical agreement and execution statuses of xTBloom CPU and CUDA.}
\begin{tabular}{@{}>{\RaggedRight\arraybackslash}p{\dimexpr 0.300000\linewidth-0.600000\tabcolsep\relax}>{\RaggedRight\arraybackslash}p{\dimexpr 0.700000\linewidth-1.400000\tabcolsep\relax}@{}}
\toprule
\bfseries Metric & \bfseries Result \\
\midrule
Common-success systems & 19,937 \\
Both energy and force criteria satisfied & 19,936 (99.995\%) \\
SCC nonconvergence in both backends & 37 \\
CPU success, CUDA failure & 13 \\
CUDA success, CPU failure & 13 \\
Maximum energy difference / Eh & 2.40\ensuremath{\times}10\textsuperscript{-2} \\
Maximum force-component difference / (Eh/bohr) & 2.90\ensuremath{\times}10\textsuperscript{-1} \\
\bottomrule
\end{tabular}
\endgroup
\end{table}

\hypertarget{fs-doxcnVSRKPRl09Vjj0C1gqF790d}{}
The CPU and CUDA results for the outlier configuration corresponded to two different self-consistent SCC solutions. Exchanging the converged terminal states as initial guesses allowed both backends to reproduce each solution branch. Energy and force differences within each branch satisfied the original comparison thresholds.

\hypertarget{fs-doxcnSq01MA9eeQdA6yehqmQFMd}{}
Both backends completed the finite-difference tests for all 32 QM9 systems and 29/32 OMol25 systems. Every examined force component of the completed systems satisfied the test threshold. The other 3 OMol25 systems terminated because of invalid input or SCC nonconvergence. The selected systems, errors and failure statuses are listed in the SI.

\FloatBarrier
\Needspace{5\baselineskip}
\hypertarget{fs-doxcncBj1Ch5j3M4PDf81LoxEdf}{}
\subsection{GPU performance}

\hypertarget{fs-doxcnMi83OPqCUr9Ko5ir3J0wvg}{}
We first evaluated the full implementation on a queue of distinct molecules and on batches of conformers with one topology. Ablations then examined the contributions and operating regimes of the execution strategies.

\Needspace{5\baselineskip}
\hypertarget{fs-doxcngm35MbnoVc014uToQVGC0o}{}
\subsubsection{Performance of the full implementation}

\hypertarget{fs-doxcnKniYevVudfjym3A8t3xdMc}{}
On the near-size OMol25 queue of distinct molecules, xTBloom CUDA at \(B=256\) completed all 2,048 configurations in each of the three measured traversals and achieved 71.13 configurations s\textsuperscript{-1} (Figure 3a). Primary evaluation time decreased from 83.802 s on its single-core CPU backend to 28.792 s on CUDA, corresponding to a 2.91-fold throughput increase. This comparison establishes a full-evaluation throughput benefit for the tested queue of similarly sized neutral singlets. AO sorting and batch-size effects are examined in Section 5.2.2.

\begin{figure}[!htbp]
\centering
\hypertarget{fs-doxcnNRuFhvgW7VIO9gFPQaU4Mg}{}
\hypertarget{fs-doxcnPok5PUYlaozBSreqAM68zb}{}
\includegraphics[width=\linewidth,height=.66\textheight,keepaspectratio]{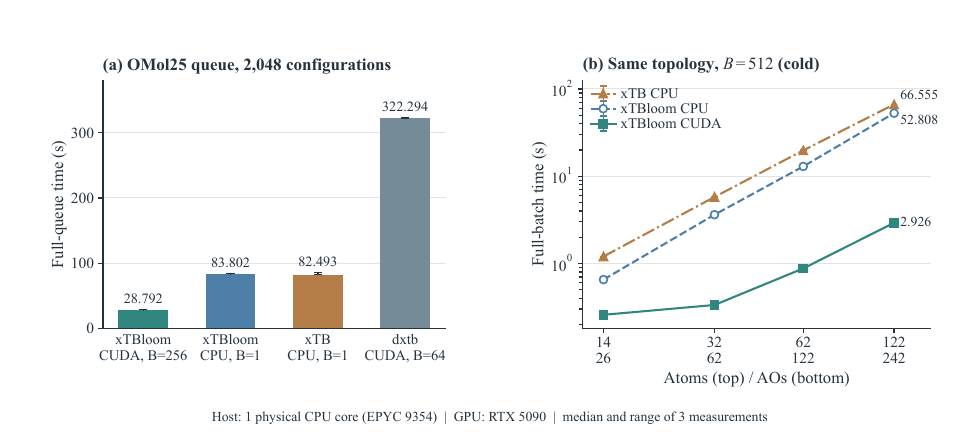}
\caption*{\hypertarget{fs-doxcndlmn2rOwfLPomSZrIDTdHe}{}\hypertarget{fs-doxcnLVwIVes05FpQnrIcWGB1ff}{}Figure 3. Energy-and-force evaluation on two workloads. (a) Primary evaluation time for the full near-size OMol25 queue of 2,048 configurations; batch sizes are shown in the panel. Timing includes per-batch preparation, transfers, calculation, synchronization and host-result publication; AO sorting was completed beforehand. (b) Full-batch time for 512 same-topology geometries at each molecular size, using a cold electronic state. Contexts and descriptors were prepared before timing; all energies and forces were available on the host when timing ended. xTB CPU evaluated the geometries sequentially. The atom-count and time axes in (b) are logarithmic. Bars and points show medians, and whiskers show minimum--maximum ranges over three full-queue traversals in (a) and three runs in (b). Each host process used one EPYC 9354 physical core; GPU runs used one RTX 5090. See SI S4.2 and S6 for timing protocols.}
\end{figure}

\hypertarget{fs-doxcnmv3q8Ivbissy1Wy7QqQMiI}{}
Under the stated numerical settings, primary times for single-core \termremoved{native} xTB and dxtb at \(B=64\) were 2.87 and 11.19 times the xTBloom CUDA time, respectively. The fastest complete dxtb configuration was \(B=64\), and its \(B=256\) runs exhausted GPU memory. Table S4b lists every configuration. Table S4c reports cross-program energy and force differences and sensitivity to a stricter \termremoved{native} xTB reference.

\hypertarget{fs-doxcny4BsqbhEcsXWOj3R1SlXMe}{}
Figure 3b gives the single-core comparison for locally perturbed conformers with a common topology. At \(B=512\) with cold electronic states, median xTBloom CUDA batch times were 0.876 and 2.926 s for the 62- and 122-atom systems. The corresponding ratios of sequential \termremoved{native} xTB time to CUDA time were 22.55 and 22.75. xTBloom CPU batch times were 12.942 and 52.808 s. Every path was restricted to one physical CPU core. The full size and batch-size curves are shown in Figure S4.

\Needspace{5\baselineskip}
\hypertarget{fs-doxcnDplvit3ocoKN5zrgUq03Gd}{}
\subsubsection{Ablation experiments}

\hypertarget{fs-doxcnj5NzxqALEqAcUUnQOMb7Ee}{}
We used controlled comparisons of AO sorting, primary AO-matrix storage and SCC active-capacity scheduling to examine how execution choices and batch size affect evaluation cost. AO sorting used the near-size queue from Section 5.2.1. SCC scheduling used controlled small-AO batches, followed by scans over AO count, solver-bucket capacity and active fraction to characterize its operating regime.

\Needspace{5\baselineskip}
\hypertarget{fs-doxcn0wo9wXHz0W09QBLjMyYoXg}{}
\paragraph*{AO sorting of the near-size queue}

\hypertarget{fs-doxcn3jadVQ86KiIFWgguAyT6Ne}{}
Improving batch composition through AO sorting reduced evaluation cost for the tested queue.%
\hypertarget{fs-doxcnBNnHI75oZucbIvzGbd4pxf}{} Using the full \(B=256\) configuration as the reference, we disabled AO sorting on the same near-size queue and tested the batch-size sensitivity of original-order batching (Figure 4 and Table S4b).

\hypertarget{fs-doxcnZWpFQfXpUY20EFz609ZS7c}{}
For fixed inputs and \(B=256\), primary time increased from 28.792 s with the full implementation to 38.094 s with AO sorting disabled. AO sorting therefore reduced primary evaluation time by 24.4\% relative to original-order batching on this queue.

\hypertarget{fs-doxcnaQAMH8U5Z5JQLo4t9VB6Ig}{}
With AO sorting disabled, original-order batches of 1, 64 and 256 achieved 6.97, 27.63 and 53.76 configurations s\textsuperscript{-1}, respectively. Throughput at \(B=1\) was below the single-core CPU baseline. Batches of 64 and 256 exceeded that baseline, showing the effect of per-batch workload on measured GPU throughput.

\hypertarget{fs-doxcnJh9Kso1yfY8UuNyHpfHSxc}{}
Across these configurations, the maximum differences from the first measured single-core CPU traversal were \(9.52\times10^{-13}\) Eh in energy and \(4.18\times10^{-13}\) Eh/bohr in force components. Per-configuration statuses, convergence flags and SCC iteration counts matched.

\begin{figure}[!htbp]
\centering
\hypertarget{fs-doxcn5XU6wFxNhKXdDfUpSIMm2g}{}
\includegraphics[width=.52\linewidth,height=.66\textheight,keepaspectratio]{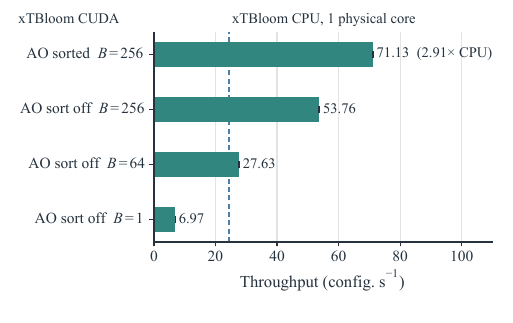}
\caption*{\hypertarget{fs-doxcngzabwgVUhOPOXy8TaIjFDb}{}Figure 4. Throughput in the AO-sorting ablation and batch-size scan. The first bar is the full AO-sorted \(B=256\) xTBloom CUDA implementation in Figure 3a; the others use original-order batches without AO sorting. Bars and labels show medians over three full-queue traversals; whiskers show minimum--maximum ranges. The dashed line is xTBloom CPU throughput on one physical CPU core. Inputs and timing follow Section 4.2; complete results are in Table S4b.}
\end{figure}

\hypertarget{fs-doxcn1b1Fg98G1Wse0Vi2DVWDbe}{}
A supplementary diagnostic used 256 wide-size OMol25 configurations that both CPU and GPU evaluations completed successfully. With FRESH electronic states and host inputs and outputs, median end-to-end time was 8.964 s for sequential single-core CPU evaluation and 29.525 s for original-order GPU batching at \(B=256\). This batch contained 100 exact-AO solver groups and had a maximum SCC count of 95 iterations.

\hypertarget{fs-doxcnx1XVkVIunoLn91tUFdzTzg}{}
At the same nominal batch size, grouping by exact AO count can produce many groups containing only a few systems. The batched solver interface submits matrices of a common dimension, so small groups provide limited parallel work compared with large same-topology groups. The long tail of SCC iteration counts can also extend batch completion time. The general masked path freezes the electronic-state updates of terminated members while retaining solver-group capacity, and the batch continues until its last active member finishes SCC. Full diagnostics and supplementary sorting comparisons are provided in the SI.

\Needspace{5\baselineskip}
\hypertarget{fs-doxcnexcmhzclqmFjScuYeIiflb}{}
\paragraph*{Storage and performance effects of the native ragged layout}

\hypertarget{fs-doxcn5cmAgWKMn8OFGqzmeQhOGc}{}
We separated the effects of batch organization and matrix storage by comparing natural-size storage with storage padded to the maximum AO count in each batch under the same input and software settings. Both paths retained batch membership, numerical kernels, actual-dimension eigensolves and exact-AO groups. Only the storage and indexing of the primary AO matrices changed. The comparison covered the near-size queue and both AO-sorted and strongly mixed batches of the same wide-size queue. Full energy-and-force evaluation time and GPU memory were measured independently, as detailed in SI S5.4.

\hypertarget{fs-doxcnOzZZxYv3GxjfRKdK6rSXKc}{}
Table 3 summarizes the three paired comparisons. For strongly mixed batches, the NVML-sampled peak decreased from 11,572 to 3,802 MiB, and natural-size storage was faster in all three full-evaluation traversals.

\begin{table}[H]
\begingroup
\fontsize{9.5}{12.8}\selectfont
\setlength{\tabcolsep}{3pt}
\renewcommand{\arraystretch}{1.18}
\setlength{\LTcapwidth}{\linewidth}
\hypertarget{fs-doxcnuzXcpWpEtcFHMiOoTMELTd}{}
\caption*{\hypertarget{fs-doxcncZVeX7ECyQCkUxD0ZuJWQb}{}Table 3. Changes in storage and evaluation time for native ragged storage relative to storage padded to the largest AO count in each batch.}
\begin{tabular}{@{}>{\RaggedRight\arraybackslash}p{\dimexpr 0.361111\linewidth-2.166667\tabcolsep\relax}>{\RaggedRight\arraybackslash}p{\dimexpr 0.208333\linewidth-1.250000\tabcolsep\relax}>{\RaggedRight\arraybackslash}p{\dimexpr 0.208333\linewidth-1.250000\tabcolsep\relax}>{\RaggedRight\arraybackslash}p{\dimexpr 0.222222\linewidth-1.333333\tabcolsep\relax}@{}}
\toprule
\bfseries Queue and batching (count, AO range) & \bfseries Primary AO-matrix capacity reduction / \% & \bfseries Sampled GPU-memory peak reduction / \% & \bfseries Full E+F time reduction / \% \\
\midrule
Near-size, AO-sorted (2,048, 100--119) & \added{2.16} & \added{1.03} & \added{0.19} \\
Wide-size, AO-sorted (1,317, 8--180) & \added{34.36} & \added{30.88} & \added{0.38} \\
Wide-size, strongly mixed (1,317, 8--180) & \added{75.90} & \added{67.14} & \added{4.02} \\
\bottomrule
\end{tabular}
\endgroup
\end{table}

\hypertarget{fs-doxcn5bkrwbI5dBqi0Zfmik5pen}{}
Reductions use padded storage as the baseline and are calculated as (padded minus native ragged)/padded. Evaluation times are medians of three full energy-and-force traversals. Absolute values and individual traversals are reported in Table S5e.

\hypertarget{fs-doxcnYQz0uiXspuoZvmlFexSrch}{}
For the tested wide-size queue, native ragged storage reduced redundant matrix capacity and GPU-memory demand.

\Needspace{5\baselineskip}
\hypertarget{fs-doxcnbwPQFoc74Hoq4KVLcAymwc}{}
\paragraph*{Scheduling the SCC active set}

\hypertarget{fs-doxcnC4FFGPWxjf8Zm0vTnf8Qkc}{}
Reducing the capacity of subsequent active solves can reduce redundant work. The resulting performance benefit depends on whether the saved computation offsets reorganization and dispatch overhead.%
\hypertarget{fs-doxcnk1D9rBp0kCZbXQBLGHIKfd}{} To test execution scheduling above the per-system SCC states, we compared a device dispatch chain that submits solves at the active capacity with a CUDA Graph that retains the full capacity on controlled small-AO workloads. The paths used the same batch inputs and initial states. Both retained per-system convergence decisions, mixing histories, termination statuses and result mappings.

\hypertarget{fs-doxcnSirP5pJ7zPlUTbQF6tWoQ2}{}
The test used mixed batches of H\textsubscript{2}, He, LiH and CH\textsubscript{2} at \(B=1\), 8, 32 and 128. Fully, half- and quarter-active initial states were constructed by setting selected systems' initial iteration counters to the limit. The paths executed the same number of SCC loops and matched in final energies, densities, populations, iteration counts and statuses. Initially fully active configurations were also compared with a sequential CPU reference. Figure 5 shows the relative latencies.

\hypertarget{fs-doxcn2Eeho8FtIqMN6Esh7t308g}{}
For partially active batches, the dispatch chain reduced subsequent solve costs associated with terminated slots. At \(B=32\), the fixed-capacity/dispatch time ratios were 1.29 and 1.53 for the half- and quarter-active configurations. Within this controlled workload and at a fixed batch size, lower initial active fractions produced a larger relative benefit from reducing the submitted solver capacity.

\hypertarget{fs-doxcnGJfftKGaXOHbI33LoewVFc}{}
The initially fully active paths had similar latencies. Fixed-capacity/dispatch time ratios of 0.95 at \(B=128\) and 0.96 at \(B=1\) indicate slightly slower active-capacity dispatch. These configurations offer less redundant solve work to remove, making reorganization and dispatch costs more prominent. The extended scan showed that the benefit depends jointly on AO count, solver-bucket capacity and active fraction. For AO dimensions of 122 and 180, \(B=256\) and 1--2 active members, single-iteration component latency decreased by approximately 16.2\%--27.0\%. Reorganization and dispatch costs dominated most other tested points (SI S5.3).

\begin{figure}[!htbp]
\centering
\hypertarget{fs-doxcnnJ9iIzYcvKaWZ7IveN7k3d}{}
\includegraphics[width=.72\linewidth,height=.66\textheight,keepaspectratio]{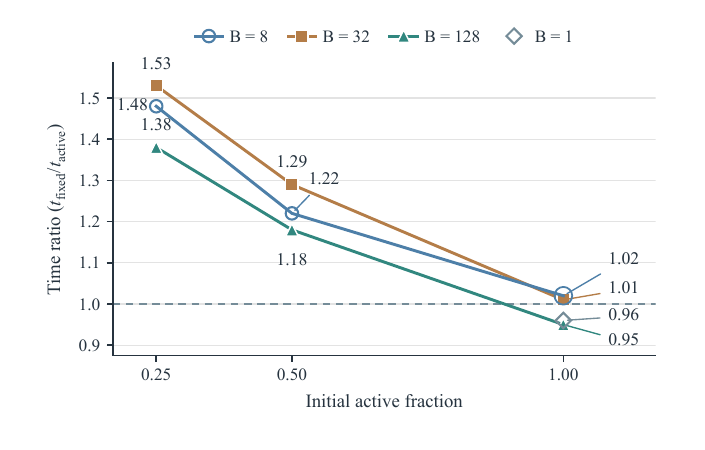}
\caption*{\hypertarget{fs-doxcnnAfsKc8TIWw8jdXUDYV7Ah}{}Figure 5. Relative SCC-loop latency of the active-capacity dispatch chain and fixed-capacity CUDA Graph at different initial active fractions. The ordinate is mean Graph time divided by mean dispatch time; values above 1 favor dispatch. \(B\) denotes batch size; only the initially fully active case was measured at \(B=1\). Each configuration was measured 50 times. Timing covers the SCC loop through batch termination and excludes input preparation, result downloads and force calculation. The dashed line marks equal time. Details are given in SI S5.1 and Table S5a.}
\end{figure}

\FloatBarrier
\Needspace{5\baselineskip}
\hypertarget{fs-doxcnSuky9jIws1KMZo9tVCOdfe}{}
\subsection{Per-system failure isolation and operational coverage of difficult inputs}

\hypertarget{fs-doxcnXoENfFfHGqLOBO5XXsh6Mc}{}
Successful results remained independently available when another batch member failed.%
\hypertarget{fs-doxcnBTOb3Y51ukrOVxlp2aPMKb}{} A healthy system and an SCC-nonconvergent system were evaluated together on the CPU host path and the CUDA mixed-memory path. The failed member returned a per-system failure status, with its energy, force and charge outputs filled with NaNs. Differences between separate and joint evaluations of the healthy member satisfied the predefined tolerances. Settings and results are given in SI S2.2.

\hypertarget{fs-doxcnciVPLDGlzNS4cDdEKPlxKf}{}
We further evaluated operational coverage on the independent set of 2,000 difficult OMol25 configurations using per-system runs. xTBloom CUDA completed 1,981 systems, a success rate of 99.05\%. The other 19 terminated through SCC nonconvergence, and the GPU-memory-exhaustion count was zero. Figure 6b shows failure categories for the comparison programs. Full status counts are reported in Table S3b.

\begin{figure}[!htbp]
\centering
\hypertarget{fs-doxcnOuimuJ2JgRRurNMZ5WCMeb}{}
\includegraphics[width=\linewidth,height=.66\textheight,keepaspectratio]{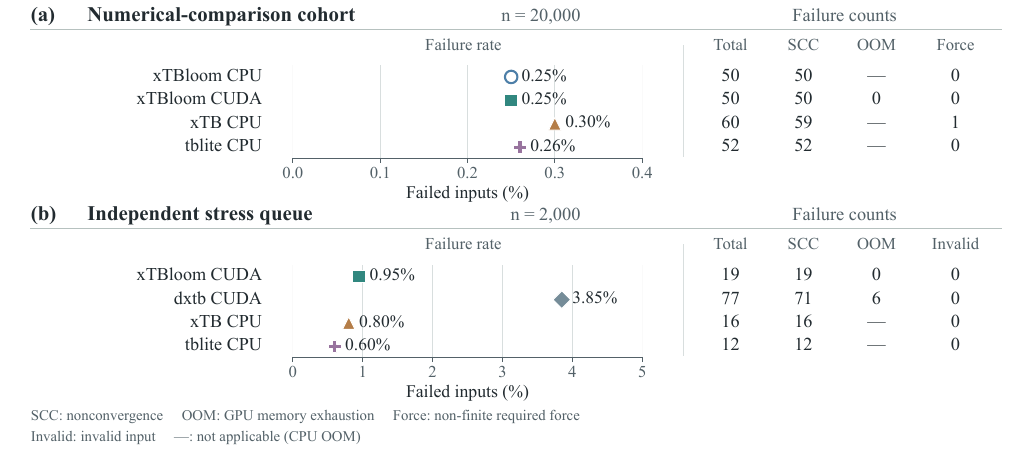}
\caption*{\hypertarget{fs-doxcnOEMBu3g5iMu7WoKaA294Te}{}Figure 6. Failure rates and counts in two input sets. (a) The numerical-comparison cohort (\(n=20{,}000\)); (b) an independent set of 2,000 difficult OMol25 configurations. Points show the percentage of failed inputs, and columns give total failures and recorded categories. SCC denotes self-consistent charge nonconvergence; OOM denotes GPU memory exhaustion. Force in (a) denotes a non-finite required force, and Invalid in (b) denotes invalid input. A dash marks GPU OOM as not applicable to CPU runs. The horizontal axes use different ranges. Dataset composition is listed in Table 1; category counts are in SI Tables S1e(d) and S3b.}
\end{figure}

\hypertarget{fs-doxcnGHWps2AB3fozqJ8IrUyVrb}{}
These tests characterize the completion coverage and failure composition of each engine on difficult inputs. xTBloom failures in this stress set were associated with the SCC convergence tail. dxtb also encountered GPU memory exhaustion, indicating that electronic-state behavior and execution resources both affect the observed failure categories.

\FloatBarrier
\Needspace{5\baselineskip}
\hypertarget{fs-doxcniFv6xF3BwGJ3FIVQ5h8FEf}{}
\subsection{QM/MM application results and cross-window acceleration}

\hypertarget{fs-doxcn9zYJJpP8yujHhmIzIJsX06}{}
The LAMMPS integration used single-GPU cross-window batches to complete both the uncorrected GFN2-xTB QM/MM and DPA4C-DPRc umbrella-sampling calculations. Each calculation comprised 48 windows, with three 250 ps production trials per window and 36 ns of aggregate production sampling.

\hypertarget{fs-doxcn6K88nwHXjTH1TWofseXPmO}{}
Single-window and 48-window evaluations of the same configuration agreed in energy and forces, with a maximum force-component difference of approximately \(4.4\times10^{-11}\) kcal mol\textsuperscript{-1} Å\textsuperscript{-1}. The DPRc production model passed 5 ps microcanonical (NVE) checks at three anchor windows both before and after equilibration. Together with the two completed sampling calculations, these checks show that batch electronic-structure evaluation can be integrated with periodic embedding, a learned correction potential and dynamical propagation.

\hypertarget{fs-doxcn2uWJjQhGlTZlMEVDsdq0og}{}
Figure 7a shows the central free-energy profiles. After each curve was independently aligned to \(A(0)=0\), the root-mean-square errors (RMSEs) relative to the digitized gwTP-PBE0 reference~\citep{Giese_JPhysChemA_2022_v126_p8519} were 6.61 kcal/mol for bare QM/MM and 0.866 kcal/mol for the DPRc-corrected profile over 233 shared grid points. The two principal DPRc peak-to-reactant free-energy differences were 9.21 and 16.45 kcal/mol as descriptive central values, differing from the reference centers by +0.17 and \(-0.33\) kcal/mol. A local difference of approximately 3.1 kcal/mol remained near \(\xi\approx1.53\) Å. Statistical diagnostics are given in SI S9.

\begin{figure}[H]
\centering
\hypertarget{fs-doxcnxWBmA0IGpbrYycT0T68hoe}{}
\includegraphics[width=\linewidth,height=.66\textheight,keepaspectratio]{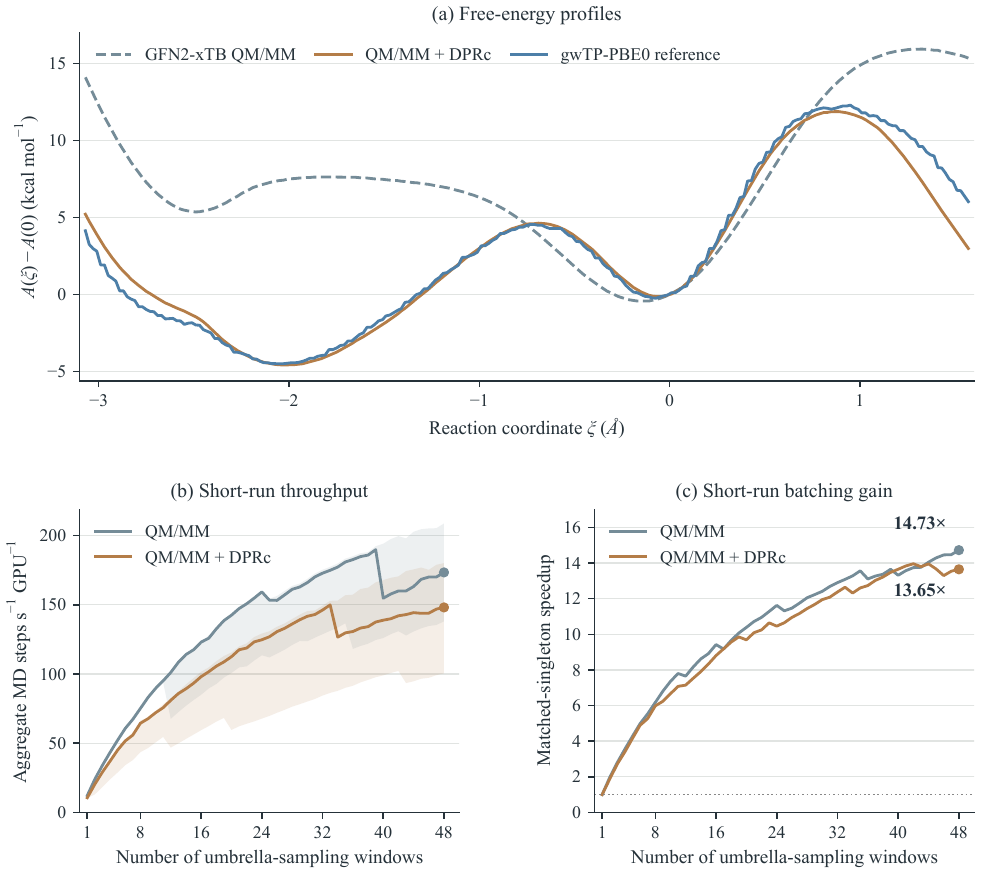}
\caption*{\hypertarget{fs-doxcn6LLydc6zl2fyzgNuefM7ld}{}Figure 7. QM/MM free-energy profiles and cross-window performance. (a) Central profiles for bare GFN2-xTB QM/MM, the retrained DPRc model and the digitized gwTP-PBE0 reference, each aligned to \(A(0)=0\). (b,c) Single-GPU short-run timings as a function of the number of umbrella-sampling windows. (b) Aggregate throughput; lines and shading show the median and minimum--maximum range over five consecutive segments of the same trajectory. (c) Speedup in total elapsed time relative to sequential execution of the same windows. The comparison grid, protocol differences and statistical scope are defined in SI S9.3; timing scopes are given in SI S7.1.}
\end{figure}

\hypertarget{fs-doxcnxdwdta8XB2NaL1zQg9I00c}{}
Application performance was measured in the single-GPU short synchronized runs (Figure 7b,c). At 48 windows, QM/MM and QM/MM+DPRc achieved aggregate throughputs of 173.35 and 148.06 MD steps s\textsuperscript{-1} GPU\textsuperscript{-1}. The corresponding speedups over sequential execution of the same windows were 14.73 and 13.65, reflecting the full coupled LAMMPS batching scheme. The three production trials recorded 131.08--133.23 MD steps s\textsuperscript{-1} GPU\textsuperscript{-1}, calculated from the sums of production-segment wall clocks within each trial. The two timing scopes are defined in SI S7.1.

\FloatBarrier
\Needspace{9\baselineskip}
\hypertarget{fs-doxcnAW4g8jCYjkoyruq8Oonmce}{}
\section{Conclusions}

\hypertarget{fs-doxcnOIuKUFEDz72gq6xm7B9V2f}{}
xTBloom organizes independent GFN-xTB energy-and-force evaluations into GPU batches for repeated electronic-structure workloads. The growing demand from molecular simulation, conformational sampling and scientific data generation makes sustained execution of these independent tasks an important throughput problem. Batch electronic-structure evaluation provides a route for connecting these calculations to larger molecular-simulation workloads.

\hypertarget{fs-doxcnIVwad0RB8mmN3dwXyMSIkg}{}
The implementation combines AO-sorted task organization with per-system SCC histories and failure states, compatible data reuse and common simulation interfaces. Under the tested restricted common-orbital settings and common-success definitions, xTBloom showed close agreement with tblite and between its CPU and CUDA backends. Comparisons involving \termremoved{native} xTB contained threshold exceedances. On the representative near-size OMol25 workload, full GPU evaluation achieved approximately 2.9 times its single-core CPU throughput. The execution studies show that the benefit depends on batch composition, convergence behavior and active fraction through the combined costs of task organization, SCC scheduling and reuse.

\hypertarget{fs-doxcn4Ic2QmPlAwesXGBqXp4Nzg}{}
Integration with LAMMPS supported periodic QM/MM dynamics and completed umbrella sampling with both GFN2-xTB and DPA4C-DPRc correction. The free-energy calculations demonstrate application of the common batch interface to the baseline and corrected potentials. In matched short synchronized tests with 48 windows on one GPU, cross-window batching increased aggregate throughput by approximately an order of magnitude over sequential execution of the same windows. This integration extends the measured benefit to a simulation workflow requiring repeated electronic-structure evaluations.

\hypertarget{fs-doxcnwadKP3bCOgIcWbffowJ5xb}{}
Overall, this work demonstrates the potential of GPU batching to improve throughput for GFN-xTB energy-and-force evaluation and provides a computational basis for future larger-scale QM/MM simulations incorporating learned correction potentials.

\FloatBarrier
\Needspace{9\baselineskip}
\hypertarget{fs-doxcnnVYvfAaefHKVMWIQq3rgGh}{}
\section*{Code Availability}

\hypertarget{fs-doxcnv5xjA65cJKxoP78BQWFo3c}{}
Source code, build instructions and interface documentation are available from the xTBloom repository at \nolinkurl{https://github.com/jinzhezenggroup/xtbloom}. The cross-window coordinator and periodic QM/MM coupling are available from the LAMMPS-DPRc repository at \nolinkurl{https://github.com/jinzhezenggroup/lammps-dprc}.

\FloatBarrier
\begin{acknowledgement}
J.Z. was supported by the Advanced Materials - National Science and
Technology Major Project (No.~2025ZD0618702) and the National Natural
Science Foundation of China (Grant No.~22603140).
T.Z. was supported by the National Natural Science Foundation of China
(Grant Nos.~92461313 and 22361132538), the AI for Science Program of the
Shanghai Municipal Commission of Economy and Informatization (Grant
No.~2025-GZL-RGZN-BTBX-01004), and the Open Research Fund of Suzhou
Laboratory (No.~SZLAB-1508-2024-TS017).
Q.G. was supported by the National Natural Science Foundation of China
(Grant No.~12504285), the Advanced Materials - National Science and
Technology Major Project (No.~2025ZD0618401), and the Natural Science
Foundation of Jiangsu Province (Grant No.~BK20250472).
J.Z. and Q.G. are both supported by the Fundamental Research Funds for the Central Universities.
The AI-driven experiments, simulations, and model training were performed on the robotic AI-Scientist platform of the Chinese Academy of Sciences.
The numerical calculations were performed on
the supercomputing system in the Supercomputing Center of the University of Science and
Technology of China.
\end{acknowledgement}

\begin{suppinfo}
Benchmark data and numerical settings; GPU and single-core CPU timing
protocols; batch scheduling, failure handling, and memory usage; periodic
QM/MM umbrella sampling and free-energy analysis; and data sources.
\end{suppinfo}

\nocite{Chodera_JChemTheoryComput_2016_v12_p1799,Flyvbjerg_JChemPhys_1989_v91_p461,Geyer_StatSci_1992_v7_p473}
\bibliography{zenglib}

\begin{mcitethebibliography}{64}
\bibitem[Lin and Truhlar(2007)Lin, and
  Truhlar]{Lin_TheorChemAcc_2007_v117_p185}
Lin,~H.; Truhlar,~D.~G. {QM/MM: what have we learned, where are we, and where
  do we go from here?} \emph{Theor Chem Acc} \textbf{2007}, \emph{117},
  185--199\relax
\mciteBstWouldAddEndPuncttrue
\mciteSetBstMidEndSepPunct{\mcitedefaultmidpunct}
{\mcitedefaultendpunct}{\mcitedefaultseppunct}\relax
\EndOfBibitem
\end{mcitethebibliography}

\begin{tocentry}
\centering
\includegraphics[width=\linewidth,height=4.25cm,keepaspectratio]{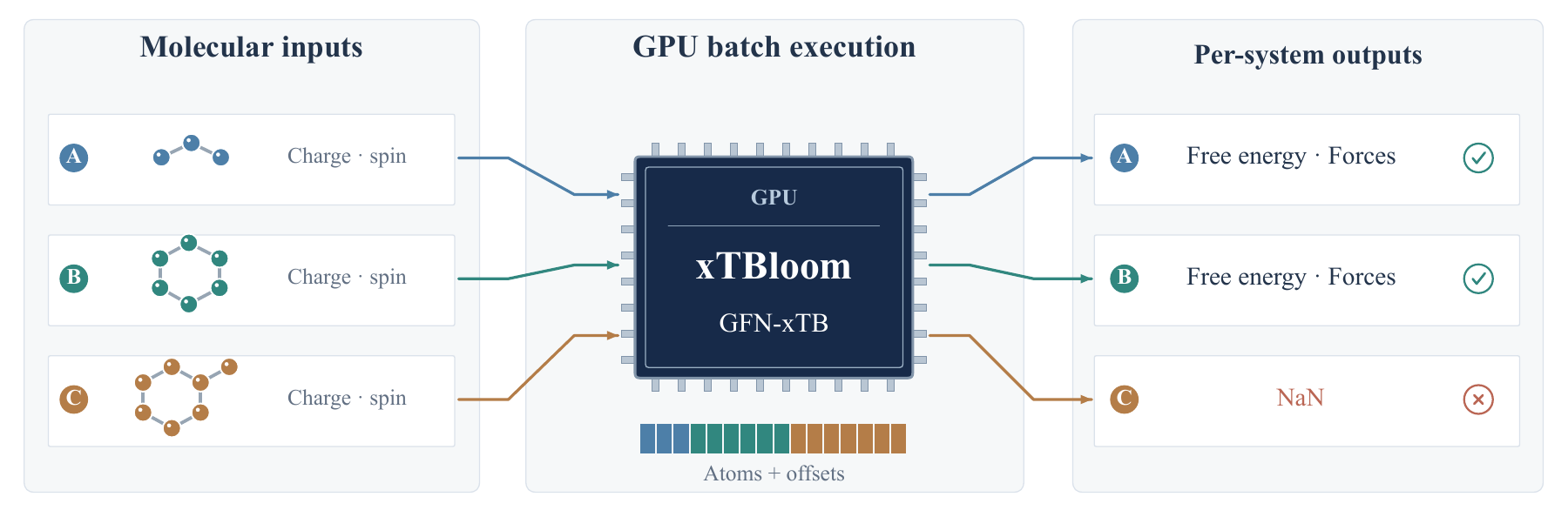}
\end{tocentry}
\end{document}


\renewcommand{\theequation}{S\arabic{equation}}
\hypertarget{fs-doxcnWC7QhqDyOqX2v7XCHebsVc}{}
This Supporting Information describes the sample composition, numerical tests, performance configurations, execution implementation, QM/MM simulations, and statistical methods used in the main text. References are numbered independently in this document. Summary values are reported with units and appropriate significant figures. Raw timing measurements retain their full precision, and all statistics and pass criteria are evaluated using unrounded data.

\FloatBarrier
\Needspace{9\baselineskip}
\hypertarget{fs-doxcn8EfMQlyWVZfAN8iEX3Vn5f}{}
\section*{S1. Datasets and numerical comparisons}
\addcontentsline{toc}{section}{S1. Datasets and numerical comparisons}

\FloatBarrier
\Needspace{8\baselineskip}
\hypertarget{fs-doxcnuTIVlcFAY0gFOjI0MmXQzh}{}
\subsection*{S1.1 Sample composition}
\addcontentsline{toc}{subsection}{S1.1 Sample composition}

\hypertarget{fs-doxcnyuCveCrvsRVCaiI1GSt4je}{}
The numerical comparison used QM9 and OMol25 data~\citep{Ramakrishnan_SciData_2014_v1_p140022,Levine_arXiv_2025_p2505.08762}. We selected 10,000 QM9 configurations, 8,000 configurations from the main OMol25 set and 2,000 configurations from an OMol25 numerical stress subset, giving 20,000 unique inputs. The numerical stress subset is included in the aggregate statistics in this section. The execution stress test in S3 is documented separately. Table S1a summarizes the sample composition.

\hypertarget{fs-doxcn2z4Wf9Jp9SBa4g7g8XQRkB}{}
This section reports sample composition and unweighted statistics.

\begingroup
\fontsize{9.2}{12.4}\selectfont
\setlength{\tabcolsep}{3pt}
\renewcommand{\arraystretch}{1.18}
\setlength{\LTcapwidth}{\linewidth}
\hypertarget{fs-doxcncMEJAVw7wl4VLS5BFyqFMc}{}
\begin{longtable}{@{}>{\RaggedRight\arraybackslash}p{\dimexpr 0.200000\linewidth-2.000000\tabcolsep\relax}>{\RaggedRight\arraybackslash}p{\dimexpr 0.105263\linewidth-1.052632\tabcolsep\relax}>{\RaggedRight\arraybackslash}p{\dimexpr 0.105263\linewidth-1.052632\tabcolsep\relax}>{\RaggedRight\arraybackslash}p{\dimexpr 0.105263\linewidth-1.052632\tabcolsep\relax}>{\RaggedRight\arraybackslash}p{\dimexpr 0.242105\linewidth-2.421053\tabcolsep\relax}>{\RaggedRight\arraybackslash}p{\dimexpr 0.242105\linewidth-2.421053\tabcolsep\relax}@{}}
\caption*{\hypertarget{fs-doxcnboxzKOsTzbiPZaRoAf8Eeb}{}Table S1a. Sample composition for the numerical comparison.}\\
\toprule
\bfseries Data group & \bfseries Configu\-rations & \bfseries Closed-shell & \bfseries Open-shell & \bfseries Atoms, median/\allowbreak{}maximum & \bfseries AOs, median/\allowbreak{}maximum \\
\midrule
\endfirsthead
\toprule
\bfseries Data group & \bfseries Configu\-rations & \bfseries Closed-shell & \bfseries Open-shell & \bfseries Atoms, median/\allowbreak{}maximum & \bfseries AOs, median/\allowbreak{}maximum \\
\midrule
\endhead
\midrule
\multicolumn{6}{r}{\footnotesize Continued on next page}\\
\endfoot
\bottomrule
\endlastfoot
QM9 main set & 10,000 & 10,000 & 0 & 18 / 29 & 45 / 56 \\
OMol25 main set & 8,000 & 6,439 & 1,561 & 41 / 350 & 114 / 884 \\
OMol25 numerical stress subset & 2,000 & 1,428 & 572 & 61 / 329 & 173.500 / 957 \\
Total & 20,000 & 17,867 & 2,133 & 19 / 350 & 46 / 957 \\
\end{longtable}
\endgroup

\hypertarget{fs-doxcnjcOXrHePwe7TSD8pMs5fXe}{}
An input was classified as open-shell when its specified number of unpaired electrons was greater than zero. The statistics for the selected samples are unweighted.

\hypertarget{fs-doxcnLZi7dqXXgkImHhncgMVqNe}{}
Each input was evaluated with xTBloom CPU, native xTB CPU~\citep{Bannwarth_WiresComputMolSci_2021_v11_pe1493}, tblite CPU~\citep{Katbashev_JPhysChemA_2025_v129_p2667} and xTBloom CUDA, producing 80,000 execution records. Failed runs are retained in the status statistics for all inputs. Numerical differences are calculated on the corresponding common-success sets.

\FloatBarrier
\Needspace{8\baselineskip}
\hypertarget{fs-doxcnhzFjVNsZAzLgSCOoq3Il67}{}
\subsection*{S1.2 Calculation settings and difference statistics}
\addcontentsline{toc}{subsection}{S1.2 Calculation settings and difference statistics}

\hypertarget{fs-doxcnOHBoLK8iTDybuL3p3oiZ1c}{}
The comparison used the nonperiodic molecular GFN2-xTB model~\citep{Bannwarth_JChemTheoryComput_2019_v15_p1652}. Spin-polarization extensions were disabled in every program. xTBloom CPU/CUDA used restricted mode with an explicit single-channel representation, and tblite was used without a spin-polarization container. Open-shell inputs retained their specified charge, multiplicity and nonzero number of unpaired electrons and were treated with restricted open-shell common orbitals. The compared quantities were the electronic free energy at finite electronic temperature and the analytic forces.

\begingroup
\fontsize{10}{13.5}\selectfont
\setlength{\tabcolsep}{3pt}
\renewcommand{\arraystretch}{1.18}
\setlength{\LTcapwidth}{\linewidth}
\hypertarget{fs-doxcnaqYvFCU6wZ5LjiEUisrrEg}{}
\begin{longtable}{@{}>{\RaggedRight\arraybackslash}p{\dimexpr 0.300000\linewidth-0.600000\tabcolsep\relax}>{\RaggedRight\arraybackslash}p{\dimexpr 0.700000\linewidth-1.400000\tabcolsep\relax}@{}}
\caption*{\hypertarget{fs-doxcnEv9womEdBrqVGPLNxO2VMg}{}Table S1b. Settings for the numerical comparison of 20,000 configurations.}\\
\toprule
\bfseries Item & \bfseries Setting \\
\midrule
\endfirsthead
\toprule
\bfseries Item & \bfseries Setting \\
\midrule
\endhead
\midrule
\multicolumn{2}{r}{\footnotesize Continued on next page}\\
\endfoot
\bottomrule
\endlastfoot
Model and program versions & \added{GFN2-xTB, native xTB 6.7.1, tblite 0.7.0(C API 700)}\added{.} \\
Spin polarization & Disabled in all programs in this section. No spin-polarization container was attached to tblite. xTBloom CPU/CUDA used restricted mode with an explicit single-channel representation. \\
Common-orbital occupations & N\ensuremath{\alpha}=(N\textsubscript{e}+u)/2 and N\ensuremath{\beta}=(N\textsubscript{e}\ensuremath{-}u)/2. Two sets of Fermi\textendash{}Dirac occupations are determined on the same spatial orbital spectrum and summed to construct the total density. The input specifies the number of unpaired electrons u. Each program retains its own numerical occupation solver. \\
GFN2 parameter source & xTBloom GFN2 parameters were generated from the structured export of tblite 0.7.0. Native xTB used the parameters built into version 6.7.1. \\
D4 parameter source & The xTBloom D4 fitting and damping coefficients are included in the tblite export described above. The D4 reference polarizability/C6 tables were generated with dftd4. Program-specific cutoffs are given below. \\
Initial electronic guess & Each system was cold-started independently, with no wavefunction reuse between systems. xTBloom FRESH used SAD, and tblite used its default SAD guess. Native xTB 6.7.1 retained its default Goedecker/EEQ guess and its SAD fallback on failure. Each program used its own initial-guess settings. \\
Electronic temperature & 300 K, kBT = 9.500\ensuremath{\times}10\textsuperscript{-4} E\textsubscript{h} \\
Maximum SCC iterations & 500 \\
xTBloom free-energy-change/state-residual RMS tolerances & 10\textsuperscript{-10} E\textsubscript{h} / 10\textsuperscript{-8} \\
xTB and tblite accuracy & 10\textsuperscript{-4} \\
xTBloom execution & B = 32, host input/output, FRESH initial state \\
Reference-program execution & Per-system calls, one CPU thread, single-threaded BLAS \\
Cross-program comparison thresholds & |\ensuremath{\Delta}E| \ensuremath{\leq} 5\ensuremath{\times}10\textsuperscript{-7} E\textsubscript{h}, max|\ensuremath{\Delta}F| \ensuremath{\leq} 5\ensuremath{\times}10\textsuperscript{-6} E\textsubscript{h} bohr\textsuperscript{-1} \\
CPU\textendash{}CUDA comparison thresholds & |\ensuremath{\Delta}E| \ensuremath{\leq} 10\textsuperscript{-6} E\textsubscript{h}, max|\ensuremath{\Delta}F| \ensuremath{\leq} 10\textsuperscript{-6} E\textsubscript{h} bohr\textsuperscript{-1} \\
D4 cutoffs & Hard molecular cutoffs in xTBloom/tblite, with tblite switch width = 0 bohr \\
Numerical libraries and interpreter & LP64 OpenBLAS 0.3.30 and Python 3.12.3. tblite used its separately bundled numerical library. \\
GPU and toolchain & NVIDIA RTX 5090, driver 580.95.05, CUDA 12.9.1, nvcc 12.9.86, sm120, GCC 13.3 \\
\end{longtable}
\endgroup

\hypertarget{fs-doxcneCWNYLTIYS48Vfw5fi03Kh}{}
The D4 reference data were generated with dftd4~\citep{Caldeweyher_JChemPhys_2019_v150_p154122} and used in the corresponding parameter tables.

\hypertarget{fs-doxcng3wewMZL1bqqFwH0RmXicl}{}
The molecular D4 coordination-number, two-body and three-body cutoff radii in xTBloom and tblite were 30, 50 and 25 bohr, respectively. The coordination-number and three-body cutoff radii in native xTB were both 40 bohr. The two-body path in its SCC dispersion-matrix routine \code{build_wdispmat} used a radius of 72 bohr. Each program retained its own SCC residual definition.

\hypertarget{fs-doxcnDkDyN9AB9uSX0l9le0UfFg}{}
For the common-orbital, explicit single-channel GFN2-xTB path used here, the SCC state vector contains all shell charges, followed by 3 dipole components and 6 stored quadrupole components per atom. Charges, dipoles and quadrupoles use e, e a\textsubscript{0} and e a\textsubscript{0}\textsuperscript{2}, respectively. The state vector stores the numerical values obtained by dividing each quantity by its atomic unit. With Nsh shells and Nat atoms, its dimension is n=Nsh+9Nat.

\hypertarget{fs-doxcn7SvnsbwzyB4DLdyiTwGFpb}{}
\begin{equation}
\boldsymbol{\mathbf r^{(k)}=\mathbf x_{\mathrm{out}}^{(k)}-\mathbf x_{\mathrm{in}}^{(k)},\qquad R_{\mathrm{RMS}}^{(k)}=\sqrt{\frac{1}{n}\sum_{j=1}^{n}(r_j^{(k)})^2}.}
\end{equation}

\hypertarget{fs-doxcnmguw064l138rwqjD0aObPd}{}
The input state constructs the Hamiltonian for the current iteration. The output state is obtained from that iteration's orbital occupations and density-matrix update. The residual is calculated before mixing, with equal weight for every stored component. No additional field scaling or off-diagonal multiplicity weighting is applied. The public API field \code{charge_tolerance} specifies the RMS threshold for this complete state vector, and \code{energy_tolerance} specifies the threshold for the change in the complete electronic free energy. The stopping conditions are

\hypertarget{fs-doxcnyVoLMbxsSptZMFWdAKc4Nh}{}
\begin{equation}
\boldsymbol{R_{\mathrm{RMS}}^{(k)}<\varepsilon_R,\qquad |\mathcal F^{(k)}-\mathcal F^{(k-1)}|<\varepsilon_F.}
\end{equation}

\hypertarget{fs-doxcnE7to7CM8V0Uc1tKZKQ5Upg}{}
Convergence requires both conditions to hold. The previous free-energy value is initialized to zero for the first iteration and subsequently taken from the preceding iteration. These definitions agree with Table S1b. CPU and CUDA use the same final stopping conditions.

\hypertarget{fs-doxcn6AnPXUEJIJaKavYrHHHRzb}{}
The energy difference is the absolute difference in total energy. For each configuration, the force difference is first reduced to the largest absolute difference over all 3N Cartesian components. P50, P95, P99 and the maximum are then calculated across configurations. Quantiles use linear interpolation. A joint pass requires both the energy and force thresholds to be met. All three CPU program pairs use the same set of 19,897 configurations for which all three programs succeeded. The CPU\textendash{}CUDA comparison uses 19,937 configurations for which both backends succeeded.

\begingroup
\fontsize{8.8}{11.9}\selectfont
\setlength{\tabcolsep}{3pt}
\renewcommand{\arraystretch}{1.18}
\setlength{\LTcapwidth}{\linewidth}
\hypertarget{fs-doxcnB5TTMUraz56yLwRpBC72dh}{}
\begin{longtable}{@{}>{\RaggedRight\arraybackslash}p{\dimexpr 0.162679\linewidth-1.952153\tabcolsep\relax}>{\RaggedRight\arraybackslash}p{\dimexpr 0.148325\linewidth-1.779904\tabcolsep\relax}>{\RaggedRight\arraybackslash}p{\dimexpr 0.148325\linewidth-1.779904\tabcolsep\relax}>{\RaggedRight\arraybackslash}p{\dimexpr 0.148325\linewidth-1.779904\tabcolsep\relax}>{\RaggedRight\arraybackslash}p{\dimexpr 0.148325\linewidth-1.779904\tabcolsep\relax}>{\RaggedRight\arraybackslash}p{\dimexpr 0.148325\linewidth-1.779904\tabcolsep\relax}>{\RaggedRight\arraybackslash}p{\dimexpr 0.095694\linewidth-1.148325\tabcolsep\relax}@{}}
\caption*{\hypertarget{fs-doxcnUN7y3QLEKBDZzyyRZf1Fob}{}Table S1c. Complete error statistics for the three CPU program pairs, with N = 19,897 in every row.}\\
\toprule
\bfseries Program pair & \bfseries Quantity & \bfseries P50 & \bfseries P95 & \bfseries P99 & \bfseries Maximum difference & \bfseries Configu\-rations exceeding threshold \\
\midrule
\endfirsthead
\toprule
\bfseries Program pair & \bfseries Quantity & \bfseries P50 & \bfseries P95 & \bfseries P99 & \bfseries Maximum difference & \bfseries Configu\-rations exceeding threshold \\
\midrule
\endhead
\midrule
\multicolumn{7}{r}{\footnotesize Continued on next page}\\
\endfoot
\bottomrule
\endlastfoot
xTBloom\textendash{}xTB & Energy / E\textsubscript{h} & 1.779\ensuremath{\times}10\textsuperscript{-8} & 2.941\ensuremath{\times}10\textsuperscript{-5} & 7.871\ensuremath{\times}10\textsuperscript{-5} & 6.359 & 3684 \\
xTBloom\textendash{}xTB & Force / E\textsubscript{h} bohr\textsuperscript{-1} & 5.128\ensuremath{\times}10\textsuperscript{-9} & 6.390\ensuremath{\times}10\textsuperscript{-6} & 1.288\ensuremath{\times}10\textsuperscript{-5} & 2.213\ensuremath{\times}10\textsuperscript{-1} & 1441 \\
xTBloom\textendash{}tblite & Energy / E\textsubscript{h} & 4.619\ensuremath{\times}10\textsuperscript{-14} & 2.264\ensuremath{\times}10\textsuperscript{-10} & 3.920\ensuremath{\times}10\textsuperscript{-9} & 9.142\ensuremath{\times}10\textsuperscript{-3} & 3 \\
xTBloom\textendash{}tblite & Force / E\textsubscript{h} bohr\textsuperscript{-1} & 3.244\ensuremath{\times}10\textsuperscript{-9} & 1.482\ensuremath{\times}10\textsuperscript{-8} & 3.267\ensuremath{\times}10\textsuperscript{-8} & 2.914\ensuremath{\times}10\textsuperscript{-2} & 3 \\
xTB\textendash{}tblite & Energy / E\textsubscript{h} & 1.778\ensuremath{\times}10\textsuperscript{-8} & 2.941\ensuremath{\times}10\textsuperscript{-5} & 7.869\ensuremath{\times}10\textsuperscript{-5} & 6.359 & 3682 \\
xTB\textendash{}tblite & Force / E\textsubscript{h} bohr\textsuperscript{-1} & 3.717\ensuremath{\times}10\textsuperscript{-9} & 6.389\ensuremath{\times}10\textsuperscript{-6} & 1.282\ensuremath{\times}10\textsuperscript{-5} & 2.213\ensuremath{\times}10\textsuperscript{-1} & 1440 \\
\end{longtable}
\endgroup

\begin{figure}[!htbp]
\centering
\hypertarget{fs-doxcnMRCa6wgKiDAYf2JiuTGsJd}{}
\includegraphics[width=\linewidth,height=.66\textheight,keepaspectratio]{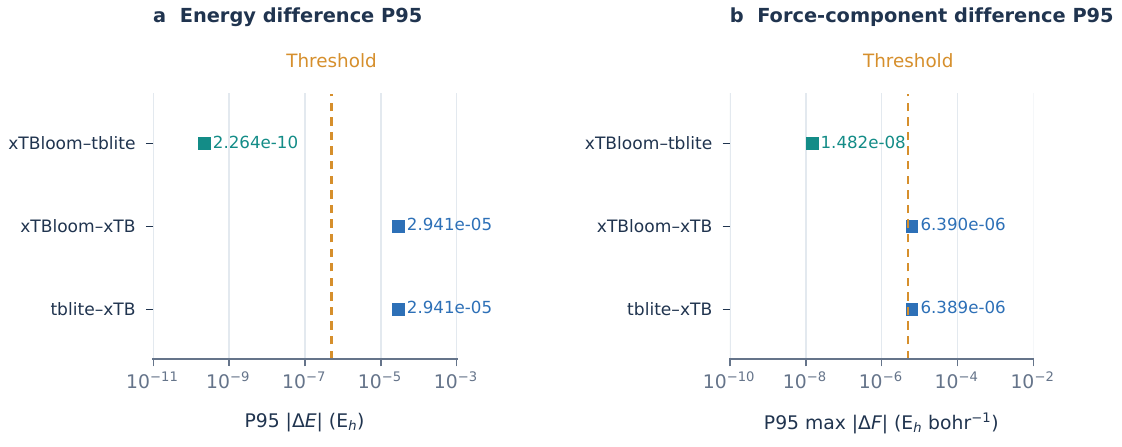}
\caption*{\hypertarget{fs-doxcneonwtS42voAzovXnF82JLd}{}Figure S1. P95 numerical differences between CPU programs. (a) Absolute energy difference per configuration. (b) Maximum absolute force-component difference per configuration. All three comparisons use the same 19,897 common-success configurations. The horizontal axes are logarithmic. Dashed lines mark the comparison thresholds of 5\ensuremath{\times}10\textsuperscript{-7} E\textsubscript{h} and 5\ensuremath{\times}10\textsuperscript{-6} E\textsubscript{h} bohr\textsuperscript{-1}, respectively. Table S1c gives all quantiles, maximum differences and exceedance counts.}
\end{figure}

\begingroup
\fontsize{9.2}{12.4}\selectfont
\setlength{\tabcolsep}{3pt}
\renewcommand{\arraystretch}{1.18}
\setlength{\LTcapwidth}{\linewidth}
\hypertarget{fs-doxcnd7zRIpSPK8OM9SPUaCJ6Sc}{}
\begin{longtable}{@{}>{\RaggedRight\arraybackslash}p{\dimexpr 0.179245\linewidth-1.792453\tabcolsep\relax}>{\RaggedRight\arraybackslash}p{\dimexpr 0.169811\linewidth-1.698113\tabcolsep\relax}>{\RaggedRight\arraybackslash}p{\dimexpr 0.169811\linewidth-1.698113\tabcolsep\relax}>{\RaggedRight\arraybackslash}p{\dimexpr 0.169811\linewidth-1.698113\tabcolsep\relax}>{\RaggedRight\arraybackslash}p{\dimexpr 0.169811\linewidth-1.698113\tabcolsep\relax}>{\RaggedRight\arraybackslash}p{\dimexpr 0.141509\linewidth-1.415094\tabcolsep\relax}@{}}
\caption*{\hypertarget{fs-doxcnqwRm0eakm0GQCoJxlAXuAc}{}Table S1d. CPU\textendash{}CUDA numerical differences and backend execution status. (a) Errors on the common-success set, N = 19,937.}\\
\toprule
\bfseries Quantity & \bfseries P50 & \bfseries P95 & \bfseries P99 & \bfseries Maximum difference & \bfseries Configu\-rations exceeding threshold \\
\midrule
\endfirsthead
\toprule
\bfseries Quantity & \bfseries P50 & \bfseries P95 & \bfseries P99 & \bfseries Maximum difference & \bfseries Configu\-rations exceeding threshold \\
\midrule
\endhead
\midrule
\multicolumn{6}{r}{\footnotesize Continued on next page}\\
\endfoot
\bottomrule
\endlastfoot
Energy / E\textsubscript{h} & 3.908\ensuremath{\times}10\textsuperscript{-14} & 1.620\ensuremath{\times}10\textsuperscript{-12} & 3.752\ensuremath{\times}10\textsuperscript{-12} & 2.395\ensuremath{\times}10\textsuperscript{-2} & 1 \\
Force / E\textsubscript{h} bohr\textsuperscript{-1} & 3.018\ensuremath{\times}10\textsuperscript{-15} & 4.808\ensuremath{\times}10\textsuperscript{-13} & 5.465\ensuremath{\times}10\textsuperscript{-11} & 2.899\ensuremath{\times}10\textsuperscript{-1} & 1 \\
\end{longtable}
\endgroup

\begingroup
\fontsize{10}{13.5}\selectfont
\setlength{\tabcolsep}{3pt}
\renewcommand{\arraystretch}{1.18}
\setlength{\LTcapwidth}{\linewidth}
\hypertarget{fs-doxcnT8h06qyi8gbX7eUOvSuUud}{}
\begin{longtable}{@{}>{\RaggedRight\arraybackslash}p{\dimexpr 0.363636\linewidth-1.454545\tabcolsep\relax}>{\RaggedRight\arraybackslash}p{\dimexpr 0.363636\linewidth-1.454545\tabcolsep\relax}>{\RaggedRight\arraybackslash}p{\dimexpr 0.272727\linewidth-1.090909\tabcolsep\relax}@{}}
\caption*{\hypertarget{fs-doxcnu7xRvBf7ZjazhRouJs5IBf}{}Table S1d (continued). (b) Backend status combinations for all 20,000 inputs.}\\
\toprule
\bfseries CPU status & \bfseries CUDA status & \bfseries Configu\-rations \\
\midrule
\endfirsthead
\toprule
\bfseries CPU status & \bfseries CUDA status & \bfseries Configu\-rations \\
\midrule
\endhead
\midrule
\multicolumn{3}{r}{\footnotesize Continued on next page}\\
\endfoot
\bottomrule
\endlastfoot
Success & Success & 19,937 \\
SCC not converged & SCC not converged & 37 \\
SCC not converged & Success & 13 \\
Success & SCC not converged & 13 \\
\end{longtable}
\endgroup

\FloatBarrier
\Needspace{8\baselineskip}
\hypertarget{fs-doxcn748mJaWrf67onP0qd3DlWh}{}
\subsection*{S1.3 Joint passes and outlier configurations}
\addcontentsline{toc}{subsection}{S1.3 Joint passes and outlier configurations}

\begingroup
\fontsize{9.5}{12.8}\selectfont
\setlength{\tabcolsep}{3pt}
\renewcommand{\arraystretch}{1.18}
\setlength{\LTcapwidth}{\linewidth}
\hypertarget{fs-doxcnq0bwuVwgbAL1ZWXqYisjae}{}
\begin{longtable}{@{}>{\RaggedRight\arraybackslash}p{\dimexpr 0.396226\linewidth-2.377358\tabcolsep\relax}>{\RaggedRight\arraybackslash}p{\dimexpr 0.198113\linewidth-1.188679\tabcolsep\relax}>{\RaggedRight\arraybackslash}p{\dimexpr 0.216981\linewidth-1.301887\tabcolsep\relax}>{\RaggedRight\arraybackslash}p{\dimexpr 0.188679\linewidth-1.132075\tabcolsep\relax}@{}}
\caption*{\hypertarget{fs-doxcniHaWd2PO6kkUwNww5BEvAH}{}Table S1e. Joint comparison results and outlier configurations. (a) Joint energy-and-force pass counts.}\\
\toprule
\bfseries Comparison & \bfseries Common-success configu\-rations & \bfseries Joint-pass configu\-rations & \bfseries Joint pass rate \\
\midrule
\endfirsthead
\toprule
\bfseries Comparison & \bfseries Common-success configu\-rations & \bfseries Joint-pass configu\-rations & \bfseries Joint pass rate \\
\midrule
\endhead
\midrule
\multicolumn{4}{r}{\footnotesize Continued on next page}\\
\endfoot
\bottomrule
\endlastfoot
xTBloom CPU\textendash{}tblite CPU & 19,897 & 19,894 & 99.985\% \\
xTBloom CPU\textendash{}CUDA & 19,937 & 19,936 & 99.995\% \\
All three pairs of the three CPU programs & 19,897 & 16,118 & 81.007\% \\
\end{longtable}
\endgroup

\hypertarget{fs-doxcnADyTfVVm6atYAAye7Ly87f}{}
Pairwise classification of the three CPU programs gave 16,212 configurations for which all three pairs met the energy threshold and 18,455 for which all three met the force threshold. For energy, xTBloom, xTB and tblite were the sole outlier in 1, 3,680 and 1 configurations, respectively. Another 1 configuration had three mutually distinct results, and 2 showed nontransitive threshold relations. The corresponding force counts were 1, 1,438, 1, 1 and 1. A sole outlier means that only the pair formed by the other two programs passed. A nontransitive threshold relation means that exactly two program pairs passed.

\begingroup
\fontsize{9.5}{12.8}\selectfont
\setlength{\tabcolsep}{3pt}
\renewcommand{\arraystretch}{1.18}
\setlength{\LTcapwidth}{\linewidth}
\hypertarget{fs-doxcnwHg6ndASFZSgJUVMchUJtf}{}
\begin{longtable}{@{}>{\RaggedRight\arraybackslash}p{\dimexpr 0.354545\linewidth-2.127273\tabcolsep\relax}>{\RaggedRight\arraybackslash}p{\dimexpr 0.200000\linewidth-1.200000\tabcolsep\relax}>{\RaggedRight\arraybackslash}p{\dimexpr 0.209091\linewidth-1.254545\tabcolsep\relax}>{\RaggedRight\arraybackslash}p{\dimexpr 0.236364\linewidth-1.418182\tabcolsep\relax}@{}}
\caption*{\hypertarget{fs-doxcnsytlGS5qffXd0jNcJOmcMh}{}Table S1e (continued). (b) The three configurations exceeding the xTBloom\textendash{}tblite comparison thresholds and their three-program differences.}\\
\toprule
\bfseries Configu\-ration ID & \bfseries Program pair & \bfseries |\ensuremath{\Delta}E| / E\textsubscript{h} & \bfseries max|\ensuremath{\Delta}F| / E\textsubscript{h} bohr\textsuperscript{-1} \\
\midrule
\endfirsthead
\toprule
\bfseries Configu\-ration ID & \bfseries Program pair & \bfseries |\ensuremath{\Delta}E| / E\textsubscript{h} & \bfseries max|\ensuremath{\Delta}F| / E\textsubscript{h} bohr\textsuperscript{-1} \\
\midrule
\endhead
\midrule
\multicolumn{4}{r}{\footnotesize Continued on next page}\\
\endfoot
\bottomrule
\endlastfoot
\code{CO_6464156197254019571837842} & xTBloom\textendash{}xTB & 1.424\ensuremath{\times}10\textsuperscript{-4} & 1.430\ensuremath{\times}10\textsuperscript{-3} \\
\code{CO_6464156197254019571837842} & xTBloom\textendash{}tblite & 1.423\ensuremath{\times}10\textsuperscript{-4} & 1.430\ensuremath{\times}10\textsuperscript{-3} \\
\code{CO_6464156197254019571837842} & xTB\textendash{}tblite & 1.257\ensuremath{\times}10\textsuperscript{-8} & 7.855\ensuremath{\times}10\textsuperscript{-9} \\
\code{CO_1728990913863245799454246} & xTBloom\textendash{}xTB & 9.189\ensuremath{\times}10\textsuperscript{-3} & 4.770\ensuremath{\times}10\textsuperscript{-3} \\
\code{CO_1728990913863245799454246} & xTBloom\textendash{}tblite & 9.142\ensuremath{\times}10\textsuperscript{-3} & 4.770\ensuremath{\times}10\textsuperscript{-3} \\
\code{CO_1728990913863245799454246} & xTB\textendash{}tblite & 4.772\ensuremath{\times}10\textsuperscript{-5} & 6.538\ensuremath{\times}10\textsuperscript{-6} \\
\code{CO_6758694963201820714974302} & xTBloom\textendash{}xTB & 1.046\ensuremath{\times}10\textsuperscript{-8} & 2.810\ensuremath{\times}10\textsuperscript{-8} \\
\code{CO_6758694963201820714974302} & xTBloom\textendash{}tblite & 2.802\ensuremath{\times}10\textsuperscript{-3} & 2.914\ensuremath{\times}10\textsuperscript{-2} \\
\code{CO_6758694963201820714974302} & xTB\textendash{}tblite & 2.802\ensuremath{\times}10\textsuperscript{-3} & 2.914\ensuremath{\times}10\textsuperscript{-2} \\
\end{longtable}
\endgroup

\hypertarget{fs-doxcnWqTK1pEe702TsNmxxhxu9c}{}
\code{CO_6464156197254019571837842} is a 20-atom configuration from the main set with charge +3, 4 unpaired electrons and Pd. It was classified as an xTBloom-only outlier. \code{CO_1728990913863245799454246} is a 116-atom, closed-shell configuration from the main set with charge \ensuremath{-}2 and Ca, Si and Te. All three programs gave mutually distinct results. \code{CO_6758694963201820714974302} is a 4-atom, closed-shell configuration from the numerical stress subset with charge +3 and Al, O and H. It was classified as a tblite-only outlier.

\begingroup
\fontsize{9.5}{12.8}\selectfont
\setlength{\tabcolsep}{3pt}
\renewcommand{\arraystretch}{1.18}
\setlength{\LTcapwidth}{\linewidth}
\hypertarget{fs-doxcnbRqqao67AgHqdHMNA29rHd}{}
\begin{longtable}{@{}>{\RaggedRight\arraybackslash}p{\dimexpr 0.290598\linewidth-2.324786\tabcolsep\relax}>{\RaggedRight\arraybackslash}p{\dimexpr 0.188034\linewidth-1.504274\tabcolsep\relax}>{\RaggedRight\arraybackslash}p{\dimexpr 0.188034\linewidth-1.504274\tabcolsep\relax}>{\RaggedRight\arraybackslash}p{\dimexpr 0.145299\linewidth-1.162393\tabcolsep\relax}>{\RaggedRight\arraybackslash}p{\dimexpr 0.188034\linewidth-1.504274\tabcolsep\relax}@{}}
\caption*{\hypertarget{fs-doxcnTQDfH9DW4aXAlQSEV9l7Kf}{}Table S1e (continued). (c) The single configuration that succeeded on both CPU and CUDA and exceeded the comparison thresholds.}\\
\toprule
\bfseries Configu\-ration ID & \bfseries CPU energy / E\textsubscript{h} & \bfseries CUDA energy / E\textsubscript{h} & \bfseries |\ensuremath{\Delta}E| / E\textsubscript{h} & \bfseries max|\ensuremath{\Delta}F| / E\textsubscript{h} bohr\textsuperscript{-1} \\
\midrule
\endfirsthead
\toprule
\bfseries Configu\-ration ID & \bfseries CPU energy / E\textsubscript{h} & \bfseries CUDA energy / E\textsubscript{h} & \bfseries |\ensuremath{\Delta}E| / E\textsubscript{h} & \bfseries max|\ensuremath{\Delta}F| / E\textsubscript{h} bohr\textsuperscript{-1} \\
\midrule
\endhead
\midrule
\multicolumn{5}{r}{\footnotesize Continued on next page}\\
\endfoot
\bottomrule
\endlastfoot
\code{CO_1112237736323506124052726} & -64.895 & -64.919 & 2.395\ensuremath{\times}10\textsuperscript{-2} & 2.899\ensuremath{\times}10\textsuperscript{-1} \\
\end{longtable}
\endgroup

\hypertarget{fs-doxcnL9coBmipBOmbHdnjAxiaLf}{}
This configuration belongs to the main OMol25 set and contains 23 atoms with charge +8 and a closed-shell specification. CPU and CUDA required 463 and 484 SCC iterations, respectively. tblite did not converge within 500 iterations, so this configuration contributes only to the common-success backend comparison.

\hypertarget{fs-doxcnluygJwIIwIAGGcJmFVVJah}{}
For configuration \code{CO_1112237736323506124052726}, a diagnostic build of the original numerical-comparison version was used to inject the shell charges, atomic charges, atomic dipoles and quadrupoles from the converged CPU and CUDA cold-start final states into the original B=32 batch. Each initial-guess type was supplied to both backends, with the mixing history reinitialized. The diagnostic used host input/output, a 300 K electronic temperature, a 500-iteration SCC limit, a free-energy-change tolerance of 10\textsuperscript{-10} E\textsubscript{h} and an SCC state-residual RMS tolerance of 10\textsuperscript{-8}. Both backends converged to the corresponding branch from each initial-guess type. Within each branch, the energy and maximum force-component differences met the CPU\textendash{}CUDA thresholds in Table S1b.

\begin{table}[H]
\begingroup
\fontsize{9.2}{12.4}\selectfont
\setlength{\tabcolsep}{3pt}
\renewcommand{\arraystretch}{1.18}
\setlength{\LTcapwidth}{\linewidth}
\hypertarget{fs-doxcnWT5iHW3fRtcw99HdxOhIQd}{}
\caption*{\hypertarget{fs-doxcnAKt8KJCowZdpvKpmyRlQbd}{}Table S1e (continued). Final-state initial-guess exchange for the configuration that succeeded on both CPU and CUDA and exceeded the thresholds.}
\begin{tabular}{@{}>{\RaggedRight\arraybackslash}p{\dimexpr 0.181818\linewidth-1.818182\tabcolsep\relax}>{\RaggedRight\arraybackslash}p{\dimexpr 0.181818\linewidth-1.818182\tabcolsep\relax}>{\RaggedRight\arraybackslash}p{\dimexpr 0.181818\linewidth-1.818182\tabcolsep\relax}>{\RaggedRight\arraybackslash}p{\dimexpr 0.145455\linewidth-1.454545\tabcolsep\relax}>{\RaggedRight\arraybackslash}p{\dimexpr 0.172727\linewidth-1.727273\tabcolsep\relax}>{\RaggedRight\arraybackslash}p{\dimexpr 0.136364\linewidth-1.363636\tabcolsep\relax}@{}}
\toprule
\bfseries Shared initial-guess source & \bfseries CPU energy / E\textsubscript{h} & \bfseries CUDA energy / E\textsubscript{h} & \bfseries \added{|\ensuremath{\Delta}E| / E\textsubscript{h}} & \bfseries \added{max|\ensuremath{\Delta}F| / (E\textsubscript{h} bohr\textsuperscript{-1})} & \bfseries SCC iterations, CPU/\allowbreak{}CUDA \\
\midrule
CPU cold-start final state & \added{-64.89522334858208} & \added{-64.89522334858238} & \added{2.984\ensuremath{\times}10\textsuperscript{-13}} & \added{1.254\ensuremath{\times}10\textsuperscript{-13}} & \added{6/6} \\
CUDA cold-start final state & \added{-64.91917719009682} & \added{-64.91917719009554} & \added{1.279\ensuremath{\times}10\textsuperscript{-12}} & \added{3.557\ensuremath{\times}10\textsuperscript{-13}} & \added{7/7} \\
\bottomrule
\end{tabular}
\endgroup
\end{table}

\begingroup
\fontsize{9.5}{12.8}\selectfont
\setlength{\tabcolsep}{3pt}
\renewcommand{\arraystretch}{1.18}
\setlength{\LTcapwidth}{\linewidth}
\hypertarget{fs-doxcnxFxbFfycN5IgkYKtPiiOEg}{}
\begin{longtable}{@{}>{\RaggedRight\arraybackslash}p{\dimexpr 0.200000\linewidth-1.600000\tabcolsep\relax}>{\RaggedRight\arraybackslash}p{\dimexpr 0.172727\linewidth-1.381818\tabcolsep\relax}>{\RaggedRight\arraybackslash}p{\dimexpr 0.209091\linewidth-1.672727\tabcolsep\relax}>{\RaggedRight\arraybackslash}p{\dimexpr 0.218182\linewidth-1.745455\tabcolsep\relax}>{\RaggedRight\arraybackslash}p{\dimexpr 0.200000\linewidth-1.600000\tabcolsep\relax}@{}}
\caption*{\hypertarget{fs-doxcnGWADsOLnbodouftT3xtOUd}{}Table S1e (continued). (d) Complete execution status by program.}\\
\toprule
\bfseries Program & \bfseries QM9 main set & \bfseries OMol25 main set & \bfseries OMol25 numerical stress subset & \bfseries Total \\
\midrule
\endfirsthead
\toprule
\bfseries Program & \bfseries QM9 main set & \bfseries OMol25 main set & \bfseries OMol25 numerical stress subset & \bfseries Total \\
\midrule
\endhead
\midrule
\multicolumn{5}{r}{\footnotesize Continued on next page}\\
\endfoot
\bottomrule
\endlastfoot
xTBloom CPU & 10000/0/0 & 7961/39/0 & 1989/11/0 & 19950/50/0 \\
xTB CPU & 10000/0/0 & 7954/45/1 & 1986/14/0 & 19940/59/1 \\
tblite CPU & 10000/0/0 & 7961/39/0 & 1987/13/0 & 19948/52/0 \\
xTBloom CUDA & 10000/0/0 & 7962/38/0 & 1988/12/0 & 19950/50/0 \\
\end{longtable}
\endgroup

\hypertarget{fs-doxcnqib2RCJEgfD8hJJ2wnuoze}{}
Each cell in Table S1e(d) lists the numbers of successful runs, SCC nonconvergence cases and nonfinite required outputs, in that order. The 80,000 execution records contain 212 failures involving 111 distinct configurations. Of these failures, 211 were SCC nonconvergence cases, and 1 was a nonfinite force output from native xTB for configuration \code{CO_5802592217901689389998917}.

\begingroup
\fontsize{9.2}{12.4}\selectfont
\setlength{\tabcolsep}{3pt}
\renewcommand{\arraystretch}{1.18}
\setlength{\LTcapwidth}{\linewidth}
\hypertarget{fs-doxcneGMN0jxOwKgDisD1wkVxYf}{}
\begin{longtable}{@{}>{\RaggedRight\arraybackslash}p{\dimexpr 0.123810\linewidth-1.238095\tabcolsep\relax}>{\RaggedRight\arraybackslash}p{\dimexpr 0.200000\linewidth-2.000000\tabcolsep\relax}>{\RaggedRight\arraybackslash}p{\dimexpr 0.190476\linewidth-1.904762\tabcolsep\relax}>{\RaggedRight\arraybackslash}p{\dimexpr 0.123810\linewidth-1.238095\tabcolsep\relax}>{\RaggedRight\arraybackslash}p{\dimexpr 0.171429\linewidth-1.714286\tabcolsep\relax}>{\RaggedRight\arraybackslash}p{\dimexpr 0.190476\linewidth-1.904762\tabcolsep\relax}@{}}
\caption*{\hypertarget{fs-doxcnjTU7eNnDdJIWBPwec2gJnh}{}Table S1f. Numerical consistency stratified by closed-shell and restricted open-shell inputs.}\\
\toprule
\bfseries Category & \bfseries Program pair & \bfseries Joint pass/\allowbreak{}common success & \bfseries Pass rate & \bfseries \added{P95 |\ensuremath{\Delta}E| / E\textsubscript{h}} & \bfseries \added{P95 max|\ensuremath{\Delta}F| / (E\textsubscript{h} bohr\textsuperscript{-1})} \\
\midrule
\endfirsthead
\toprule
\bfseries Category & \bfseries Program pair & \bfseries Joint pass/\allowbreak{}common success & \bfseries Pass rate & \bfseries \added{P95 |\ensuremath{\Delta}E| / E\textsubscript{h}} & \bfseries \added{P95 max|\ensuremath{\Delta}F| / (E\textsubscript{h} bohr\textsuperscript{-1})} \\
\midrule
\endhead
\midrule
\multicolumn{6}{r}{\footnotesize Continued on next page}\\
\endfoot
\bottomrule
\endlastfoot
Closed-shell & \added{xTBloom\textendash{}xTB} & \added{14,740/17,782} & \added{82.893\%} & \added{2.950\ensuremath{\times}10\textsuperscript{-5}} & \added{6.500\ensuremath{\times}10\textsuperscript{-6}} \\
Closed-shell & \added{xTBloom\textendash{}tblite} & \added{17,780/17,782} & \added{99.989\%} & \added{1.341\ensuremath{\times}10\textsuperscript{-11}} & \added{1.198\ensuremath{\times}10\textsuperscript{-8}} \\
Closed-shell & \added{xTB\textendash{}tblite} & \added{14,740/17,782} & \added{82.893\%} & \added{2.956\ensuremath{\times}10\textsuperscript{-5}} & \added{6.503\ensuremath{\times}10\textsuperscript{-6}} \\
Restricted open-shell & \added{xTBloom\textendash{}xTB} & \added{1,379/2,115} & \added{65.201\%} & \added{2.797\ensuremath{\times}10\textsuperscript{-5}} & \added{5.681\ensuremath{\times}10\textsuperscript{-6}} \\
Restricted open-shell & \added{xTBloom\textendash{}tblite} & \added{2,114/2,115} & \added{99.953\%} & \added{3.152\ensuremath{\times}10\textsuperscript{-9}} & \added{2.937\ensuremath{\times}10\textsuperscript{-8}} \\
Restricted open-shell & \added{xTB\textendash{}tblite} & \added{1,381/2,115} & \added{65.296\%} & \added{2.755\ensuremath{\times}10\textsuperscript{-5}} & \added{5.673\ensuremath{\times}10\textsuperscript{-6}} \\
Closed-shell & \added{xTBloom CPU\textendash{}CUDA} & \added{17,818/17,819} & \added{99.994\%} & \added{1.478\ensuremath{\times}10\textsuperscript{-12}} & \added{2.391\ensuremath{\times}10\textsuperscript{-13}} \\
Restricted open-shell & \added{xTBloom CPU\textendash{}CUDA} & \added{2,118/2,118} & \added{100.000\%} & \added{2.046\ensuremath{\times}10\textsuperscript{-12}} & \added{2.573\ensuremath{\times}10\textsuperscript{-12}} \\
\end{longtable}
\endgroup

\hypertarget{fs-doxcnRLKeZo1qEdWUGgIjSm4woe}{}
Closed-shell and restricted open-shell inputs are defined by u=0 and u>0 unpaired electrons, respectively. The three CPU program pairs use the three-program common-success set, with closed-shell/restricted open-shell denominators of 17,782/2,115 and energy and force thresholds of 5\ensuremath{\times}10\textsuperscript{-7} E\textsubscript{h} and 5\ensuremath{\times}10\textsuperscript{-6} E\textsubscript{h} bohr\textsuperscript{-1}, respectively. CPU\textendash{}CUDA uses the backend common-success set, with corresponding denominators of 17,819/2,118 and a threshold of 10\textsuperscript{-6} for both quantities in their respective atomic units. Force differences are reduced to the maximum absolute Cartesian-component difference within each system. P95 is then calculated by linear quantiles over the common-success systems in each group. A joint pass requires both energy and force to meet their thresholds. Unsuccessful systems remain in the status statistics in Table S1e. This table summarizes the per-system results for the three-program and CPU\textendash{}CUDA common-success sets.

\FloatBarrier
\Needspace{9\baselineskip}
\hypertarget{fs-doxcnsRvJt7oh5HPQpE5JEnlHxh}{}
\section*{S2. Analytic forces and failure isolation}
\addcontentsline{toc}{section}{S2. Analytic forces and failure isolation}

\FloatBarrier
\Needspace{8\baselineskip}
\hypertarget{fs-doxcn3ToeaSfEawCeRbnPYpgrPd}{}
\subsection*{S2.1 Finite-difference method and results}
\addcontentsline{toc}{subsection}{S2.1 Finite-difference method and results}

\hypertarget{fs-doxcnSbWL04IdeuXNhaVVsvldXe}{}
The finite-difference test selected 32 configurations from each of QM9 and OMol25 and fixed 8 Cartesian coordinates per configuration. Second-order central finite-difference forces were calculated with h = 5\ensuremath{\times}10\textsuperscript{-4}, 10\textsuperscript{-3} and 2\ensuremath{\times}10\textsuperscript{-3} bohr. Every displaced point was evaluated independently from a FRESH initial state, with an electronic temperature of 300 K, a maximum of 500 SCC iterations, a free-energy-change tolerance of 10\textsuperscript{-8} E\textsubscript{h} and an SCC state RMS tolerance of 10\textsuperscript{-6}. The absolute analytic-versus-finite-difference force threshold was 10\textsuperscript{-5} E\textsubscript{h} bohr\textsuperscript{-1}. This test checks each program's energy derivatives against its own analytic forces. For OMol25, 29/32 configurations completed, giving 696 checks. Table S2a reports the maximum error over the three step sizes.

\hypertarget{fs-doxcn5a7VHLs9cYc1RA4NvoOync}{}
\begin{equation}
F_i^{\mathrm{FD}}(h)=-\frac{E(\mathbf R+h\mathbf e_i)-E(\mathbf R-h\mathbf e_i)}{2h},\qquad \varepsilon_i(h)=\left|F_i^{\mathrm{analytic}}-F_i^{\mathrm{FD}}(h)\right|.
\end{equation}

\begingroup
\fontsize{8.2}{11.1}\selectfont
\setlength{\tabcolsep}{3pt}
\renewcommand{\arraystretch}{1.18}
\setlength{\LTcapwidth}{\linewidth}
\hypertarget{fs-doxcnhNCXueaLeGvLHbumJHRl1f}{}
\begin{longtable}{@{}>{\RaggedRight\arraybackslash}p{\dimexpr 0.109091\linewidth-1.527273\tabcolsep\relax}>{\RaggedRight\arraybackslash}p{\dimexpr 0.090909\linewidth-1.272727\tabcolsep\relax}>{\RaggedRight\arraybackslash}p{\dimexpr 0.109091\linewidth-1.527273\tabcolsep\relax}>{\RaggedRight\arraybackslash}p{\dimexpr 0.109091\linewidth-1.527273\tabcolsep\relax}>{\RaggedRight\arraybackslash}p{\dimexpr 0.109091\linewidth-1.527273\tabcolsep\relax}>{\RaggedRight\arraybackslash}p{\dimexpr 0.131818\linewidth-1.845455\tabcolsep\relax}>{\RaggedRight\arraybackslash}p{\dimexpr 0.109091\linewidth-1.527273\tabcolsep\relax}>{\RaggedRight\arraybackslash}p{\dimexpr 0.231818\linewidth-3.245455\tabcolsep\relax}@{}}
\caption*{\hypertarget{fs-doxcnsHkg0x2DJVvN2YxviGhvpf}{}Table S2a. Finite-difference checks of analytic forces.}\\
\toprule
\bfseries Dataset & \bfseries Backend & \bfseries Selected configu\-rations & \bfseries Completed configu\-rations & \bfseries Finite-difference checks & \bfseries Configu\-rations with execution failure & \bfseries Numerical threshold exceed\-ances & \bfseries Maximum error / E\textsubscript{h} bohr\textsuperscript{-1} \\
\midrule
\endfirsthead
\toprule
\bfseries Dataset & \bfseries Backend & \bfseries Selected configu\-rations & \bfseries Completed configu\-rations & \bfseries Finite-difference checks & \bfseries Configu\-rations with execution failure & \bfseries Numerical threshold exceed\-ances & \bfseries Maximum error / E\textsubscript{h} bohr\textsuperscript{-1} \\
\midrule
\endhead
\midrule
\multicolumn{8}{r}{\footnotesize Continued on next page}\\
\endfoot
\bottomrule
\endlastfoot
QM9 & CPU & 32 & 32 & 768 & 0 & 0 & 1.374\ensuremath{\times}10\textsuperscript{-6} \\
QM9 & CUDA & 32 & 32 & 768 & 0 & 0 & 1.374\ensuremath{\times}10\textsuperscript{-6} \\
OMol25 & CPU & 32 & 29 & 696 & 3 & 0 & 2.347\ensuremath{\times}10\textsuperscript{-6} \\
OMol25 & CUDA & 32 & 29 & 696 & 3 & 0 & 2.347\ensuremath{\times}10\textsuperscript{-6} \\
\end{longtable}
\endgroup

\hypertarget{fs-doxcn9nECPeWv1OSyxR9wnwv6cc}{}
Each completed configuration contributes 8 coordinates at 3 step sizes, giving 24 finite-difference checks. The three incomplete OMol25 configurations terminated during evaluation of the undisplaced structure. Table S2b lists the causes. CPU and CUDA failed on the same configurations with the same failure categories.

\begingroup
\fontsize{10}{13.5}\selectfont
\setlength{\tabcolsep}{3pt}
\renewcommand{\arraystretch}{1.18}
\setlength{\LTcapwidth}{\linewidth}
\hypertarget{fs-doxcnhnsz7KoqTpeheehB9e8lHg}{}
\begin{longtable}{@{}>{\RaggedRight\arraybackslash}p{\dimexpr 0.400000\linewidth-1.600000\tabcolsep\relax}>{\RaggedRight\arraybackslash}p{\dimexpr 0.163636\linewidth-0.654545\tabcolsep\relax}>{\RaggedRight\arraybackslash}p{\dimexpr 0.436364\linewidth-1.745455\tabcolsep\relax}@{}}
\caption*{\hypertarget{fs-doxcn7jhMRIFp6aTOLOoi50eptg}{}Table S2b. OMol25 configurations that did not complete the finite-difference test.}\\
\toprule
\bfseries Configu\-ration ID & \bfseries Returned status & \bfseries Cause \\
\midrule
\endfirsthead
\toprule
\bfseries Configu\-ration ID & \bfseries Returned status & \bfseries Cause \\
\midrule
\endhead
\midrule
\multicolumn{3}{r}{\footnotesize Continued on next page}\\
\endfoot
\bottomrule
\endlastfoot
\code{CO_1057867831392699108473564} & Invalid input & Incompatible parity of total and unpaired electron counts \\
\code{CO_2681531878022357052243628} & SCC not converged & Reached the 500-iteration SCC limit \\
\code{CO_8298694747446508682846328} & SCC not converged & Reached the 500-iteration SCC limit \\
\end{longtable}
\endgroup

\hypertarget{fs-doxcnc1bLcRWQeucfpbQ2MZc88g}{}
The finite-difference calculations ran on an NVIDIA RTX PRO 6000 Blackwell Workstation Edition with driver 595.91.07 and CUDA toolkit 12.9.1. CPU and BLAS execution were single-threaded.

\FloatBarrier
\Needspace{8\baselineskip}
\hypertarget{fs-doxcnq1zqItNJVOVdv2F7dBVwNg}{}
\subsection*{S2.2 Controlled mixed batch}
\addcontentsline{toc}{subsection}{S2.2 Controlled mixed batch}

\hypertarget{fs-doxcnnoXxL8We85ZbiV0UcXmsJD}{}
The controlled mixed batch contained the healthy configuration \code{CO_6325012143019994842243708} with 13 atoms and the SCC-nonconverged configuration \code{CO_9343753771713800272817479} with 48 atoms. Each configuration was first evaluated separately. They were then submitted together as a ragged batch in healthy-then-failed order, requesting energy, forces and atomic charges. CPU used host descriptors, and CUDA used mixed descriptors. Energy, forces and charge use E\textsubscript{h}, E\textsubscript{h} bohr\textsuperscript{-1} and e, respectively. For the healthy configuration, the absolute tolerance between separate and batched results was 10\textsuperscript{-10} for each quantity.

\begingroup
\fontsize{8.8}{11.9}\selectfont
\setlength{\tabcolsep}{3pt}
\renewcommand{\arraystretch}{1.18}
\setlength{\LTcapwidth}{\linewidth}
\hypertarget{fs-doxcnUeLPrWpEoy4JVcviZND7Ug}{}
\begin{longtable}{@{}>{\RaggedRight\arraybackslash}p{\dimexpr 0.140625\linewidth-1.687500\tabcolsep\relax}>{\RaggedRight\arraybackslash}p{\dimexpr 0.125000\linewidth-1.500000\tabcolsep\relax}>{\RaggedRight\arraybackslash}p{\dimexpr 0.140625\linewidth-1.687500\tabcolsep\relax}>{\RaggedRight\arraybackslash}p{\dimexpr 0.132812\linewidth-1.593750\tabcolsep\relax}>{\RaggedRight\arraybackslash}p{\dimexpr 0.179688\linewidth-2.156250\tabcolsep\relax}>{\RaggedRight\arraybackslash}p{\dimexpr 0.156250\linewidth-1.875000\tabcolsep\relax}>{\RaggedRight\arraybackslash}p{\dimexpr 0.125000\linewidth-1.500000\tabcolsep\relax}@{}}
\caption*{\hypertarget{fs-doxcn9vqVp9EvjZHL7Zd9GiSAvV}{}Table S2c. Batched results for the healthy and failed configurations.}\\
\toprule
\bfseries Execution path & \bfseries Healthy/\allowbreak{}failed status codes & \bfseries Healthy/\allowbreak{}failed SCC iterations & \bfseries Healthy energy difference / E\textsubscript{h} & \bfseries Healthy maximum force difference / E\textsubscript{h} bohr\textsuperscript{-1} & \bfseries Healthy maximum charge difference / e & \bfseries Failed E/\allowbreak{}F/\allowbreak{}charge \\
\midrule
\endfirsthead
\toprule
\bfseries Execution path & \bfseries Healthy/\allowbreak{}failed status codes & \bfseries Healthy/\allowbreak{}failed SCC iterations & \bfseries Healthy energy difference / E\textsubscript{h} & \bfseries Healthy maximum force difference / E\textsubscript{h} bohr\textsuperscript{-1} & \bfseries Healthy maximum charge difference / e & \bfseries Failed E/\allowbreak{}F/\allowbreak{}charge \\
\midrule
\endhead
\midrule
\multicolumn{7}{r}{\footnotesize Continued on next page}\\
\endfoot
\bottomrule
\endlastfoot
CPU host & 0 / 7 & 17 / 500 & 0 & 0 & 0 & All NaN \\
CUDA mixed & 0 / 7 & 17 / 500 & 0 & 3.123\ensuremath{\times}10\textsuperscript{-17} & 0 & All NaN \\
\end{longtable}
\endgroup

\hypertarget{fs-doxcnDUPdKtQNzj4ydKF3mugwie}{}
Per-system status codes 0, 7 and 8 denote success, SCC nonconvergence and eigensolver failure, respectively. The call return code and per-system statuses were checked separately.

\hypertarget{fs-doxcnM0B8ka7su691wZUvsIPm2g}{}
The two-member experiment ran on an NVIDIA RTX PRO 6000 Blackwell Workstation Edition with driver 595.91.07 and CUDA toolkit 12.9.1. Both CPU and CUDA mixed mode were tested with separate per-system calls and ragged batching. Table S2c reports member statuses, convergence iteration counts and the output state of the failed system.

\hypertarget{fs-doxcnOoh1rQE3adf8EaSlFZ8L5f}{}
SCC nonconvergence and eigensolver failures attributable to an individual system are reported through member statuses within a successful call. Input-validation, resource-allocation and device/stream runtime errors are call-level failures. A call-level error detected before caller-output commit leaves all result buffers and result flags unchanged. A catastrophic CUDA error after output commit begins returns an internal error and may leave partially written outputs.

\hypertarget{fs-doxcn1AclEjO56CcCrujX3xsXwc}{}
This experiment verifies member-level isolation of SCC nonconvergence within a successful batched call and NaN filling of the failed member's E/F/charge slices.

\hypertarget{fs-doxcnLbNyZilwzwPDhmoF6Zzb6f}{}
FRESH calls reinitialize the electronic state. Strict WARM is available only when the preceding batch converged completely and the electronic-state identity is compatible.

\FloatBarrier
\Needspace{8\baselineskip}
\hypertarget{fs-doxcnAHbbmh3HYb1sBvOt0Xxjcd}{}
\subsection*{S2.3 Tested configurations and coordinate indices}
\addcontentsline{toc}{subsection}{S2.3 Tested configurations and coordinate indices}

\hypertarget{fs-doxcn1hYCQACpZQAeRtz4EaEvzh}{}
We selected 32 configurations from each of QM9 and OMol25 with 8 fixed Cartesian test coordinates per configuration. CPU and CUDA used identical selections. Table S2d lists all 64 configuration identifiers and coordinate indices. The indices use zero-based flattened Cartesian numbering, with 3i, 3i+1 and 3i+2 corresponding to the x, y and z components of zero-based atom i. The list includes configurations that were attempted but did not complete the test. Tables S2a\textendash{}S2b give execution statuses and error statistics.

\begingroup
\fontsize{9.5}{12.8}\selectfont
\setlength{\tabcolsep}{3pt}
\renewcommand{\arraystretch}{1.18}
\setlength{\LTcapwidth}{\linewidth}
\hypertarget{fs-doxcnjvW6SSWOYtvzbwblAHnyjC}{}
\begin{longtable}{@{}>{\RaggedRight\arraybackslash}p{\dimexpr 0.097561\linewidth-0.390244\tabcolsep\relax}>{\RaggedRight\arraybackslash}p{\dimexpr 0.414634\linewidth-1.658537\tabcolsep\relax}>{\RaggedRight\arraybackslash}p{\dimexpr 0.487805\linewidth-1.951220\tabcolsep\relax}@{}}
\caption*{\hypertarget{fs-doxcnhtVNVAvRMjk3Gs6LC2uhRg}{}Table S2d. Configurations and eight coordinate indices used for finite-difference checks.}\\
\toprule
\bfseries Dataset & \bfseries Configu\-ration ID & \bfseries Tested coordinate indices \\
\midrule
\endfirsthead
\toprule
\bfseries Dataset & \bfseries Configu\-ration ID & \bfseries Tested coordinate indices \\
\midrule
\endhead
\midrule
\multicolumn{3}{r}{\footnotesize Continued on next page}\\
\endfoot
\bottomrule
\endlastfoot
\added{QM9} & \added{\code{qm9_007458}} & \added{61, 16, 22, 5, 48, 56, 34, 45} \\
\added{QM9} & \added{\code{qm9_133330}} & \added{4, 0, 27, 12, 25, 3, 6, 20} \\
\added{QM9} & \added{\code{qm9_040597}} & \added{30, 12, 46, 45, 37, 43, 23, 11} \\
\added{QM9} & \added{\code{qm9_116072}} & \added{51, 34, 11, 19, 33, 28, 41, 12} \\
\added{QM9} & \added{\code{qm9_089964}} & \added{43, 9, 3, 2, 52, 17, 12, 45} \\
\added{QM9} & \added{\code{qm9_046915}} & \added{53, 15, 6, 22, 3, 38, 5, 24} \\
\added{QM9} & \added{\code{qm9_045237}} & \added{26, 11, 13, 14, 30, 50, 46, 7} \\
\added{QM9} & \added{\code{qm9_052822}} & \added{37, 18, 47, 10, 16, 5, 0, 9} \\
\added{QM9} & \added{\code{qm9_050920}} & \added{40, 35, 27, 0, 42, 21, 17, 44} \\
\added{QM9} & \added{\code{qm9_119581}} & \added{5, 47, 42, 49, 34, 11, 15, 16} \\
\added{QM9} & \added{\code{qm9_015609}} & \added{18, 20, 13, 32, 36, 14, 23, 31} \\
\added{QM9} & \added{\code{qm9_078582}} & \added{38, 6, 28, 16, 34, 12, 23, 18} \\
\added{QM9} & \added{\code{qm9_048091}} & \added{8, 17, 43, 15, 27, 30, 34, 20} \\
\added{QM9} & \added{\code{qm9_079879}} & \added{21, 9, 54, 28, 10, 46, 56, 13} \\
\added{QM9} & \added{\code{qm9_014071}} & \added{11, 38, 16, 23, 39, 49, 17, 36} \\
\added{QM9} & \added{\code{qm9_025488}} & \added{31, 34, 41, 2, 12, 4, 29, 16} \\
\added{QM9} & \added{\code{qm9_112807}} & \added{37, 68, 32, 26, 59, 56, 65, 31} \\
\added{QM9} & \added{\code{qm9_030586}} & \added{10, 3, 40, 35, 44, 20, 6, 37} \\
\added{QM9} & \added{\code{qm9_086470}} & \added{20, 36, 24, 28, 10, 31, 12, 26} \\
\added{QM9} & \added{\code{qm9_086490}} & \added{41, 54, 13, 23, 22, 49, 44, 35} \\
\added{QM9} & \added{\code{qm9_023086}} & \added{25, 8, 43, 28, 4, 2, 32, 29} \\
\added{QM9} & \added{\code{qm9_071259}} & \added{21, 16, 26, 52, 0, 36, 47, 32} \\
\added{QM9} & \added{\code{qm9_052541}} & \added{39, 25, 7, 13, 20, 12, 16, 1} \\
\added{QM9} & \added{\code{qm9_044157}} & \added{31, 16, 7, 37, 33, 10, 23, 47} \\
\added{QM9} & \added{\code{qm9_070467}} & \added{3, 38, 46, 29, 9, 16, 1, 40} \\
\added{QM9} & \added{\code{qm9_078347}} & \added{54, 31, 1, 22, 35, 2, 10, 24} \\
\added{QM9} & \added{\code{qm9_064306}} & \added{36, 34, 52, 50, 38, 22, 29, 60} \\
\added{QM9} & \added{\code{qm9_127663}} & \added{1, 13, 25, 24, 0, 33, 4, 12} \\
\added{QM9} & \added{\code{qm9_043656}} & \added{13, 12, 35, 2, 26, 20, 16, 8} \\
\added{QM9} & \added{\code{qm9_067177}} & \added{43, 41, 0, 15, 30, 6, 4, 44} \\
\added{QM9} & \added{\code{qm9_128216}} & \added{16, 44, 46, 17, 24, 15, 54, 11} \\
\added{QM9} & \added{\code{qm9_089801}} & \added{42, 2, 12, 4, 30, 20, 13, 53} \\
\added{OMol25} & \added{\code{CO_1136431244543114064425896}} & \added{21, 20, 9, 29, 5, 13, 28, 14} \\
\added{OMol25} & \added{\code{CO_6189934124259438229170486}} & \added{182, 168, 119, 89, 81, 160, 113, 140} \\
\added{OMol25} & \added{\code{CO_1057867831392699108473564}} & \added{39, 12, 6, 21, 7, 27, 10, 16} \\
\added{OMol25} & \added{\code{CO_1050662051056006357859002}} & \added{7, 28, 9, 15, 6, 31, 4, 13} \\
\added{OMol25} & \added{\code{CO_2681531878022357052243628}} & \added{103, 270, 2, 69, 100, 345, 286, 74} \\
\added{OMol25} & \added{\code{CO_8848039806788650743307131}} & \added{56, 34, 60, 15, 48, 64, 17, 59} \\
\added{OMol25} & \added{\code{CO_3071188985923719149160631}} & \added{241, 164, 71, 45, 124, 294, 178, 87} \\
\added{OMol25} & \added{\code{CO_8298694747446508682846328}} & \added{193, 31, 245, 47, 184, 23, 120, 166} \\
\added{OMol25} & \added{\code{CO_6454588750467394776830157}} & \added{238, 21, 147, 175, 109, 66, 131, 228} \\
\added{OMol25} & \added{\code{CO_5377121089962908969148665}} & \added{9, 5, 63, 50, 45, 4, 8, 11} \\
\added{OMol25} & \added{\code{CO_6817148436676158642166489}} & \added{31, 25, 121, 54, 16, 101, 110, 82} \\
\added{OMol25} & \added{\code{CO_1882940930435698542755778}} & \added{143, 46, 61, 152, 288, 24, 359, 181} \\
\added{OMol25} & \added{\code{CO_1100507435607172621851273}} & \added{171, 8, 164, 58, 169, 153, 170, 136} \\
\added{OMol25} & \added{\code{CO_7791841176610489324341915}} & \added{101, 6, 127, 252, 21, 303, 102, 220} \\
\added{OMol25} & \added{\code{CO_1344796200747384825951391}} & \added{65, 22, 137, 10, 35, 109, 71, 136} \\
\added{OMol25} & \added{\code{CO_5095567109539540080386136}} & \added{201, 80, 167, 48, 209, 210, 183, 161} \\
\added{OMol25} & \added{\code{CO_4825248385191983697351688}} & \added{6, 22, 12, 2, 8, 10, 18, 7} \\
\added{OMol25} & \added{\code{CO_1564102652018849082053013}} & \added{52, 182, 19, 188, 140, 194, 227, 42} \\
\added{OMol25} & \added{\code{CO_1715193214842408051371520}} & \added{43, 79, 123, 132, 121, 73, 125, 107} \\
\added{OMol25} & \added{\code{CO_6944664227163535839419051}} & \added{315, 106, 204, 77, 111, 235, 107, 229} \\
\added{OMol25} & \added{\code{CO_1781181311303345976285781}} & \added{159, 228, 129, 229, 73, 126, 101, 70} \\
\added{OMol25} & \added{\code{CO_5271660410412731654148096}} & \added{37, 35, 29, 21, 30, 31, 40, 8} \\
\added{OMol25} & \added{\code{CO_1550988080625443085478744}} & \added{25, 28, 24, 37, 33, 8, 30, 6} \\
\added{OMol25} & \added{\code{CO_1249657292598048773718361}} & \added{56, 41, 77, 1, 95, 67, 3, 102} \\
\added{OMol25} & \added{\code{CO_1150036650524048691512988}} & \added{33, 116, 140, 32, 120, 90, 138, 95} \\
\added{OMol25} & \added{\code{CO_1163209408718741788791476}} & \added{16, 60, 52, 29, 51, 2, 12, 19} \\
\added{OMol25} & \added{\code{CO_1318147124456168892614118}} & \added{60, 71, 102, 64, 96, 88, 17, 160} \\
\added{OMol25} & \added{\code{CO_6527503939528439100361211}} & \added{36, 18, 25, 23, 3, 9, 43, 20} \\
\added{OMol25} & \added{\code{CO_7289135314389910106135448}} & \added{184, 331, 134, 344, 146, 277, 358, 38} \\
\added{OMol25} & \added{\code{CO_1091021934094564144828708}} & \added{31, 23, 17, 33, 35, 21, 6, 38} \\
\added{OMol25} & \added{\code{CO_1261546634852081578532461}} & \added{151, 57, 224, 76, 292, 150, 216, 50} \\
\added{OMol25} & \added{\code{CO_1098924348742000803824720}} & \added{23, 21, 41, 15, 25, 48, 44, 1} \\
\end{longtable}
\endgroup

\FloatBarrier
\Needspace{8\baselineskip}
\hypertarget{fs-doxcnEe92nB5ETC8HLPRkgtTRHe}{}
\subsection*{S2.4 Multiple-step-size diagnostics for three QM9 force components}
\addcontentsline{toc}{subsection}{S2.4 Multiple-step-size diagnostics for three QM9 force components}

\hypertarget{fs-doxcnuK7WkGuWplJmKvutAlmhue}{}
Among the 71 QM9 force cases classified as native-xTB-only outliers in the three-program CPU comparison, we selected the three configurations with the largest xTB\textendash{}tblite force differences. These were \code{qm9_066596}, \code{qm9_020345} and \code{qm9_108978}, with force components at flattened indices 25, 1 and 10, respectively. CPU cold-start calculations used step sizes of 10\textsuperscript{-3}, 5\ensuremath{\times}10\textsuperscript{-4} and 10\textsuperscript{-4} bohr under two settings, comprising the original numerical settings and a request for stricter SCC convergence. They produced 144 energy-point records and 54 derivative checks. Native xTB passed 0/18 checks in this diagnostic panel.

\hypertarget{fs-doxcneGNveLhvFInDwrjx87xWJg}{}
The second panel set the xTBloom free-energy-change and SCC state RMS tolerances to 10\textsuperscript{-12} E\textsubscript{h} and 10\textsuperscript{-10}, respectively, and requested accuracy = 10\textsuperscript{-6} from both reference programs. The public xTB interface clamps this request to 10\textsuperscript{-4}, giving the same effective xTB accuracy in both panels. All energy points were evaluated successfully. Repeated evaluations of the central point gave zero energy and force differences. The diagnostic error threshold remained 10\textsuperscript{-5} E\textsubscript{h} bohr\textsuperscript{-1}.

\begin{table}[H]
\begingroup
\fontsize{10}{13.5}\selectfont
\setlength{\tabcolsep}{3pt}
\renewcommand{\arraystretch}{1.18}
\setlength{\LTcapwidth}{\linewidth}
\hypertarget{fs-doxcn7qDrBMiHRlopvtZ9xc9Sr8}{}
\caption*{\hypertarget{fs-doxcnrmToH5xvmyCbjCKRK5mC5g}{}Table S2e. Derivative diagnostics for three selected QM9 force components.}
\begin{tabular}{@{}>{\RaggedRight\arraybackslash}p{\dimexpr 0.281818\linewidth-1.127273\tabcolsep\relax}>{\RaggedRight\arraybackslash}p{\dimexpr 0.227273\linewidth-0.909091\tabcolsep\relax}>{\RaggedRight\arraybackslash}p{\dimexpr 0.490909\linewidth-1.963636\tabcolsep\relax}@{}}
\toprule
\bfseries Program & \bfseries Passes/\allowbreak{}checks & \bfseries Maximum finite-difference error / E\textsubscript{h} bohr\textsuperscript{-1} \\
\midrule
xTBloom CPU & 18 / 18 & 2.209\ensuremath{\times}10\textsuperscript{-7} \\
tblite CPU & 18 / 18 & 2.211\ensuremath{\times}10\textsuperscript{-7} \\
native xTB CPU & 0 / 18 & 3.537\ensuremath{\times}10\textsuperscript{-2} \\
\bottomrule
\end{tabular}
\endgroup
\end{table}

\FloatBarrier
\Needspace{9\baselineskip}
\hypertarget{fs-doxcnanXYaPd4dFuEI1D0spusld}{}
\section*{S3. Difficult configurations and memory diagnostics}
\addcontentsline{toc}{section}{S3. Difficult configurations and memory diagnostics}

\FloatBarrier
\Needspace{8\baselineskip}
\hypertarget{fs-doxcnEyMitgWVIJ29c4jLaeo82b}{}
\subsection*{S3.1 Independent execution stress test}
\addcontentsline{toc}{subsection}{S3.1 Independent execution stress test}

\hypertarget{fs-doxcnHJtVU337UFld0MBC8uZopc}{}
The execution stress test used 2,000 OMol25 configurations~\citep{Levine_arXiv_2025_p2505.08762}. Table S3a summarizes the categories, counted according to a predefined primary-category assignment rule.

\begingroup
\fontsize{10}{13.5}\selectfont
\setlength{\tabcolsep}{3pt}
\renewcommand{\arraystretch}{1.18}
\setlength{\LTcapwidth}{\linewidth}
\hypertarget{fs-doxcndvQldittXgEc8H3P4wZHih}{}
\begin{longtable}{@{}>{\RaggedRight\arraybackslash}p{\dimexpr 0.300000\linewidth-0.600000\tabcolsep\relax}>{\RaggedRight\arraybackslash}p{\dimexpr 0.700000\linewidth-1.400000\tabcolsep\relax}@{}}
\caption*{\hypertarget{fs-doxcnsLJecyx8cQJ2iQZJ8jOUic}{}Table S3a. Category composition of the execution stress set.}\\
\toprule
\bfseries Stress category & \bfseries Configu\-rations \\
\midrule
\endfirsthead
\toprule
\bfseries Stress category & \bfseries Configu\-rations \\
\midrule
\endhead
\midrule
\multicolumn{2}{r}{\footnotesize Continued on next page}\\
\endfoot
\bottomrule
\endlastfoot
Open-shell & 600 \\
Charged & 600 \\
Metal/heavy elements & 300 \\
Large systems & 200 \\
Environment & 150 \\
Reactivity/distortion & 150 \\
Total & 2,000 \\
\end{longtable}
\endgroup

\hypertarget{fs-doxcnNMTXapqNqAleEImconOzfb}{}
The test evaluated GFN2-xTB energies and forces at an electronic temperature of 300 K, with a maximum of 500 SCC iterations and reference-program accuracy = 10\textsuperscript{-4}. xTBloom and dxtb each used one RTX 5090 with B = 1. Native xTB and tblite used per-system calls on one CPU thread. The CPU was an AMD EPYC 9354, and the interpreter was Python 3.10.19.

\begingroup
\fontsize{9.2}{12.4}\selectfont
\setlength{\tabcolsep}{3pt}
\renewcommand{\arraystretch}{1.18}
\setlength{\LTcapwidth}{\linewidth}
\hypertarget{fs-doxcn6pJiMn16KiOJ2fHTJIyrR3}{}
\begin{longtable}{@{}>{\RaggedRight\arraybackslash}p{\dimexpr 0.281818\linewidth-2.818182\tabcolsep\relax}>{\RaggedRight\arraybackslash}p{\dimexpr 0.136364\linewidth-1.363636\tabcolsep\relax}>{\RaggedRight\arraybackslash}p{\dimexpr 0.181818\linewidth-1.818182\tabcolsep\relax}>{\RaggedRight\arraybackslash}p{\dimexpr 0.118182\linewidth-1.181818\tabcolsep\relax}>{\RaggedRight\arraybackslash}p{\dimexpr 0.145455\linewidth-1.454545\tabcolsep\relax}>{\RaggedRight\arraybackslash}p{\dimexpr 0.136364\linewidth-1.363636\tabcolsep\relax}@{}}
\caption*{\hypertarget{fs-doxcnoF2nsPYQqDOqOKvF5wPiK8}{}Table S3b. Complete execution status for the stress set.}\\
\toprule
\bfseries Program & \bfseries Success & \bfseries SCC not converged & \bfseries OOM & \bfseries Invalid input & \bfseries Success rate \\
\midrule
\endfirsthead
\toprule
\bfseries Program & \bfseries Success & \bfseries SCC not converged & \bfseries OOM & \bfseries Invalid input & \bfseries Success rate \\
\midrule
\endhead
\midrule
\multicolumn{6}{r}{\footnotesize Continued on next page}\\
\endfoot
\bottomrule
\endlastfoot
xTBloom CUDA & 1,981 & 19 & 0 & 0 & 99.050\% \\
dxtb CUDA & 1,923 & 71 & 6 & 0 & 96.150\% \\
native xTB CPU & 1,984 & 16 & \textemdash{} & 0 & 99.200\% \\
tblite CPU & 1,988 & 12 & \textemdash{} & 0 & 99.400\% \\
\end{longtable}
\endgroup

\hypertarget{fs-doxcntuEL7nTxOmH46LazTdAcXb}{}
The two GPU programs shared 1,913 successful configurations, the two CPU programs shared 1,979, and all four programs shared 1,905.

\FloatBarrier
\Needspace{8\baselineskip}
\hypertarget{fs-doxcnalzWFiSpyv7T1tB80EL8Wc}{}
\subsection*{S3.2 Size-stratified execution tests on H200}
\addcontentsline{toc}{subsection}{S3.2 Size-stratified execution tests on H200}

\hypertarget{fs-doxcnY442CZZw7c5H9JI9diNAJe}{}
The supplementary H200 test divided the variable-size queue into three AO-dimension ranges and also evaluated the complete mixed queue in its original order. xTBloom used B=1, 64 and 256, and dxtb used B=1. Each execution configuration completed three measured traversals. Throughput was calculated as the successful-system count divided by traversal time, including the cost of failures. This test provides a size-stratified H200 diagnostic, reported separately from the main RTX 5090 performance results.

\begin{figure}[p]
\centering
\hypertarget{fs-doxcnuv09LN6bgYOz5LDD3gr1Ug}{}
\includegraphics[width=\linewidth,height=.84\textheight,keepaspectratio]{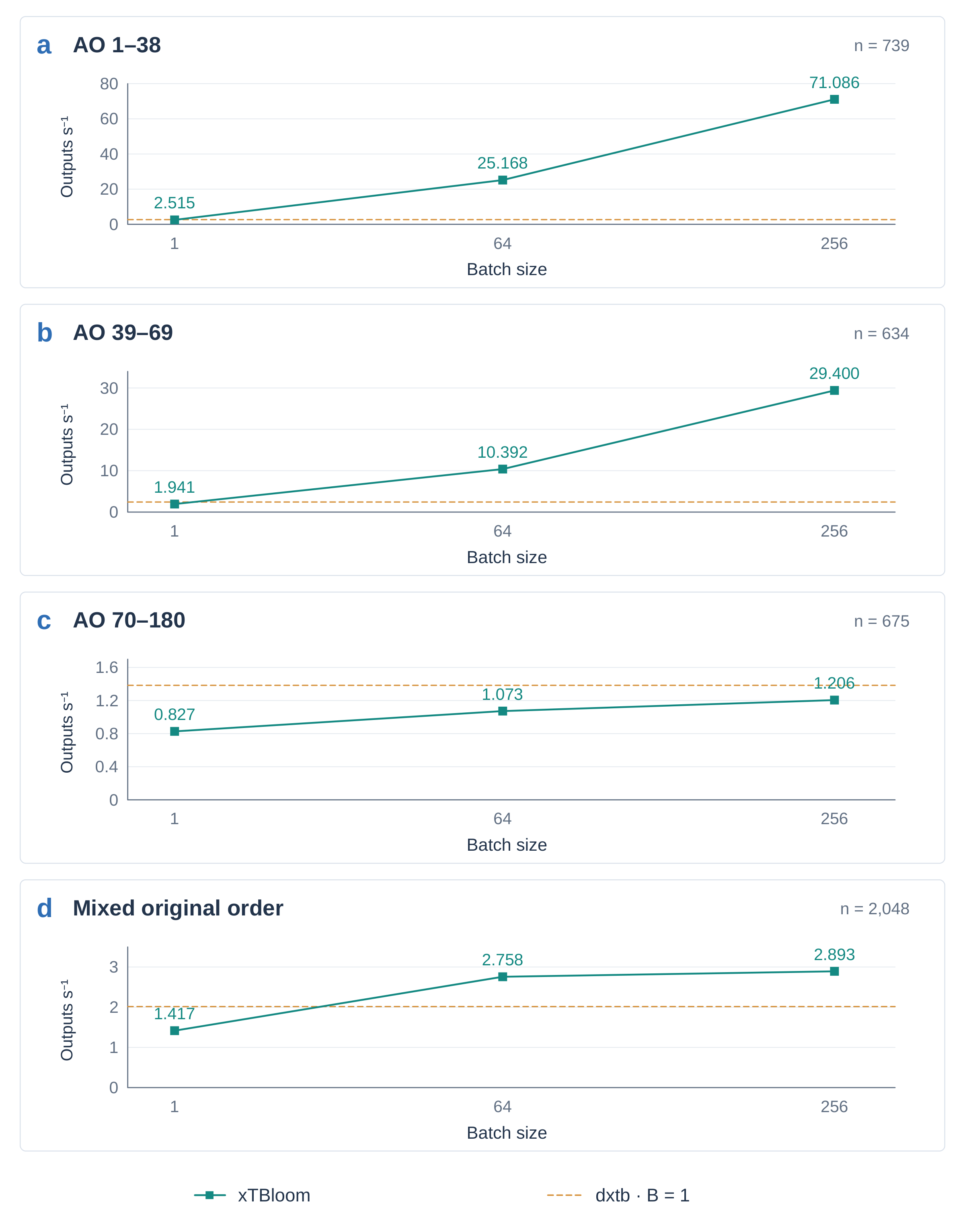}
\caption*{\hypertarget{fs-doxcnMDl8LRWTn61b5ebvr84xsh}{}Figure S2. Size-stratified throughput for the heterogeneous H200 queue. (a\textendash{}d) AO counts of 1\textendash{}38, 39\textendash{}69 and 70\textendash{}180, followed by the complete mixed queue in its original order, containing 739, 634, 675 and 2,048 configurations, respectively. Solid lines show xTBloom throughput across batch sizes, and dashed lines show dxtb B=1. The vertical-axis ranges differ between panels. Values are medians over three measured traversals.}
\end{figure}

\FloatBarrier
\Needspace{8\baselineskip}
\hypertarget{fs-doxcnDdUavQNOOIe1rmvIHE0Vmf}{}
\subsection*{S3.3 dxtb GPU memory for the near-size queue}
\addcontentsline{toc}{subsection}{S3.3 dxtb GPU memory for the near-size queue}

\hypertarget{fs-doxcnCBDfi5Yy9EUeWgGDRC1djd}{}
The dxtb memory diagnostic used the same 2,048 near-size configurations and calculation settings as S4. The atom-padding ratio is the sum of padded atom slots across batches divided by the total number of real atoms. Memory is reported in GiB as the largest allocated-memory record from the PyTorch allocator over the three measured traversals.

\begingroup
\fontsize{9.5}{12.8}\selectfont
\setlength{\tabcolsep}{3pt}
\renewcommand{\arraystretch}{1.18}
\setlength{\LTcapwidth}{\linewidth}
\hypertarget{fs-doxcn5ZdoKOZFSIMry24GMU0D6d}{}
\begin{longtable}{@{}>{\RaggedRight\arraybackslash}p{\dimexpr 0.327273\linewidth-1.963636\tabcolsep\relax}>{\RaggedRight\arraybackslash}p{\dimexpr 0.163636\linewidth-0.981818\tabcolsep\relax}>{\RaggedRight\arraybackslash}p{\dimexpr 0.281818\linewidth-1.690909\tabcolsep\relax}>{\RaggedRight\arraybackslash}p{\dimexpr 0.227273\linewidth-1.363636\tabcolsep\relax}@{}}
\caption*{\hypertarget{fs-doxcnrkgydKpj9xMoPnOrDGmsOf}{}Table S3c. dxtb atom padding, memory allocation and queue completion.}\\
\toprule
\bfseries Configu\-ration & \bfseries Atom-padding ratio & \bfseries Maximum recorded allocated memory / GiB & \bfseries Successful configu\-rations per traversal \\
\midrule
\endfirsthead
\toprule
\bfseries Configu\-ration & \bfseries Atom-padding ratio & \bfseries Maximum recorded allocated memory / GiB & \bfseries Successful configu\-rations per traversal \\
\midrule
\endhead
\midrule
\multicolumn{4}{r}{\footnotesize Continued on next page}\\
\endfoot
\bottomrule
\endlastfoot
dxtb GPU batch 1 & 1.000 & 0.137 & 2048/2048 \\
dxtb GPU batch 64 & 1.261 & 10.176 & 2048/2048 \\
dxtb GPU batch 128 & 1.300 & 20.246 & 2048/2048 \\
dxtb GPU batch 256, original order & 1.322 & 30.154 & 0/2048 \\
dxtb GPU batch 256, AO-sorted & 1.266 & 30.309 & 256/2048 \\
\end{longtable}
\endgroup

\hypertarget{fs-doxcn8T896ML3xZ5Vk9Ra2cGFPc}{}
B = 1, 64 and 128 completed the full queue. At B = 256, the original-order and AO-sorted configurations encountered OOM after publishing 0 and 256 successful results per traversal, respectively. Their memory values report allocation before failure and exclude the unsuccessful allocation request.

\begin{figure}[!htbp]
\centering
\hypertarget{fs-doxcnwhvLbP2gdmAilzbYLWLZUe}{}
\includegraphics[width=\linewidth,height=.66\textheight,keepaspectratio]{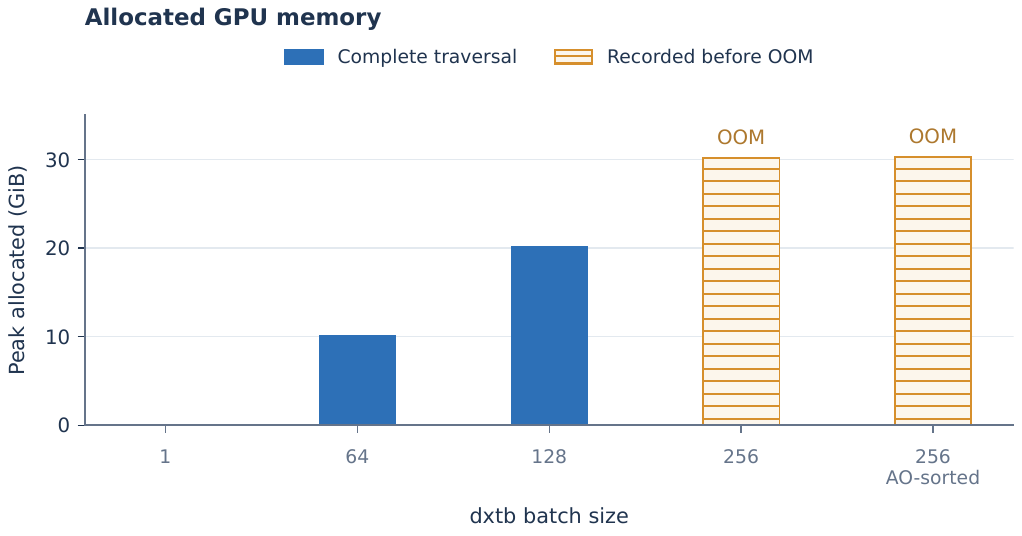}
\caption*{\hypertarget{fs-doxcnEqoqTdLzzxnOcyXt3tU8Eb}{}Figure S3. Allocated dxtb GPU memory for the near-size queue. Solid bars denote configurations that completed the queue. Orange hatched bars denote the maximum allocation recorded before OOM. Table S3c gives batch sizes, atom-padding ratios and successful-configuration counts.}
\end{figure}

\clearpage
\Needspace{9\baselineskip}
\hypertarget{fs-doxcnv0SCfAQD46gHsWVFfOc47b}{}
\section*{S4. Main near-size performance benchmark}
\addcontentsline{toc}{section}{S4. Main near-size performance benchmark}

\FloatBarrier
\Needspace{8\baselineskip}
\hypertarget{fs-doxcnxkF2VpxorDdzhAC2foilAg}{}
\subsection*{S4.1 Fixed queue and queue-wide AO sorting}
\addcontentsline{toc}{subsection}{S4.1 Fixed queue and queue-wide AO sorting}

\hypertarget{fs-doxcnz8hVkm1M2N3iFZthGQjuSc}{}
The main near-size performance benchmark used 2,048 \code{geom_orca6} configurations from the OMol25 train 4M catalog. All were marked eligible, had charge 0 and multiplicity 1, contained no metals, belonged to the \code{light_main_group} element category and had 100\textendash{}119 GFN2 AOs.

\hypertarget{fs-doxcn2CLIlrJzeOC2F3YQyQqxBb}{}
This section gives the selection criteria and summary statistics for the 2,048 configurations.

\hypertarget{fs-doxcnF8jYqhtQq4q4EkjqtCxfgc}{}
The fixed queue contained 27\textendash{}58 atoms per configuration, with a mean of 41.974 and 85,963 atoms in total. The mean AO count was 110.644, the sample standard deviation was 5.637 and the coefficient of variation was 5.095\%. The default policy sorts the entire queue by AO count before forming batches. In this benchmark, the runner performed the same queue-wide AO sort and submitted B = 256 batches. AO-sorting ablations retained the input order and measured B = 1, 64 and 256. Outputs were restored to their original correspondence by configuration ID.

\FloatBarrier
\Needspace{8\baselineskip}
\hypertarget{fs-doxcnAABgzq6CeJOLxwOqOQL7zb}{}
\subsection*{S4.2 Hardware, evaluation settings and timing}
\addcontentsline{toc}{subsection}{S4.2 Hardware, evaluation settings and timing}

\hypertarget{fs-doxcntQiWbN7a5ws1Bfgo6xYghc}{}
The benchmarks ran on a dual-socket AMD EPYC 9354 host. Each compute process or GPU-driving host process was pinned to one physical core, with OMP, BLAS and relevant PyTorch thread counts set to 1. GPU configurations used one NVIDIA RTX 5090 with approximately 31.800 GiB of reported memory. xTBloom and external programs were measured separately on different physical GPUs of the same model and different pinned CPU cores.

\begingroup
\fontsize{10}{13.5}\selectfont
\setlength{\tabcolsep}{3pt}
\renewcommand{\arraystretch}{1.18}
\setlength{\LTcapwidth}{\linewidth}
\hypertarget{fs-doxcnLz905gFfjXsZvDuQHX5igb}{}
\begin{longtable}{@{}>{\RaggedRight\arraybackslash}p{\dimexpr 0.300000\linewidth-0.600000\tabcolsep\relax}>{\RaggedRight\arraybackslash}p{\dimexpr 0.700000\linewidth-1.400000\tabcolsep\relax}@{}}
\caption*{\hypertarget{fs-doxcnn3KEirzWStCGgKF9ppcr8g}{}Table S4a. Calculation controls and measurement settings for the main performance benchmark.}\\
\toprule
\bfseries Item & \bfseries Setting \\
\midrule
\endfirsthead
\toprule
\bfseries Item & \bfseries Setting \\
\midrule
\endhead
\midrule
\multicolumn{2}{r}{\footnotesize Continued on next page}\\
\endfoot
\bottomrule
\endlastfoot
Common settings & GFN2-xTB, binary64, energy and forces, electronic temperature 300 K, maximum 500 SCC iterations, FRESH initial state \\
xTBloom & Public CPU/CUDA API, host input/output, SCC free-energy-change tolerance 10\textsuperscript{-8} E\textsubscript{h}, state-residual RMS tolerance 10\textsuperscript{-6}, native analytic forces \\
native xTB & libxtb 6.7.1, public C API, accuracy = 1.0, a new Calculator/Results per batch, forces obtained by negating gradients \\
dxtb & dxtb 0.4.0, PyTorch 2.10.0+cu128, float64, \code{x_atol} = 10\textsuperscript{-4}, \code{x_atol_max} = 10\textsuperscript{-5}, \code{f_atol} = 10\textsuperscript{-4}, \code{force_convergence} = True, \code{fermi_etemp} = 300 K, singlepoint with autograd forces \\
Build and runtime environment & Release, CUDA toolkit 12.9.1, sm120, LP64 OpenBLAS 0.3.30, Python 3.10.19 \\
Repetition protocol & One full-queue warmup and three measured traversals per configuration, with 2,048 structures per traversal \\
Primary timing & Sum of per-batch preparation, synchronous computation and complete host-output publication \\
Traversal wall clock & Also includes per-batch resource release, Python GC and progress handling \\
\end{longtable}
\endgroup

\hypertarget{fs-doxcnekfryfDLr6rUrVoOJIhBrf}{}
Native xTB accuracy controls both SCC stopping criteria and integral screening. At accuracy = 1.0, the energy-change threshold is approximately 10\textsuperscript{-6} E\textsubscript{h}, and the shell-charge and multipole residual threshold is approximately 10\textsuperscript{-4}. The tested dxtb path determines convergence from the L2 norm and maximum component of the unmixed Fock fixed-point residual, controlled by \code{x_atol} and \code{x_atol_max}, respectively. \code{f_atol} is a retained configuration field. The dxtb method is described in \citep{Friede_JChemPhys_2024_v161_p062501}.

\hypertarget{fs-doxcnPY6WZHkBHjuUOq0d9MN12f}{}
One xTBloom context was reused between batches within each traversal, while every compute call remained FRESH. Preparation included the batch adapter, descriptors, input processing, required uploads, native-context creation, temporary-context destruction and reuse switching. Computation included SCC, force evaluation and completion synchronization. Publication included transfer of results to the host, removal of padding and complete output publication. Primary timing is the sum of these three stages. Input loading, AO sorting, Executor construction, initial module or library loading, final destruction of the retained context and compressed result storage were outside both primary timing and traversal wall clock. Per-batch Calculator or model creation in native xTB and dxtb was included in preparation. Post-batch dxtb garbage collection was included in traversal wall clock, with primary timing ending before that cleanup. Throughput was calculated as 2,048 divided by the median primary time over three measurements.

\hypertarget{fs-doxcn4oOM7ioZOFdFgdrkK7dUAf}{}
Each of the 11 configurations underwent one warmup and three measured traversals, giving 44 traversals and 90,112 target-configuration records in total. The measured runs comprised 33 traversals and 67,584 target-configuration records. Inputs were fixed for every configuration. Failed outputs had their energy and complete force slice filled with NaN.

\FloatBarrier
\Needspace{8\baselineskip}
\hypertarget{fs-doxcn8YxxIVM05Sv49FUGnmRrDd}{}
\subsection*{S4.3 All configurations and timing results}
\addcontentsline{toc}{subsection}{S4.3 All configurations and timing results}

\begingroup
\fontsize{9.5}{12.8}\selectfont
\setlength{\tabcolsep}{3pt}
\renewcommand{\arraystretch}{1.18}
\setlength{\LTcapwidth}{\linewidth}
\hypertarget{fs-doxcn99b1imXCciINyUYB8ZyJMc}{}
\begin{longtable}{@{}>{\RaggedRight\arraybackslash}p{\dimexpr 0.288136\linewidth-2.305085\tabcolsep\relax}>{\RaggedRight\arraybackslash}p{\dimexpr 0.254237\linewidth-2.033898\tabcolsep\relax}>{\RaggedRight\arraybackslash}p{\dimexpr 0.152542\linewidth-1.220339\tabcolsep\relax}>{\RaggedRight\arraybackslash}p{\dimexpr 0.161017\linewidth-1.288136\tabcolsep\relax}>{\RaggedRight\arraybackslash}p{\dimexpr 0.144068\linewidth-1.152542\tabcolsep\relax}@{}}
\caption*{\hypertarget{fs-doxcn687iRBqnmKdt5C1pceih0c}{}Table S4b. Measured results for all 11 performance configurations.}\\
\toprule
\bfseries Configu\-ration & \bfseries Primary time, median [minimum\textendash{}maximum] / s & \bfseries Median wall clock / s & \bfseries Throughput / configu\-rations s\textsuperscript{-1} & \bfseries Successes/\allowbreak{}attempts \\
\midrule
\endfirsthead
\toprule
\bfseries Configu\-ration & \bfseries Primary time, median [minimum\textendash{}maximum] / s & \bfseries Median wall clock / s & \bfseries Throughput / configu\-rations s\textsuperscript{-1} & \bfseries Successes/\allowbreak{}attempts \\
\midrule
\endhead
\midrule
\multicolumn{5}{r}{\footnotesize Continued on next page}\\
\endfoot
\bottomrule
\endlastfoot
xTBloom CUDA default policy, B=256 (AO-sorted) & 28.792 [28.730\textendash{}28.793] & 28.859 & 71.130 & 6144/6144 \\
xTBloom CUDA ablation, B=256 (AO sorting disabled) & 38.094 [38.034\textendash{}38.171] & 38.160 & 53.760 & 6144/6144 \\
xTBloom CUDA, B=64 (AO sorting disabled) & 74.133 [73.667\textendash{}74.271] & 74.204 & 27.630 & 6144/6144 \\
xTB CPU, single core & 82.493 [82.374\textendash{}85.488] & 82.756 & 24.830 & 6144/6144 \\
xTBloom CPU, single core & 83.802 [83.776\textendash{}83.891] & 84.006 & 24.440 & 6144/6144 \\
xTBloom CUDA, B=1 (AO sorting disabled) & 293.822 [293.637\textendash{}294.759] & 294.067 & 6.970 & 6144/6144 \\
dxtb GPU batch 64 & 322.294 [321.827\textendash{}322.703] & 327.654 & 6.350 & 6144/6144 \\
dxtb GPU batch 128 & 403.220 [394.534\textendash{}405.795] & 406.697 & 5.080 & 6144/6144 \\
dxtb GPU batch 1 & 813.846 [810.560\textendash{}826.274] & 1198.143 & 2.520 & 6144/6144 \\
dxtb GPU batch 256, original order & OOM & 141.580 & \textemdash{} & 0/6144 \\
dxtb GPU batch 256, AO-sorted & OOM & 188.357 & \textemdash{} & 768/6144 \\
\end{longtable}
\endgroup

\hypertarget{fs-doxcnlMFnx4ekjaBWNQfP9W7VDg}{}
Successes/attempts use the 6,144 target records from three measured traversals as the denominator. For the two OOM configurations, the wall-clock column records time spent on the failed attempts, and full-queue throughput is left blank. The other nine configurations completed all 2,048 structures in every measured traversal.

\hypertarget{fs-doxcnEbX1ceDyFXNVSOabU4eYAe}{}
With the same fixed inputs and software implementation, the B = 256 primary times were 28.792 s with default AO sorting and 38.094 s with sorting disabled. The default policy reduced evaluation time by 24.4\% relative to this ablation. The single-core xTBloom CPU primary time was 83.802 s, giving a throughput ratio of 2.91 for the default AO-sorted GPU configuration.

\hypertarget{fs-doxcnl95Y3GSeiJW0Z3M9mb2Oje}{}
Across all 20 traversals of the five xTBloom configurations, the maximum differences from the first measured CPU output were 9.521\ensuremath{\times}10\textsuperscript{-13} E\textsubscript{h} in energy and 4.179\ensuremath{\times}10\textsuperscript{-13} E\textsubscript{h} bohr\textsuperscript{-1} in any force component. Per-configuration statuses, convergence flags and SCC iteration counts were identical.

\FloatBarrier
\Needspace{8\baselineskip}
\hypertarget{fs-doxcnR92LEvrxdWrkjm6m55RSTg}{}
\subsection*{S4.4 Sensitivity to external-reference settings}
\addcontentsline{toc}{subsection}{S4.4 Sensitivity to external-reference settings}

\hypertarget{fs-doxcnuc2JmRE3eyiONxh6OGvH9b}{}
We used the same cross-program absolute thresholds as S1, comprising 5\ensuremath{\times}10\textsuperscript{-7} E\textsubscript{h} for energy and 5\ensuremath{\times}10\textsuperscript{-6} E\textsubscript{h} bohr\textsuperscript{-1} for forces. Relative to the first measured native xTB output at accuracy = 1.0, each traversal of the five xTBloom configurations jointly passed 1,382/2,048 configurations. The complete dxtb configurations passed 0/2,048. The accuracy=10\textsuperscript{-4} reference was used for sensitivity diagnostics. Throughputs in Table S4b correspond to the specified program settings and do not constitute an equal-accuracy comparison.

\hypertarget{fs-doxcnibZlR7mkHHUjmpV8S2k6Gg}{}
We additionally evaluated the same inputs once on the CPU with the same native xTB library at accuracy = 10\textsuperscript{-4}. The maximum iteration count, electronic temperature and FRESH settings were unchanged. These results provide the numerical comparison in Table S4c. Table S4b retains the original performance measurements at accuracy = 1.0. Joint-pass counts use all 2,048 target configurations as the denominator, with successful-pair counts reported separately.

\begingroup
\fontsize{8.8}{11.9}\selectfont
\setlength{\tabcolsep}{3pt}
\renewcommand{\arraystretch}{1.18}
\setlength{\LTcapwidth}{\linewidth}
\hypertarget{fs-doxcnv7WzRtdSxYIpnptkfjZDtc}{}
\begin{longtable}{@{}>{\RaggedRight\arraybackslash}p{\dimexpr 0.193798\linewidth-2.325581\tabcolsep\relax}>{\RaggedRight\arraybackslash}p{\dimexpr 0.116279\linewidth-1.395349\tabcolsep\relax}>{\RaggedRight\arraybackslash}p{\dimexpr 0.108527\linewidth-1.302326\tabcolsep\relax}>{\RaggedRight\arraybackslash}p{\dimexpr 0.108527\linewidth-1.302326\tabcolsep\relax}>{\RaggedRight\arraybackslash}p{\dimexpr 0.116279\linewidth-1.395349\tabcolsep\relax}>{\RaggedRight\arraybackslash}p{\dimexpr 0.162791\linewidth-1.953488\tabcolsep\relax}>{\RaggedRight\arraybackslash}p{\dimexpr 0.193798\linewidth-2.325581\tabcolsep\relax}@{}}
\caption*{\hypertarget{fs-doxcnAuGydKQkJZOux7dOVbhgag}{}Table S4c. Numerical comparison against the native xTB reference at accuracy = 10\textsuperscript{-4}.}\\
\toprule
\bfseries Configu\-ration & \bfseries Successful pairs & \bfseries Energy exceed\-ances & \bfseries Force exceed\-ances & \bfseries Joint passes & \bfseries Maximum |\ensuremath{\Delta}E| / E\textsubscript{h} & \bfseries Maximum |\ensuremath{\Delta}F| / E\textsubscript{h} bohr\textsuperscript{-1} \\
\midrule
\endfirsthead
\toprule
\bfseries Configu\-ration & \bfseries Successful pairs & \bfseries Energy exceed\-ances & \bfseries Force exceed\-ances & \bfseries Joint passes & \bfseries Maximum |\ensuremath{\Delta}E| / E\textsubscript{h} & \bfseries Maximum |\ensuremath{\Delta}F| / E\textsubscript{h} bohr\textsuperscript{-1} \\
\midrule
\endhead
\midrule
\multicolumn{7}{r}{\footnotesize Continued on next page}\\
\endfoot
\bottomrule
\endlastfoot
Five xTBloom configurations & 2048/2048 & 384 & 0 & 1664/2048 & 2.116\ensuremath{\times}10\textsuperscript{-5} & 4.725\ensuremath{\times}10\textsuperscript{-6} \\
xTB CPU accuracy 1.0 & 2048/2048 & 0 & 369 & 1679/2048 & 2.089\ensuremath{\times}10\textsuperscript{-8} & 2.240\ensuremath{\times}10\textsuperscript{-5} \\
dxtb batch 1/64/128 & 2048/2048 & 2048 & 1 & 0/2048 & 3.043\ensuremath{\times}10\textsuperscript{-6} & 1.387\ensuremath{\times}10\textsuperscript{-5} \\
dxtb batch 256, AO-sorted & 256/2048 & 256 & 0 & 0/2048 & 2.638\ensuremath{\times}10\textsuperscript{-6} & 3.928\ensuremath{\times}10\textsuperscript{-6} \\
dxtb batch 256, original order & 0/2048 & \textemdash{} & \textemdash{} & 0/2048 & No paired output & No paired output \\
\end{longtable}
\endgroup

\hypertarget{fs-doxcnYLXgx8lKuDifBG6IHAWK4b}{}
Against the stricter reference, all xTBloom force components met the comparison threshold, while 384 configurations exceeded the energy threshold. Comparing the two native xTB accuracy settings gave 369 configurations exceeding the force threshold. The complete dxtb configurations had 2,048 energy exceedances and 1 force exceedance, with a maximum energy difference of 3.043\ensuremath{\times}10\textsuperscript{-6} Eh. Pass counts were calculated per configuration using joint energy and force thresholds. S4.2 defines the stopping criteria and residuals for each program. The timing ratios in Table S4b correspond to these numerical settings.

\FloatBarrier
\Needspace{9\baselineskip}
\hypertarget{fs-doxcnB0iWMu2ov1iYZY1Ntefvib}{}
\section*{S5. AO sorting, SCC scheduling and supplementary variable-size experiments}
\addcontentsline{toc}{section}{S5. AO sorting, SCC scheduling and supplementary variable-size experiments}

\FloatBarrier
\Needspace{8\baselineskip}
\hypertarget{fs-doxcnNDRZNUEaXMAapz032lvgWc}{}
\subsection*{S5.1 Protocols for the two ablation studies}
\addcontentsline{toc}{subsection}{S5.1 Protocols for the two ablation studies}

\hypertarget{fs-doxcnlLAyzh0WvIN9I1Noq2jrEb}{}
The AO-sorting comparison used the 2,048 near-size OMol25 configurations from the main performance benchmark. Inputs, neutral-singlet constraints, software implementation, electronic temperature, SCC tolerances and timing boundaries were fixed. The complete default AO-sorted scheme at B=256 served as the reference. Sorting alone was disabled for the paired ablation. Additional measurements with sorting disabled used B=1, 64 and 256. The benchmark runner sorted the queue before the calls. Each configuration completed one full-queue warmup and three measured traversals. All outputs were aligned by original configuration identity to compare energies, complete force blocks, statuses, convergence flags and SCC iteration counts.

\hypertarget{fs-doxcnN5QxWnzKIr77eoQUT0Psce}{}
The SCC-scheduling comparison used small-AO, neutral, restricted-spin batches formed by cycling through H\textsubscript{2}, He, LiH and CH\textsubscript{2}. The maximum SCC iteration count was 8, the mixing-history length was 3 and the electronic-temperature parameter was 0. D4, explicit point charges and periodic embedding were disabled in this controlled SCC loop. An active-capacity scheduling chain and a fixed-capacity device-tail Graph were constructed from the same input and initial state. Selected systems had their initial iteration counters set to the limit, causing them to exit at the first active-state check and producing initial active fractions of 1, 1/2 and 1/4. Membership was selected using the system index modulo 1, 2 or 4. At B=1, only the single fully active configuration was retained.

\hypertarget{fs-doxcnVksVncK1xJ4HOnDxCdhHCe}{}
The same initial state was restored before every measurement. Each path had three warmups and 50 measured runs. CUDA events on the same stream measured the complete SCC loop until all members terminated. Final-state download occurred after the timed interval. Paired checks compared the number of numerical updates, free energy, eigenvalues, occupations, density, weighted density, populations, multipoles, iteration counters and statuses. Initially fully active configurations were also checked against sequential CPU calculations.

\begingroup
\fontsize{10}{13.5}\selectfont
\setlength{\tabcolsep}{3pt}
\renewcommand{\arraystretch}{1.18}
\setlength{\LTcapwidth}{\linewidth}
\hypertarget{fs-doxcn08ioNH1lnKMYfvIi1GR5Nr}{}
\begin{longtable}{@{}>{\RaggedRight\arraybackslash}p{\dimexpr 0.333333\linewidth-1.333333\tabcolsep\relax}>{\RaggedRight\arraybackslash}p{\dimexpr 0.333333\linewidth-1.333333\tabcolsep\relax}>{\RaggedRight\arraybackslash}p{\dimexpr 0.333333\linewidth-1.333333\tabcolsep\relax}@{}}
\caption*{\hypertarget{fs-doxcn7zk1rSsWATUotJyeAIWbWg}{}Table S5a. Controlled-comparison protocols for AO sorting and active SCC scheduling.}\\
\toprule
\bfseries Item & \bfseries AO sorting and batch-size scan & \bfseries Active SCC scheduling \\
\midrule
\endfirsthead
\toprule
\bfseries Item & \bfseries AO sorting and batch-size scan & \bfseries Active SCC scheduling \\
\midrule
\endhead
\midrule
\multicolumn{3}{r}{\footnotesize Continued on next page}\\
\endfoot
\bottomrule
\endlastfoot
Workload & 2,048 OMol25 configurations, AO100\textendash{}119 & Batches cycling through H\textsubscript{2}/He/LiH/CH\textsubscript{2} \\
Batch size & 1, 64, 256, with the sorting pair at B=256 & 1, 8, 32, 128 \\
Comparison variable & AO-sorting switch or original-order batch size & Fixed-capacity Graph and active-capacity scheduling chain \\
Initial electronic state & FRESH & Identical initial state and preset iteration counts \\
Initial active fraction & Natural SCC evolution & 1, 1/2, 1/4 \\
Warmup/measured runs & Full queue, 1/3 & Each path, 3/50 \\
Timing statistic & Median and minimum\textendash{}maximum & Mean \\
Timing boundary & Preparation, synchronous evaluation and output publication, with sorting outside timing & CUDA events enclose the complete SCC loop. Preparation, download and force evaluation are outside the interval. \\
Resources & One EPYC9354 physical core and one RTX5090 & EPYC7K62 and one RTX5090 \\
Correctness checks & Per-configuration energy, forces, status and iterations & Complete final states and numerical-update counts for both paths, with an additional CPU reference for fully active batches \\
\end{longtable}
\endgroup

\FloatBarrier
\Needspace{8\baselineskip}
\hypertarget{fs-doxcngDK18S54LpejBLNa3msblb}{}
\subsection*{S5.2 Variable-size batches and context reuse}
\addcontentsline{toc}{subsection}{S5.2 Variable-size batches and context reuse}

\hypertarget{fs-doxcnlohjmTzheviW3aObCIK99d}{}
The variable-size experiment used the final 256 configurations of the mixed OMol25 queue in its original order. All of these configurations completed evaluation in the CPU and GPU configurations used here. Calculations used host input/output, a 300 K electronic temperature, up to 500 SCC iterations and FRESH initial states. The strict configuration set the SCC state-residual RMS tolerance to 10\textsuperscript{-6} and the free-energy-change tolerance to 10\textsuperscript{-8} Eh. Each execution configuration completed one full warmup and three measured runs. End-to-end timing included input preparation, synchronous evaluation and output publication.

\begingroup
\fontsize{9.5}{12.8}\selectfont
\setlength{\tabcolsep}{3pt}
\renewcommand{\arraystretch}{1.18}
\setlength{\LTcapwidth}{\linewidth}
\hypertarget{fs-doxcnBW9rxjCSVRFOoR0VimN0R4}{}
\begin{longtable}{@{}>{\RaggedRight\arraybackslash}p{\dimexpr 0.250000\linewidth-1.500000\tabcolsep\relax}>{\RaggedRight\arraybackslash}p{\dimexpr 0.250000\linewidth-1.500000\tabcolsep\relax}>{\RaggedRight\arraybackslash}p{\dimexpr 0.250000\linewidth-1.500000\tabcolsep\relax}>{\RaggedRight\arraybackslash}p{\dimexpr 0.250000\linewidth-1.500000\tabcolsep\relax}@{}}
\caption*{\hypertarget{fs-doxcnoiLCfWE4zlUxGhXggMiFvh}{}Table S5b. Supplementary timing for the final healthy batch of 256 variable-size configurations.}\\
\toprule
\bfseries Execution configu\-ration & \bfseries Tolerance setting & \bfseries Median end-to-end time / s & \bfseries Completed per traversal \\
\midrule
\endfirsthead
\toprule
\bfseries Execution configu\-ration & \bfseries Tolerance setting & \bfseries Median end-to-end time / s & \bfseries Completed per traversal \\
\midrule
\endhead
\midrule
\multicolumn{4}{r}{\footnotesize Continued on next page}\\
\endfoot
\bottomrule
\endlastfoot
CPU, per-system & Strict & 8.964 & 256/256 \\
GPU, original order, B=256 & Strict & 29.525 & 256/256 \\
GPU, original order, B=64 & Strict & 30.027 & 256/256 \\
GPU B=64, sorted by prior iteration count & Strict & 24.944 & 256/256 \\
GPU B=64, sorted by AO count & Strict & 26.067 & 256/256 \\
GPU, original order, B=256 & Loose & 22.762 & 256/256 \\
\end{longtable}
\endgroup

\hypertarget{fs-doxcnXJljcNSpFdRNfIyw7wRCif}{}
The loose configuration used state-residual RMS and free-energy-change tolerances of 10\textsuperscript{-4} and 10\textsuperscript{-6} Eh, respectively. Sorting by prior iteration count used iteration counts from preceding per-system GPU evaluations as a retrospective grouping comparison. Both AO sorting and iteration-count sorting were performed outside timing. Relative to original-order B=64, sorting by prior iteration count and by AO count reduced synchronous evaluation time by approximately 16.9\% and 13.2\%, respectively, using unrounded values.

\hypertarget{fs-doxcnr4dY9eiGeNPzJvXsD4Bknb}{}
The strict original-order B=256 batch contained 100 exact-AO solver groups and required up to 95 SCC iterations. The general masked path retained each group's solver capacity and continued until the final active member terminated. Equal-dimension work was distributed among many small groups, while SCC convergence also had a long tail. Both features affected execution organization for this batch.

\hypertarget{fs-doxcnn4vx7KeNRhiNVaF7ZS3bQd}{}
A separate comparison fixed the same healthy final batch and four original-order B=64 batches, changing only context ownership. With one shared context across the four batches, median end-to-end and synchronous evaluation times were 30.052 and 29.961 s, respectively. With a separate retained context per batch, they were 29.013 and 28.928 s. The two configurations produced identical energies, a maximum force-component difference of 1.390\ensuremath{\times}10\textsuperscript{-16} Eh/bohr and identical SCC iteration counts.

\FloatBarrier
\Needspace{8\baselineskip}
\hypertarget{fs-doxcnrhWuecW4dwAnhyTWbgXPuf}{}
\subsection*{S5.3 AO, capacity and active-fraction scans for active-capacity scheduling}
\addcontentsline{toc}{subsection}{S5.3 AO, capacity and active-fraction scans for active-capacity scheduling}

\hypertarget{fs-doxcnAETCjJt53DsGJEG9Jelsyf}{}
To examine the operating range of active-capacity scheduling, we compared the existing device scheduling chain with a fixed-capacity device-tail Graph on an RTX 5090. Let n denote the AO count, B the original capacity of a same-AO-dimension solver bucket, and A the number of currently active systems. The active fraction is A/B. Synthetic component probes were used for 40, 41 and 180 AOs. C\textsubscript{10}H\textsubscript{22} and C\textsubscript{20}H\textsubscript{42} supplied the 62- and 122-AO probes, respectively.

\hypertarget{fs-doxcnhGJCqrV9jMJLq2KRKxQzdf}{}
Both paths restored the same FRESH checkpoint and termination mask. Preset iteration counters controlled active membership, and at most one SCC iteration was executed. CUDA events measured that iteration's graph-execution latency, including each path's organization, dispatch and solver work. Initialization, final-state readback and comparison were performed separately. The build used GCC 13.3.0, CUDA 12.9.86, Release mode and sm120, with driver 580.95.05.

\hypertarget{fs-doxcnZUGVkbXdu8RKz4PAUcLkib}{}
The single-iteration masked-dispatch comparison measures one SCC graph execution from the same checkpoint.

\hypertarget{fs-doxcnhdfM9rAKs0JaEWIH45Dfog}{}
The initial scan contained 81 coordinates. The five AO probes used A=128, 96, 64, 32, 16, 8 and 0 at B=128. The 41-, 122- and 180-AO probes also used A=B, B/4 and max(1,B/16) at B=8, 32, 64 and 256. All five probes additionally tested A=1 and 0 at B=1. A supplementary scan then added 16 points for 122/180 AOs and B=128/256 at A=8, 4, 2 and 1. The 8 points with B=256 were independently rechecked. Together, the stages covered 95 unique coordinates and 105 stage-specific observations.

\hypertarget{fs-doxcnF9RMcpcHIxobOE5uf4QANe}{}
One warmup pair was excluded at each point. The initial and supplementary scans used four timing pairs per point, and the independent recheck used six. The paths were interleaved in AB/BA order with a fixed random seed of 515. Statistics were calculated separately for each stage. Relative cost was represented by R, the median of the paired scheduling-chain/fixed-capacity latency ratios. The prespecified candidate-benefit margin required R\ensuremath{\leq}0.900 and every timing-pair ratio below 1. The initial scan used one 5090, while the supplementary scan and independent recheck used another GPU of the same model.

\begingroup
\fontsize{9.5}{12.8}\selectfont
\setlength{\tabcolsep}{3pt}
\renewcommand{\arraystretch}{1.18}
\setlength{\LTcapwidth}{\linewidth}
\hypertarget{fs-doxcnZvLlrHcVufkwQH5Gw3F8mb}{}
\begin{longtable}{@{}>{\RaggedRight\arraybackslash}p{\dimexpr 0.162500\linewidth-1.300000\tabcolsep\relax}>{\RaggedRight\arraybackslash}p{\dimexpr 0.200000\linewidth-1.600000\tabcolsep\relax}>{\RaggedRight\arraybackslash}p{\dimexpr 0.225000\linewidth-1.800000\tabcolsep\relax}>{\RaggedRight\arraybackslash}p{\dimexpr 0.225000\linewidth-1.800000\tabcolsep\relax}>{\RaggedRight\arraybackslash}p{\dimexpr 0.187500\linewidth-1.500000\tabcolsep\relax}@{}}
\caption*{\hypertarget{fs-doxcndpw9Xdtk9OgGNjUFxiV0Md}{}Table S5c. Single-iteration SCC component latency at 122 AOs and B=128. Scheduling-chain and fixed-capacity columns report their respective median latencies. R is the median paired latency ratio, with R>1 indicating a slower scheduling chain. A=0 is the zero-active control.}\\
\toprule
\bfseries Active systems A & \bfseries Active fraction / \% & \bfseries Scheduling chain / ms & \bfseries Fixed capacity / ms & \bfseries \added{R} \\
\midrule
\endfirsthead
\toprule
\bfseries Active systems A & \bfseries Active fraction / \% & \bfseries Scheduling chain / ms & \bfseries Fixed capacity / ms & \bfseries \added{R} \\
\midrule
\endhead
\midrule
\multicolumn{5}{r}{\footnotesize Continued on next page}\\
\endfoot
\bottomrule
\endlastfoot
\added{128} & \added{100} & \added{126.787} & \added{14.016} & \added{9.045} \\
\added{96} & \added{75} & \added{97.872} & \added{13.164} & \added{7.434} \\
\added{64} & \added{50} & \added{68.734} & \added{11.984} & \added{5.735} \\
\added{32} & \added{25} & \added{39.537} & \added{11.013} & \added{3.589} \\
\added{16} & \added{12.500} & \added{24.933} & \added{10.356} & \added{2.408} \\
\added{8} & \added{6.250} & \added{17.842} & \added{10.276} & \added{1.736} \\
\added{0} & \added{0} & \added{0.012} & \added{0.012} & \added{1.000} \\
\end{longtable}
\endgroup

\hypertarget{fs-doxcnnKLXZoi39kMh5xmTpd0uDe}{}
All 71 nonzero-active coordinates in the initial scan had R greater than 1.020. There were also 10 zero-active controls.

\begingroup
\fontsize{9.2}{12.4}\selectfont
\setlength{\tabcolsep}{3pt}
\renewcommand{\arraystretch}{1.18}
\setlength{\LTcapwidth}{\linewidth}
\hypertarget{fs-doxcnGN4BsbiDdrAT9OeMEPumRb}{}
\begin{longtable}{@{}>{\RaggedRight\arraybackslash}p{\dimexpr 0.212500\linewidth-2.125000\tabcolsep\relax}>{\RaggedRight\arraybackslash}p{\dimexpr 0.137500\linewidth-1.375000\tabcolsep\relax}>{\RaggedRight\arraybackslash}p{\dimexpr 0.162500\linewidth-1.625000\tabcolsep\relax}>{\RaggedRight\arraybackslash}p{\dimexpr 0.162500\linewidth-1.625000\tabcolsep\relax}>{\RaggedRight\arraybackslash}p{\dimexpr 0.162500\linewidth-1.625000\tabcolsep\relax}>{\RaggedRight\arraybackslash}p{\dimexpr 0.162500\linewidth-1.625000\tabcolsep\relax}@{}}
\caption*{\hypertarget{fs-doxcnDQq4V3h4KJxDr8mrq7n2kd}{}Table S5d. Paired latency ratio R at low active counts. Four pairs were used in the supplementary scan and six in the independent recheck. Column headings give the active count A.}\\
\toprule
\bfseries \added{AO / B} & \bfseries Timing pairs & \bfseries \added{A=8} & \bfseries \added{A=4} & \bfseries \added{A=2} & \bfseries \added{A=1} \\
\midrule
\endfirsthead
\toprule
\bfseries \added{AO / B} & \bfseries Timing pairs & \bfseries \added{A=8} & \bfseries \added{A=4} & \bfseries \added{A=2} & \bfseries \added{A=1} \\
\midrule
\endhead
\midrule
\multicolumn{6}{r}{\footnotesize Continued on next page}\\
\endfoot
\bottomrule
\endlastfoot
\added{122 / 128} & \added{4} & \added{1.728} & \added{1.403} & \added{1.218} & \added{1.161} \\
\added{122 / 256} & \added{4} & \added{1.147} & \added{0.948} & \added{0.836} & \added{0.805} \\
\added{122 / 256} & \added{6} & \added{1.149} & \added{0.946} & \added{0.838} & \added{0.805} \\
\added{180 / 128} & \added{4} & \added{1.721} & \added{1.357} & \added{1.149} & \added{1.011} \\
\added{180 / 256} & \added{4} & \added{1.164} & \added{0.933} & \added{0.812} & \added{0.730} \\
\added{180 / 256} & \added{6} & \added{1.164} & \added{0.932} & \added{0.813} & \added{0.730} \\
\end{longtable}
\endgroup

\hypertarget{fs-doxcnmoZMfvHGRQgVRFXp7Xx3re}{}
For 122/180 AOs and B=128, the supplementary scan did not meet the candidate-benefit margin even at A=1. At B=256 and A=4, the recheck showed latency reductions of approximately 5.400\%\textendash{}6.800\%. The prespecified margin was reached at A=1 or 2, with median recheck ratios of 0.730\textendash{}0.838. These points corresponded to active fractions of 0.391\%\textendash{}0.781\% and component-latency reductions of approximately 16.200\%\textendash{}27.000\%.

\hypertarget{fs-doxcn6vTwmPIKzkNgwKQ4kv9Ymh}{}
The benefit region was concentrated at very small active sets. This supports the current restriction of active-capacity scheduling to validated small-AO layouts, with fixed-capacity execution retained for other layouts.

\FloatBarrier
\Needspace{8\baselineskip}
\hypertarget{fs-doxcnb1s1teVChJVUee15L60jhb}{}
\subsection*{S5.4 Natural-size and padded AO matrix layouts}
\addcontentsline{toc}{subsection}{S5.4 Natural-size and padded AO matrix layouts}

\hypertarget{fs-doxcnc5xVCdnb3fUZ2UtbuuAxvf}{}
To isolate the effect of the principal AO matrix layout, we compared A, the ragged natural-size n\ensuremath{\times}n layout, with B, the padded-natural layout stored as m\ensuremath{\times}m with a uniform stride, where m is the largest AO count in the batch. Both used the same experimental inputs and software implementation. Padding covered overlap and Hamiltonian matrices, dipole and quadrupole integral planes, orbital coefficients, density and energy-weighted density, and associated workspaces and analytic-force routines that consume these matrices. Required packing and retrieval were included in call cost. Both variants retained physical loops and divide-and-conquer eigensolution at the true dimension n and used the same exact-AO grouping.

\hypertarget{fs-doxcnHbv4BcEkGhMBuK4Krc4E4g}{}
The near-size queue reused the 2,048 AO100\textendash{}119 configurations. The variable-size queue retained all 1,317 configurations with AO8\textendash{}180 that satisfied the input criteria of neutrality, singlet multiplicity, no unpaired electrons, light main-group elements, no metal or heavy-element flags and AO\ensuremath{\leq}180. Selection did not use convergence or timing results. In the sorted condition, inputs were ordered by (AO, configuration identifier) and divided into B=256 batches. The near-size queue formed 8 batches, and the variable-size queue formed 5\ensuremath{\times}256+37. The strongly mixed condition reused the same variable-size set. After AO/identifier sorting, it was split into six contiguous quantile groups of sizes 220, 220, 220, 219, 219 and 219. Round-robin interleaving across the groups was followed by batching as 5\ensuremath{\times}256+37 and sorting only within each batch. Each configuration was used once. The sums of exact-AO group counts over batches were 27, 164 and 192 for the near-size, sorted variable-size and strongly mixed conditions, respectively. These are static group counts, with dynamic GPU occupancy remaining a separate quantity.

\hypertarget{fs-doxcnmfxU6hvnSlq3L0CW4Rn9Kx}{}
All A/B configurations used double-precision GFN2-xTB, host input/output, strict FRESH initial states, a 300 K electronic temperature and up to 500 SCC iterations. SCC state-residual RMS and free-energy-change tolerances were 10\textsuperscript{-6} and 10\textsuperscript{-8} Eh, respectively. CPU and BLAS execution were limited to one thread on a single physical core, with one RTX 5090. The build used CUDA 12.9.86 and Release mode. Each mode completed one full warmup and three measured traversals. The strongly mixed condition used measured orders BA, AB and BA. Primary time included per-batch preparation, synchronous complete energy-and-force evaluation and host-result publication. File reading, sorting, initial context creation and result checks were excluded. Memory was measured in separate processes, with one traversal per configuration and NVML sampling of the target process every 20 ms, independently of the measured timing runs.

\begingroup
\fontsize{8.8}{11.9}\selectfont
\setlength{\tabcolsep}{3pt}
\renewcommand{\arraystretch}{1.18}
\setlength{\LTcapwidth}{\linewidth}
\hypertarget{fs-doxcnGT3glt0ERbTjRE1ahlMwDg}{}
\begin{longtable}{@{}>{\RaggedRight\arraybackslash}p{\dimexpr 0.174699\linewidth-2.096386\tabcolsep\relax}>{\RaggedRight\arraybackslash}p{\dimexpr 0.265060\linewidth-3.180723\tabcolsep\relax}>{\RaggedRight\arraybackslash}p{\dimexpr 0.126506\linewidth-1.518072\tabcolsep\relax}>{\RaggedRight\arraybackslash}p{\dimexpr 0.108434\linewidth-1.301205\tabcolsep\relax}>{\RaggedRight\arraybackslash}p{\dimexpr 0.108434\linewidth-1.301205\tabcolsep\relax}>{\RaggedRight\arraybackslash}p{\dimexpr 0.102410\linewidth-1.228916\tabcolsep\relax}>{\RaggedRight\arraybackslash}p{\dimexpr 0.114458\linewidth-1.373494\tabcolsep\relax}@{}}
\caption*{\hypertarget{fs-doxcnYEOslGsD1bG8Aw77E8Motf}{}Table S5e. Principal AO matrix layouts within the same software implementation. A uses natural-size storage, and B uses padded storage with true-dimension solves. All reductions use B as the denominator. Times are medians and minimum\textendash{}maximum ranges over three measured traversals. Sampled peaks come from separate memory traversals. Capacity is the queue-wide sum of storage-element counts for one principal AO matrix per batch, a quantity distinct from peak GPU memory and FLOPs.}\\
\toprule
\bfseries Queue (configu\-rations, AO) & \bfseries Median E+F time [minimum, maximum] / s & \bfseries A / B sampled peak / MiB & \bfseries Principal-matrix capacity reduction / \% & \bfseries Sampled-peak reduction / \% & \bfseries Time reduction / \% & \bfseries A, B successes/\allowbreak{}attempts per traversal \\
\midrule
\endfirsthead
\toprule
\bfseries Queue (configu\-rations, AO) & \bfseries Median E+F time [minimum, maximum] / s & \bfseries A / B sampled peak / MiB & \bfseries Principal-matrix capacity reduction / \% & \bfseries Sampled-peak reduction / \% & \bfseries Time reduction / \% & \bfseries A, B successes/\allowbreak{}attempts per traversal \\
\midrule
\endhead
\midrule
\multicolumn{7}{r}{\footnotesize Continued on next page}\\
\endfoot
\bottomrule
\endlastfoot
Near-size, AO-sorted (2,048, 100\textendash{}119) & \added{A 29.339 [29.249, 30.197], B 29.395 [29.360, 29.574]} & \added{7288 / 7364} & \added{2.16} & \added{1.03} & \added{0.19} & \added{2048/2048} \\
Variable-size, AO-sorted (1,317, 8\textendash{}180) & \added{A 140.847 [140.843, 140.870], B 141.389 [141.371, 141.672]} & \added{5324 / 7702} & \added{34.36} & \added{30.88} & \added{0.38} & \added{1314/1317} \\
Variable-size, strongly mixed (1,317, 8\textendash{}180) & \added{A 82.827 [82.737, 83.006], B 86.296 [86.251, 86.543]} & \added{3802 / 11572} & \added{75.90} & \added{67.14} & \added{4.02} & \added{1314/1317} \\
\end{longtable}
\endgroup

\hypertarget{fs-doxcnic0Dv0orVk90gr14nAYuhg}{}
For the near-size, AO-sorted queue (2,048, 100\textendash{}119), the three A times were 30.197343, 29.338962 and 29.249334 s, and the B times were 29.360499, 29.394766 and 29.573740 s. Total element counts for one principal matrix were 25,136,660/25,691,392 for A/B.

\hypertarget{fs-doxcnnAR3U8miQXINSxgz0WLJjg}{}
For the variable-size, AO-sorted queue (1,317, 8\textendash{}180), the three A times were 140.842652, 140.846581 and 140.869701 s, and the B times were 141.371168, 141.388672 and 141.671975 s. Total element counts for one principal matrix were 7,177,368/10,934,480 for A/B.

\hypertarget{fs-doxcnCV6gUT0jaqJDamJbMPQVAb}{}
For the variable-size, strongly mixed queue (1,317, 8\textendash{}180), the three A times were 83.005766, 82.737267 and 82.827109 s, and the B times were 86.543091, 86.296201 and 86.250849 s. Total element counts for one principal matrix were 7,177,368/29,775,568 for A/B.

\hypertarget{fs-doxcnFyfvPIYCVoDz4rG6GOHRKh}{}
A was faster than B in all three strongly mixed runs. The median time differences were below 1\% for both AO-sorted queues. Sorted and strongly mixed conditions used the same RTX 5090 model and driver 580.95.05. A/B comparisons within each condition were matched on the same physical GPU.

\hypertarget{fs-doxcnGxjC8DUCkXve5Foz1XSusg}{}
Paired thresholds required maximum absolute differences of at most 10\textsuperscript{-10} Eh for successful-system energies, 10\textsuperscript{-10} Eh/bohr for forces and 10\textsuperscript{-10} e for precheck charges. System identifiers, atom offsets, statuses, convergence flags and SCC iteration counts also had to match. A/B prechecks and measured-output verification passed. Measured traversals requested energy and forces, with charges checked in the precheck. For the complete strongly mixed queue, precheck energy and charge differences were zero, and the maximum force difference was 2.22\ensuremath{\times}10\textsuperscript{-16} Eh/bohr. Across batching arrangements, statuses, convergence flags and iteration counts were unchanged. Maximum energy and force differences were 5.26\ensuremath{\times}10\textsuperscript{-13} Eh and 4.45\ensuremath{\times}10\textsuperscript{-13} Eh/bohr, respectively.

\hypertarget{fs-doxcn579XGE7IZuC2Gbr0EfD64g}{}
Every near-size traversal succeeded on 2,048/2,048 configurations. Both variable-size batching arrangements succeeded on 1,314/1,317. The same three configurations, \code{CO_7270261446849249421858501}, \code{CO_9343753771713800272817479} and \code{CO_9847448562207090849131680}, remained SCC-nonconverged after 500 iterations, with all requested floating-point output slices filled with NaN. Failed systems were retained in timing and in the attempted-throughput denominator.

\hypertarget{fs-doxcntIdgU1OYLDMX8Pe4ldfQtb}{}
A separate precheck of C, padded-unified, also expanded the solver dimension to m after padding storage. In the 37-configuration variable-size diagnostic, its maximum charge difference was 1.2638\ensuremath{\times}10\textsuperscript{-10} e, exceeding the prespecified threshold. C therefore did not enter measured timing or memory comparisons.

\hypertarget{fs-doxcneTb0gxC2UcQXV90HgHk9kf}{}
Two complete batches of 256 systems each passed CUDA memory checking. CUDA synchronization checking stopped after producing 512 diagnostics for the natural-size layout. Synchronization checks remain incomplete for the two padded-storage variants using natural or expanded solver dimensions.

\hypertarget{fs-doxcnnhqvRp0Z6dldGA6EitDbZg}{}
Timings in this section are reported separately from the main performance timings in S4.

\FloatBarrier
\Needspace{9\baselineskip}
\hypertarget{fs-doxcnsV2Qev1xbr9xUupI67Jrfg}{}
\section*{S6. Repeated evaluation performance for conformers with a common topology}
\addcontentsline{toc}{section}{S6. Repeated evaluation performance for conformers with a common topology}

\FloatBarrier
\Needspace{8\baselineskip}
\hypertarget{fs-doxcn9cJa25OzOm6qhpnPYUEW3d}{}
\subsection*{S6.1 Conformer generation and comparison protocol}
\addcontentsline{toc}{subsection}{S6.1 Conformer generation and comparison protocol}

\hypertarget{fs-doxcndLvt8iuQ83Vk2F1RWnb7Ph}{}
One alkane composition and topology were used at each molecular size. The first conformer in each batch retained the ideal geometry. Independent Gaussian perturbations with a standard deviation of 0.020 bohr were applied to every Cartesian coordinate of the remaining conformers. The random seed was the atom count\ensuremath{\times}1000+B. All conformers were neutral and used a restricted spin treatment. GFN2-xTB energies and forces were evaluated in double precision. Because the seed included B, the conformer lists could differ between batch sizes.\sourcebreak{}Tests ran on an AMD EPYC 9354 host with an NVIDIA RTX 5090. The host process of each engine was pinned to one physical CPU core. The xTBloom CPU, OMP, and BLAS thread counts were all 1. The PyTorch intra-op and inter-op thread counts for dxtb were also 1. The xTB and tblite adapters called their public single-system interfaces sequentially. Contexts and descriptors were prepared before timing, and the complete batch of energies and forces was available on the host when timing ended. Each configuration received one warmup and three measured runs. A single-threaded tblite calculation provided the numerical reference for each panel.

\begingroup
\fontsize{9.2}{12.4}\selectfont
\setlength{\tabcolsep}{3pt}
\renewcommand{\arraystretch}{1.18}
\setlength{\LTcapwidth}{\linewidth}
\hypertarget{fs-doxcnWxqn9HWw4XT6x8iS9qITog}{}
\begin{longtable}{@{}>{\RaggedRight\arraybackslash}p{\dimexpr 0.300000\linewidth-0.600000\tabcolsep\relax}>{\RaggedRight\arraybackslash}p{\dimexpr 0.700000\linewidth-1.400000\tabcolsep\relax}@{}}
\caption*{\hypertarget{fs-doxcnk20v6n7YB7SGLQx6N5bkie}{}Table S6a. Initialization and convergence settings for the common-topology conformer benchmark.}\\
\toprule
\bfseries Item & \bfseries Setting \\
\midrule
\endfirsthead
\toprule
\bfseries Item & \bfseries Setting \\
\midrule
\endhead
\midrule
\multicolumn{2}{r}{\footnotesize Continued on next page}\\
\endfoot
\bottomrule
\endlastfoot
Cold-electronic-state panels & B=1 and B=512. xTBloom performs FRESH initialization inside the call. xTB/tblite recreate the calculation object outside the measured interval. The dxtb call includes reset. \\
Auto-warm panel & B=128. After an untimed cold seed, xTBloom uses WARM, xTB/tblite reuse persistent objects, and dxtb uses reset/cold. \\
xTBloom SCC & State-residual RMS tolerance of 10\textsuperscript{-4}, free-energy-change tolerance of 10\textsuperscript{-6} Eh, and a maximum of 500 iterations \\
xTB and tblite & accuracy=1.0 and a maximum of 500 iterations \\
dxtb & \code{x_atol}=10\textsuperscript{-4}, \code{x_atol_max}=10\textsuperscript{-5}, \code{f_atol}=10\textsuperscript{-4}, \code{force_convergence} enabled, and a maximum of 500 iterations \\
Output comparison & Against the tblite reference in the same panel, maximum energy difference\ensuremath{\leq}2\ensuremath{\times}10\textsuperscript{-3} Eh and maximum force-component difference\ensuremath{\leq}2\ensuremath{\times}10\textsuperscript{-3} Eh/bohr \\
Repeatability check & Applies only to cold/reset calls, with tolerances of 10\textsuperscript{-10} Eh for energy and 10\textsuperscript{-8} Eh/bohr for force components \\
Reported statistics & Median and minimum\textendash{}maximum range of three measured runs \\
\end{longtable}
\endgroup

\begin{table}[H]
\begingroup
\fontsize{10}{13.5}\selectfont
\setlength{\tabcolsep}{3pt}
\renewcommand{\arraystretch}{1.18}
\setlength{\LTcapwidth}{\linewidth}
\hypertarget{fs-doxcnYpAeJWyNelBRd5rSwSRUye}{}
\caption*{\hypertarget{fs-doxcntSjOBsTtoH4hHlb94sLefh}{}Table S6b. Molecular compositions and AO dimensions in the common-topology conformer benchmark.}
\begin{tabular}{@{}>{\RaggedRight\arraybackslash}p{\dimexpr 0.333333\linewidth-1.333333\tabcolsep\relax}>{\RaggedRight\arraybackslash}p{\dimexpr 0.333333\linewidth-1.333333\tabcolsep\relax}>{\RaggedRight\arraybackslash}p{\dimexpr 0.333333\linewidth-1.333333\tabcolsep\relax}@{}}
\toprule
\bfseries Atom count & \bfseries Composition & \bfseries GFN2-xTB AO count \\
\midrule
14 & C\textsubscript{4}H\textsubscript{10} & 26 \\
32 & C\textsubscript{10}H\textsubscript{22} & 62 \\
62 & C\textsubscript{20}H\textsubscript{42} & 122 \\
122 & C\textsubscript{40}H\textsubscript{82} & 242 \\
\added{242} & \added{C\textsubscript{80}H\textsubscript{162}} & \added{482} \\
\added{362} & \added{C\textsubscript{120}H\textsubscript{242}} & \added{722} \\
\bottomrule
\end{tabular}
\endgroup
\end{table}

\clearpage
\Needspace{8\baselineskip}
\hypertarget{fs-doxcnqIdhJyI5nTjHoxInkd7rnh}{}
\subsection*{S6.2 Complete size-scaling curves and numerical checks}
\addcontentsline{toc}{subsection}{S6.2 Complete size-scaling curves and numerical checks}

\begin{figure}[H]
\centering
\hypertarget{fs-doxcnc0X4nAuNEvwNZJf920mc3e}{}
\includegraphics[width=\linewidth,height=\ifShowRevisions .74\else .775\fi \textheight,keepaspectratio]{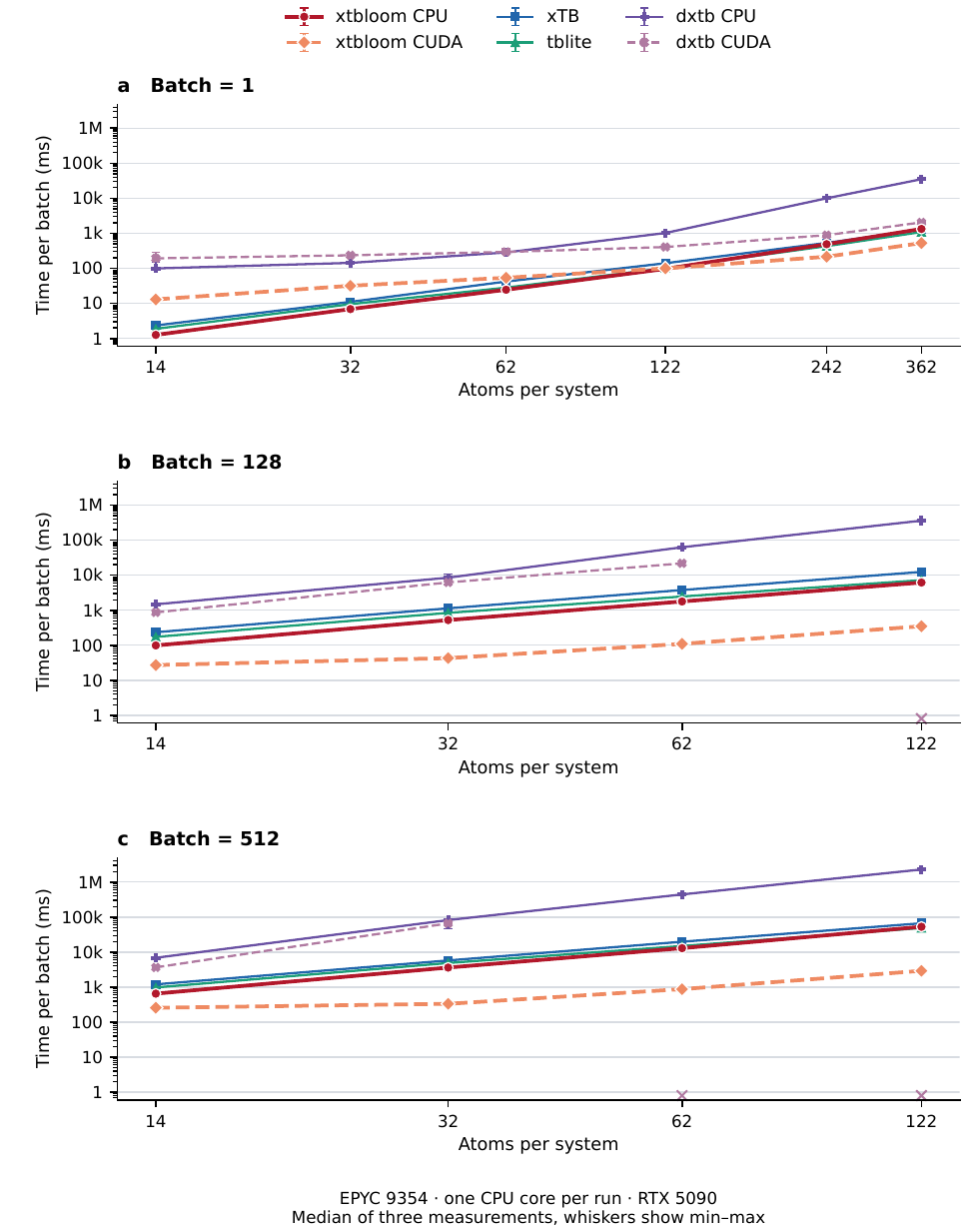}
\captionsetup{font=footnotesize}
\caption*{\hypertarget{fs-doxcnpcxZT96wQTLqtwAMvhv5xt}{}Figure S4. Energy-and-force evaluation latency for locally perturbed conformers with a common topology under a single-physical-CPU-core constraint. (a) B=1 with a cold electronic state for systems containing 14, 32, 62, 122, 242, and 362 atoms. (b) B=128 with auto-warm initialization. (c) B=512 with a cold electronic state. The latter two panels use 14, 32, 62, and 122 atoms. The GPU is an RTX 5090. Each execution path uses one physical CPU core, with OMP, BLAS, and PyTorch thread counts of 1. The vertical axis gives the latency to evaluate the complete batch. Error bars show the minimum\textendash{}maximum range of three measured runs. Gaps or crosses indicate unavailable configurations or GPU out-of-memory failures.}
\end{figure}
Across the common-topology tests, the maximum energy and force-component differences of xTBloom CPU and CUDA from the same-panel tblite reference are 4.571\ensuremath{\times}10\textsuperscript{-7} Eh and 2.932\ensuremath{\times}10\textsuperscript{-5} Eh/bohr. All timed configurations returned finite energies and forces. Results from xTBloom, native xTB, and dxtb passed the same-panel tblite reference checks. GPU out-of-memory failures prevented completion for dxtb-cuda at B=512 with 62 atoms, dxtb-cuda at B=512 with 122 atoms, and dxtb-cuda at B=128 with 122 atoms.

The xTBloom measurements use a Release build of version ab3a3bc6132c, CUDA toolkit 12.9.1, and target architecture sm120. The remaining versions are native xTB 6.7.1, tblite 0.7.0, LP64 OpenBLAS 0.3.30, dxtb 0.4.0, PyTorch 2.10.0+cu128, and tad-libcint 0.3.0.

\FloatBarrier
\Needspace{9\baselineskip}
\hypertarget{fs-doxcn8pxkeoB7zIVGOTYraUEJuf}{}
\section*{S7. QM/MM system, biases, and dynamics parameters}
\addcontentsline{toc}{section}{S7. QM/MM system, biases, and dynamics parameters}

\hypertarget{fs-doxcn0lTYMvfDuhqTXEHu1RjOLd}{}
We used the ethylene phosphate model reaction with a methoxy leaving group from \citep{Giese_JPhysChemA_2022_v126_p8519}. The periodic system contains 8,938 real atoms. Its 16-atom QM region comprises two complete molecules with composition P\textsubscript{1}O\textsubscript{5}C\textsubscript{3}H\textsubscript{7}. The environment contains 2,974 TIP4P-Ew water molecules~\citep{Horn_JChemPhys_2004_v120_p9665}. Including the virtual charge sites of the water model gives 11,912 electrostatic sites. The QM region has a total charge of \ensuremath{-}2e and uses closed-shell GFN2-xTB without link atoms. The bare QM/MM and DPRc-corrected sampling use the same classical environment and geometric restraints.

\hypertarget{fs-doxcnS92ItnFsTyoOLDXXl3VmYf}{}
The reaction coordinate is \ensuremath{\xi}=r(P,O5\ensuremath{\prime})\ensuremath{-}r(P,O2\ensuremath{\prime})=d(1,12)\ensuremath{-}d(5,1), where 1=P, 5=O2\ensuremath{\prime} is the attacking oxygen, and 12=O5\ensuremath{\prime} is the leaving oxygen. The attack angle is \ensuremath{\angle}(5,1,12). LAMMPS~\citep{Thompson_ComputPhysCommun_2022_v271_p108171} propagates the dynamics with a Langevin thermostat~\citep{Schneider_PhysRevB_1978_v17_p1302}, and SHAKE~\citep{Ryckaert_JComputPhys_1977_v23_p327} constrains the water geometry. Colvars~\citep{Fiorin_MolPhys_2013_v111_p3345} applies the harmonic biases for umbrella sampling~\citep{Torrie_JComputPhys_1977_v23_p187,Kastner_WiresComputMolSci_2011_v1_p932}. At each step, a coordinator collects inputs from 48 windows for cross-window QM/MM batching on one shared GPU. The batched work covers GFN2-xTB evaluation, classical nonbonded interactions, periodic electrostatics using the particle--particle particle--mesh (PPPM) method with fast Fourier transforms (FFTs), and DeePMD/DPRc inference when enabled. Each window retains its own coordinates, integrator, and SCC history. Periodic electrostatic preprocessing, postprocessing, and force assembly follow Section 3.4 of the main text.

\begingroup
\fontsize{10}{13.5}\selectfont
\setlength{\tabcolsep}{3pt}
\renewcommand{\arraystretch}{1.18}
\setlength{\LTcapwidth}{\linewidth}
\hypertarget{fs-doxcn55VFD8rAT4xxTaonVyGdjg}{}
\begin{longtable}{@{}>{\RaggedRight\arraybackslash}p{\dimexpr 0.300000\linewidth-0.600000\tabcolsep\relax}>{\RaggedRight\arraybackslash}p{\dimexpr 0.700000\linewidth-1.400000\tabcolsep\relax}@{}}
\caption*{\hypertarget{fs-doxcn7DmlEHSewQ1njgYD127qYe}{}Table S7. System and dynamics parameters for QM/MM umbrella sampling.}\\
\toprule
\bfseries Parameter & \bfseries Setting \\
\midrule
\endfirsthead
\toprule
\bfseries Parameter & \bfseries Setting \\
\midrule
\endhead
\midrule
\multicolumn{2}{r}{\footnotesize Continued on next page}\\
\endfoot
\bottomrule
\endlastfoot
QM region and environment & 16-atom P\textsubscript{1}O\textsubscript{5}C\textsubscript{3}H\textsubscript{7}, charge \ensuremath{-}2e, closed shell, and 2,974 TIP4P-Ew water molecules \\
Window centers & \ensuremath{-}3.100 to 1.600 Å in increments of 0.100 Å, giving 48 windows \\
Harmonic force constant for the reaction coordinate & 200 kcal mol\textsuperscript{-1} Å\textsuperscript{-2} \\
Attack-angle center and harmonic force constant & 180\ensuremath{^{\circ}}/0.050 kcal mol\textsuperscript{-1} degree\textsuperscript{-2} \\
Equilibration per window & 100 ps \\
Production per window & Three independent random-number streams, each with 250 ps \\
Total production sampling per simulation set & 36 ns \\
Time step & 1 fs \\
Thermostat & Langevin thermostat at 298 K with damping time 200 fs \\
Water constraints & SHAKE with tolerance 10\textsuperscript{-6} and a maximum of 200 iterations \\
Lennard\textendash{}Jones cutoff & 9 Å \\
Real-space electrostatic cutoff & 9 Å \\
Periodic electrostatics & Fourth-order TIP4P/PPPM, a 50\ensuremath{\times}50\ensuremath{\times}50 grid, and target accuracy 10\textsuperscript{-6} \\
Ewald splitting parameter & 3.488\ensuremath{\times}10\textsuperscript{-1} Å\textsuperscript{-1} \\
GFN2-xTB SCC control & accuracy=0.001 and a maximum of 250 iterations \\
Electronic temperature & 300 K, with kBT = 9.500\ensuremath{\times}10\textsuperscript{-4} E\textsubscript{h} for electronic occupation broadening \\
Periodic QM response parameters & kmax=(8,8,8), ksqmax=100 \\
Recording interval for the reaction coordinate and angle & 25 fs \\
Production GPU & One NVIDIA RTX PRO6000 Blackwell \\
\end{longtable}
\endgroup

\hypertarget{fs-doxcn4RVH3YmO7MLkhaSoqTRNLb}{}
The 300 K value in the table sets the numerical broadening of electronic occupations, with kBT = 9.500\ensuremath{\times}10\textsuperscript{-4} E\textsubscript{h}. The MD thermostat and WHAM analysis both use 298 K.

\FloatBarrier
\Needspace{8\baselineskip}
\hypertarget{fs-doxcncBLONTFfkI13U6gXMNIl4f}{}
\subsection*{S7.1 Single-GPU short-run cross-window timing protocol}
\addcontentsline{toc}{subsection}{S7.1 Single-GPU short-run cross-window timing protocol}

\begingroup
\fontsize{10}{13.5}\selectfont
\setlength{\tabcolsep}{3pt}
\renewcommand{\arraystretch}{1.18}
\setlength{\LTcapwidth}{\linewidth}
\hypertarget{fs-doxcnT7kMqgG12jZfE7tv23tWZb}{}
\begin{longtable}{@{}>{\RaggedRight\arraybackslash}p{\dimexpr 0.300000\linewidth-0.600000\tabcolsep\relax}>{\RaggedRight\arraybackslash}p{\dimexpr 0.700000\linewidth-1.400000\tabcolsep\relax}@{}}
\toprule
\bfseries Item & \bfseries Setting \\
\midrule
\endfirsthead
\toprule
\bfseries Item & \bfseries Setting \\
\midrule
\endhead
\midrule
\multicolumn{2}{r}{\footnotesize Continued on next page}\\
\endfoot
\bottomrule
\endlastfoot
GPU and CPU configuration & One NVIDIA RTX PRO 6000 Blackwell 96 GB and an AMD EPYC 9J14 with 48 physical cores allocated \\
Windows and parallel configuration & Every integer B=1\textendash{}48, with one MPI process and one thread per window. The remaining allocated cores are idle at small B. \\
Warmup and timing segments & 25 warmup steps followed by five consecutive 100-step segments from the same trajectory \\
Synchronized segment duration & Maximum LAMMPS loop time across windows within each segment \\
Timing boundary & Includes force evaluation, synchronization, neighbor calculations, and integration. Loading, initialization, and final file output are outside the measured interval. \\
DPRc model used for timing & DPRc model from before retraining \\
\end{longtable}
\endgroup

\hypertarget{fs-doxcnSbQtZOMrrR3B0uDu2VfMNh}{}
Let \(t^{B}_{w,r}\) denote the loop time of window w in the rth 100-step segment when B windows are batched, and let \(t^{1}_{w,r}\) denote the corresponding duration when that window runs sequentially. Segment-level aggregate throughput and paired speedup are defined as

\hypertarget{fs-doxcnNB5AsLcuxxosmJSgCfk23b}{}
\begin{equation}
\boldsymbol{R_{B,r}=\frac{100B}{\max_{w\in B}t^{B}_{w,r}}},\qquad \boldsymbol{S_B=\frac{\sum_{w\in B}\sum_{r=1}^{5}t^{1}_{w,r}}{\sum_{r=1}^{5}\max_{w\in B}t^{B}_{w,r}}}.
\end{equation}

\hypertarget{fs-doxcnb40UsH90l7gm35vMoxrCbe}{}
Throughput is reported as the median of the five consecutive segments. Shaded regions show the minimum\textendash{}maximum range. The numerator of the speedup uses the total sequential execution time for the same group of windows.

\hypertarget{fs-doxcnsCYYuGw5GgKzsoGH5Ohojg}{}
The GFN2-xTB settings in Table S7, accuracy=0.001 and a maximum of 250 iterations, apply to the QM/MM production runs.

\hypertarget{fs-doxcnhlAQMFjs7wT4qdYxWJ2b0i}{}
The PBE0 SCF, geometry-pairing, and label-assembly checks passed for all 6,560 frames. The maximum high-level/low-level geometry-pairing residual was 5.69\ensuremath{\times}10\textsuperscript{-14} Å, and the maximum classical-baseline energy residual was 5.83\ensuremath{\times}10\textsuperscript{-11} kcal mol\textsuperscript{-1}. Each of the four models was trained for 100,000 steps. All 144 window\textendash{}trial production trajectories with the retrained DPRc model reached their target step counts. The 10 paired checks, consistency between single-window and 48-window evaluation, and short NVE stability tests of this production model were recorded as passes. The maximum force-component difference between single-window and 48-window evaluation was 4.4\ensuremath{\times}10\textsuperscript{-11} kcal mol\textsuperscript{-1} Å\textsuperscript{-1}. The production model passed 5 ps NVE checks in three anchor windows both before and after equilibration.

\hypertarget{fs-doxcnns9mwq1hsfaj8AoeVGezuh}{}
In one 5 ps NVE check in the short-run timing environment, the net energy drift of one window exceeded the original threshold. The retrained production model passed the 5 ps NVE checks in three anchor windows before and after equilibration.

\begingroup
\fontsize{9.2}{12.4}\selectfont
\setlength{\tabcolsep}{3pt}
\renewcommand{\arraystretch}{1.18}
\setlength{\LTcapwidth}{\linewidth}
\hypertarget{fs-doxcn1l0p8jl2dZtX4Wv9Lpq98e}{}
\begin{longtable}{@{}>{\RaggedRight\arraybackslash}p{\dimexpr 0.127371\linewidth-1.273713\tabcolsep\relax}>{\RaggedRight\arraybackslash}p{\dimexpr 0.154472\linewidth-1.544715\tabcolsep\relax}>{\RaggedRight\arraybackslash}p{\dimexpr 0.112466\linewidth-1.124661\tabcolsep\relax}>{\RaggedRight\arraybackslash}p{\dimexpr 0.174797\linewidth-1.747967\tabcolsep\relax}>{\RaggedRight\arraybackslash}p{\dimexpr 0.215447\linewidth-2.154472\tabcolsep\relax}>{\RaggedRight\arraybackslash}p{\dimexpr 0.215447\linewidth-2.154472\tabcolsep\relax}@{}}
\caption*{\hypertarget{fs-doxcnJcSxZHfWHwUvjzgR6XTsnc}{}Table S7a. Advancement rates during DPRc production sampling. Each trial uses one RTX PRO 6000 and contains 48 windows with 250,000 steps per window. The aggregate advancement rate is the total window step count divided by the sum of segment wall times for that trial. The mean rate per window is this aggregate rate divided by 48.}\\
\toprule
\bfseries Production record & \bfseries Total steps across windows & \bfseries Segment count & \bfseries Sum of segment wall times / s & \bfseries Aggregate MD steps s\textsuperscript{-1} GPU\textsuperscript{-1} & \bfseries Mean MD steps s\textsuperscript{-1} per window \\
\midrule
\endfirsthead
\toprule
\bfseries Production record & \bfseries Total steps across windows & \bfseries Segment count & \bfseries Sum of segment wall times / s & \bfseries Aggregate MD steps s\textsuperscript{-1} GPU\textsuperscript{-1} & \bfseries Mean MD steps s\textsuperscript{-1} per window \\
\midrule
\endhead
\midrule
\multicolumn{6}{r}{\footnotesize Continued on next page}\\
\endfoot
\bottomrule
\endlastfoot
\added{trial 0} & \added{12000000} & \added{50} & \added{90849.709} & \added{132.09} & \added{2.752} \\
\added{trial 1} & \added{12000000} & \added{50} & \added{91545.798} & \added{131.08} & \added{2.731} \\
\added{trial 2} & \added{12000000} & \added{50} & \added{90069.025} & \added{133.23} & \added{2.776} \\
\end{longtable}
\endgroup

\hypertarget{fs-doxcnUqtS5rsL9GV6Mg7bi7v8od}{}
These advancement rates use the sum of wall times for 50 production segments, with intersegment scheduling waits outside the timing boundary. The short-run tests use synchronized LAMMPS loop time.

\hypertarget{fs-doxcnIYbgsvEYxfoAcA4prIlBhe}{}
The short-run timing tests use a 4 Å neighbor-list skin and check displacements every 10 steps. Both xTBloom evaluation and DeePMD inference use batch size B.

\begingroup
\fontsize{9.5}{12.8}\selectfont
\setlength{\tabcolsep}{3pt}
\renewcommand{\arraystretch}{1.18}
\setlength{\LTcapwidth}{\linewidth}
\hypertarget{fs-doxcnALd56JMCeD6VKJgMI1ipbE}{}
\begin{longtable}{@{}>{\RaggedRight\arraybackslash}p{\dimexpr 0.048780\linewidth-0.390244\tabcolsep\relax}>{\RaggedRight\arraybackslash}p{\dimexpr 0.304878\linewidth-2.439024\tabcolsep\relax}>{\RaggedRight\arraybackslash}p{\dimexpr 0.170732\linewidth-1.365854\tabcolsep\relax}>{\RaggedRight\arraybackslash}p{\dimexpr 0.304878\linewidth-2.439024\tabcolsep\relax}>{\RaggedRight\arraybackslash}p{\dimexpr 0.170732\linewidth-1.365854\tabcolsep\relax}@{}}
\caption*{\hypertarget{fs-doxcn0XxcIYpBcfAwzHARNP3Ptb}{}Table S7b. Complete batch-size scan for single-GPU short-run cross-window timing. Aggregate throughput is in MD steps s\textsuperscript{-1} GPU\textsuperscript{-1}. Entries give the median [minimum, maximum] of five consecutive segments. Paired speedup is calculated using the equation in this section. The DPRc timing uses the model from before retraining.}\\
\toprule
\bfseries \added{B} & \bfseries QM/\allowbreak{}MM aggregate throughput & \bfseries QM/\allowbreak{}MM paired speedup & \bfseries QM/\allowbreak{}MM+DPRc aggregate throughput & \bfseries QM/\allowbreak{}MM+DPRc paired speedup \\
\midrule
\endfirsthead
\toprule
\bfseries \added{B} & \bfseries QM/\allowbreak{}MM aggregate throughput & \bfseries QM/\allowbreak{}MM paired speedup & \bfseries QM/\allowbreak{}MM+DPRc aggregate throughput & \bfseries QM/\allowbreak{}MM+DPRc paired speedup \\
\midrule
\endhead
\midrule
\multicolumn{5}{r}{\footnotesize Continued on next page}\\
\endfoot
\bottomrule
\endlastfoot
\added{1} & \added{12.183 [12.048, 12.292]} & \added{1.000} & \added{10.571 [10.299, 10.608]} & \added{1.000} \\
\added{2} & \added{23.997 [23.862, 24.244]} & \added{1.970} & \added{20.381 [13.027, 20.911]} & \added{1.876} \\
\added{3} & \added{34.038 [33.819, 34.387]} & \added{2.806} & \added{29.150 [18.424, 29.277]} & \added{2.719} \\
\added{4} & \added{43.305 [43.058, 43.730]} & \added{3.560} & \added{37.010 [23.873, 37.614]} & \added{3.401} \\
\added{5} & \added{52.157 [51.654, 52.841]} & \added{4.281} & \added{45.021 [28.863, 45.647]} & \added{4.160} \\
\added{6} & \added{60.621 [60.386, 61.358]} & \added{4.988} & \added{51.607 [34.113, 52.579]} & \added{4.892} \\
\added{7} & \added{67.210 [66.844, 68.953]} & \added{5.539} & \added{55.851 [37.788, 57.623]} & \added{5.269} \\
\added{8} & \added{75.093 [74.649, 76.243]} & \added{6.174} & \added{64.389 [43.502, 66.371]} & \added{5.982} \\
\added{9} & \added{82.998 [82.179, 84.256]} & \added{6.807} & \added{67.678 [46.490, 68.517]} & \added{6.238} \\
\added{10} & \added{89.565 [88.635, 90.961]} & \added{7.356} & \added{71.998 [50.580, 73.071]} & \added{6.665} \\
\added{11} & \added{95.064 [93.782, 96.277]} & \added{7.798} & \added{75.641 [54.537, 78.168]} & \added{7.077} \\
\added{12} & \added{101.079 [67.608, 102.102]} & \added{7.653} & \added{80.968 [46.825, 82.381]} & \added{7.146} \\
\added{13} & \added{108.391 [72.455, 108.905]} & \added{8.169} & \added{85.837 [49.890, 87.218]} & \added{7.532} \\
\added{14} & \added{114.017 [77.002, 114.433]} & \added{8.620} & \added{89.271 [53.072, 92.443]} & \added{7.907} \\
\added{15} & \added{117.464 [81.147, 117.807]} & \added{8.915} & \added{93.275 [56.227, 96.900]} & \added{8.333} \\
\added{16} & \added{122.991 [84.973, 123.261]} & \added{9.422} & \added{98.032 [59.394, 101.621]} & \added{8.804} \\
\added{17} & \added{125.771 [89.522, 127.907]} & \added{9.173} & \added{101.591 [62.363, 104.711]} & \added{9.185} \\
\added{18} & \added{132.579 [92.795, 133.479]} & \added{9.661} & \added{105.602 [65.587, 109.200]} & \added{9.564} \\
\added{19} & \added{138.529 [97.403, 139.150]} & \added{10.070} & \added{108.702 [68.269, 112.963]} & \added{9.853} \\
\added{20} & \added{142.453 [100.732, 142.959]} & \added{10.388} & \added{112.464 [59.519, 115.541]} & \added{9.689} \\
\added{21} & \added{147.242 [104.767, 147.708]} & \added{10.717} & \added{117.458 [62.242, 119.834]} & \added{10.090} \\
\added{22} & \added{150.704 [105.505, 150.949]} & \added{10.951} & \added{118.731 [64.045, 122.079]} & \added{10.248} \\
\added{23} & \added{154.822 [109.734, 155.109]} & \added{11.283} & \added{123.184 [65.765, 126.706]} & \added{10.642} \\
\added{24} & \added{159.284 [113.501, 159.497]} & \added{11.624} & \added{124.698 [67.768, 129.198]} & \added{10.458} \\
\added{25} & \added{153.281 [111.950, 154.230]} & \added{11.318} & \added{126.875 [69.699, 130.856]} & \added{10.649} \\
\added{26} & \added{153.167 [114.756, 156.379]} & \added{11.469} & \added{130.584 [71.986, 134.115]} & \added{10.949} \\
\added{27} & \added{157.104 [117.556, 159.974]} & \added{11.751} & \added{132.966 [74.098, 136.531]} & \added{11.171} \\
\added{28} & \added{161.153 [120.972, 163.582]} & \added{12.053} & \added{136.103 [76.413, 139.694]} & \added{11.445} \\
\added{29} & \added{162.848 [123.833, 165.343]} & \added{12.229} & \added{139.142 [78.173, 142.844]} & \added{11.676} \\
\added{30} & \added{165.992 [123.638, 168.680]} & \added{12.413} & \added{141.670 [80.105, 145.492]} & \added{11.948} \\
\added{31} & \added{170.119 [126.285, 172.534]} & \added{12.688} & \added{142.956 [81.966, 145.893]} & \added{12.086} \\
\added{32} & \added{172.806 [128.745, 175.273]} & \added{12.901} & \added{146.179 [84.405, 149.272]} & \added{12.375} \\
\added{33} & \added{174.873 [131.162, 177.361]} & \added{13.084} & \added{149.865 [86.345, 152.841]} & \added{12.643} \\
\added{34} & \added{177.227 [133.527, 179.790]} & \added{13.281} & \added{126.579 [87.833, 153.037]} & \added{12.316} \\
\added{35} & \added{180.864 [136.632, 183.275]} & \added{13.566} & \added{129.665 [90.005, 156.887]} & \added{12.603} \\
\added{36} & \added{182.625 [118.141, 185.017]} & \added{13.108} & \added{130.673 [91.488, 157.268]} & \added{12.738} \\
\added{37} & \added{184.786 [120.227, 187.185]} & \added{13.272} & \added{133.156 [93.925, 161.398]} & \added{13.026} \\
\added{38} & \added{185.876 [121.878, 188.015]} & \added{13.373} & \added{134.166 [95.750, 162.328]} & \added{13.217} \\
\added{39} & \added{189.775 [124.553, 192.246]} & \added{13.650} & \added{137.510 [97.661, 166.173]} & \added{13.476} \\
\added{40} & \added{154.702 [125.988, 194.068]} & \added{13.313} & \added{138.839 [99.013, 167.842]} & \added{13.653} \\
\added{41} & \added{157.778 [128.780, 197.451]} & \added{13.572} & \added{139.896 [100.736, 169.479]} & \added{13.836} \\
\added{42} & \added{159.953 [130.904, 199.580]} & \added{13.743} & \added{142.032 [102.184, 170.467]} & \added{13.962} \\
\added{43} & \added{159.975 [127.845, 197.931]} & \added{13.761} & \added{142.889 [93.438, 174.348]} & \added{13.798} \\
\added{44} & \added{163.146 [130.696, 202.752]} & \added{14.050} & \added{144.289 [94.715, 175.541]} & \added{13.954} \\
\added{45} & \added{168.375 [132.997, 204.549]} & \added{14.290} & \added{143.800 [96.588, 176.887]} & \added{13.670} \\
\added{46} & \added{170.021 [134.647, 205.135]} & \added{14.458} & \added{143.914 [97.246, 175.225]} & \added{13.297} \\
\added{47} & \added{170.046 [135.396, 205.298]} & \added{14.473} & \added{146.851 [99.135, 178.868]} & \added{13.542} \\
\added{48} & \added{173.348 [137.977, 208.696]} & \added{14.725} & \added{148.062 [100.436, 180.092]} & \added{13.649} \\
\end{longtable}
\endgroup

\FloatBarrier
\Needspace{9\baselineskip}
\hypertarget{fs-doxcn7mkWUtEXtnZmceDuk59nRg}{}
\section*{S8. DPRc labels, window-level splits, and model diagnostics}
\addcontentsline{toc}{section}{S8. DPRc labels, window-level splits, and model diagnostics}

\hypertarget{fs-doxcnm1GRZUmRnskoApFs4grCOd}{}
DPRc learns the differences between high-level and low-level QM/MM energies and forces for complete periodic configurations~\citep{Zeng_JChemTheoryComput_2021_v17_p6993,Ramakrishnan_JChemTheoryComput_2015_v11_p2087}. The high-level calculations use PBE0/6-31G*~\citep{Adamo_JChemPhys_1999_v110_p6158,Hariharan_TheorChimActa_1973_v28_p213} in AmberTools with QUICK~\citep{Case_JChemInfModel_2023_v63_p6183,Cruzeiro_JChemInfModel_2021_v61_p2109}, and the low-level calculations use the xTBloom baseline matched to production simulations. Differences are evaluated for each complete periodic configuration before constructing local inputs. The QM region and complete water molecules within 6 Å are then unwrapped using the minimum-image convention. DPA4C represents the local environment, and DeePMD-kit evaluates the correction potential and its force correction~\citep{Zhang_PhysRevLett_2018_v120_p143001,Wang_ComputPhysCommun_2018_v228_p178,Li_arXiv_2026_p2608.19041,Zeng_JChemTheoryComput_2025_v21_p4375}.

\hypertarget{fs-doxcnpKzpHBQI67I8dDZFIlKncg}{}
The dataset contains 6,560 periodic configurations and is partitioned by complete umbrella windows. The negative-coordinate subset contains 1,560 frames, with window centers from \ensuremath{-}3.100 to \ensuremath{-}1.700 Å and 104 frames per window. The windows centered at \ensuremath{-}2.700 and \ensuremath{-}2.400 Å are assigned to validation and testing, respectively. Four models with random seeds 20260829, 20260830, 20260831, and 20260901 are trained from scratch for 100,000 steps each. Production trajectories use the final model with seed 20260829, selected in advance. Labels use binary64, and the descriptor and fitting network use FP32.

\begingroup
\fontsize{10}{13.5}\selectfont
\setlength{\tabcolsep}{3pt}
\renewcommand{\arraystretch}{1.18}
\setlength{\LTcapwidth}{\linewidth}
\hypertarget{fs-doxcnIvBORpg6awDziez2XiwYox}{}
\begin{longtable}{@{}>{\RaggedRight\arraybackslash}p{\dimexpr 0.300000\linewidth-0.600000\tabcolsep\relax}>{\RaggedRight\arraybackslash}p{\dimexpr 0.700000\linewidth-1.400000\tabcolsep\relax}@{}}
\caption*{\hypertarget{fs-doxcndCzphDUHlpxXhM0FQW0Wcf}{}Table S8. DPRc label and model-training configuration.}\\
\toprule
\bfseries Item & \bfseries Setting \\
\midrule
\endfirsthead
\toprule
\bfseries Item & \bfseries Setting \\
\midrule
\endhead
\midrule
\multicolumn{2}{r}{\footnotesize Continued on next page}\\
\endfoot
\bottomrule
\endlastfoot
Number of periodic labels & \added{6,560} \\
High-level method & PBE0/6-31G* QM/MM, AmberTools/QUICK \\
Low-level method & xTBloom GFN2-xTB QM/MM matched to production simulations \\
Order of differencing and localization & Evaluate differences for the complete periodic configuration, then retain the QM region and complete water molecules within 6 Å \\
Actual reaction-coordinate range of the 15-window negative-coordinate subset & [\ensuremath{-}3.073,\ensuremath{-}1.541] Å for 1,560 frames \\
Partitioning unit & Complete umbrella windows \\
Training/validation/test frame counts & \added{5,519/520/521} \\
Number of independent models & 4 \\
Training steps per model & 100,000 \\
Production model & Final primary model with seed 20260829, selected in advance \\
Held-out committee diagnostic & 521 held-out test configurations, comprising 104 frames from the negative-coordinate window subset and 417 other frames \\
\end{longtable}
\endgroup

\hypertarget{fs-doxcnrBUWbYGyZI6HJJrK6rOzVg}{}
The four-model force-vector RMS spread is computed for each QM atom and summarized over 8,336 QM atoms in 521 held-out configurations. Its mean, P95, and maximum are 0.026063, 0.046128, and 0.116216 eV/Å, respectively. This statistic describes model disagreement on the held-out set. Label errors of the primary production model are reported in Table S8a.

\hypertarget{fs-doxcngutLx0aXSKKfNPNMN6Ez0g}{}
Training uses a DPA4C descriptor with a Bessel basis, channels=32, lmax=2, \code{n_radial}=16, \code{radial_modes}=4, and rcut=6 Å. The SiLU fitting network has dimensions 192\ensuremath{\times}192\ensuremath{\times}192 and \code{resnet_dt}=false. Element types are ordered as P, O, C, H, OW, and HW. OW and HW denote water oxygen and water hydrogen. MM\textendash{}MM pairs are excluded from the descriptor and model pair calculations, and the isolated-atom energies of water oxygen and hydrogen are fixed at zero. The descriptor and fitting network use FP32, with \code{use_amp}=false and \code{enable_compile}=false. The optimizer is HybridMuon, with \code{weight_decay}=0.001 and \code{gradient_max_norm}=5. The learning rate follows a WSD schedule from 0.001 to 10\textsuperscript{-6}, with \code{warmup_ratio}=0.003, \code{warmup_start_factor}=0.2, \code{decay_type}=\code{inverse_linear}, and \code{decay_phase_ratio}=0.2. The settings are \code{training_data.batch_size}=1, \code{validation_data.batch_size}=1, and \code{validation_data.numb_btch}=16. The loss is MAE, both energy and force prefactors are 20, the virial prefactor is 0, \code{f_use_norm}=true, and \code{force_loss_use_model_mask}=true. Table S8a reports energy, QM-force, and environment-force errors for the original held-out domain of 417 frames and the additional negative-coordinate held-out window of 104 frames.

\hypertarget{fs-doxcnpv7fKoiNTYRNT7U3y4fW9e}{}
Energy MAE is the mean absolute error in the total correction energy over held-out configurations. Force MAE is the mean absolute error in Cartesian force components of the relevant atoms. Committee spread summarizes the distribution of per-QM-atom force-vector disagreement. It is reported alongside the primary model's label error as a separate evaluation quantity.

\begingroup
\fontsize{9.2}{12.4}\selectfont
\setlength{\tabcolsep}{3pt}
\renewcommand{\arraystretch}{1.18}
\setlength{\LTcapwidth}{\linewidth}
\hypertarget{fs-doxcn3iX1Siq1AycrTR6giA8dKd}{}
\begin{longtable}{@{}>{\RaggedRight\arraybackslash}p{\dimexpr 0.291328\linewidth-2.913279\tabcolsep\relax}>{\RaggedRight\arraybackslash}p{\dimexpr 0.105691\linewidth-1.056911\tabcolsep\relax}>{\RaggedRight\arraybackslash}p{\dimexpr 0.140921\linewidth-1.409214\tabcolsep\relax}>{\RaggedRight\arraybackslash}p{\dimexpr 0.140921\linewidth-1.409214\tabcolsep\relax}>{\RaggedRight\arraybackslash}p{\dimexpr 0.159892\linewidth-1.598916\tabcolsep\relax}>{\RaggedRight\arraybackslash}p{\dimexpr 0.161247\linewidth-1.612466\tabcolsep\relax}@{}}
\caption*{\hypertarget{fs-doxcnxcxYCBUCcGwMvL48B611sg}{}Table S8a. Held-out errors of the primary production model with seed 20260829. Energy errors refer to the total correction energy per configuration and are given in eV. Force errors refer to Cartesian components of the indicated atoms and are given in eV/Å.}\\
\toprule
\bfseries Held-out domain & \bfseries Configu\-ration count & \bfseries Energy MAE / eV & \bfseries Energy RMSE / eV & \bfseries QM force-component MAE / (eV/Å) & \bfseries Environment force-component MAE / (eV/Å) \\
\midrule
\endfirsthead
\toprule
\bfseries Held-out domain & \bfseries Configu\-ration count & \bfseries Energy MAE / eV & \bfseries Energy RMSE / eV & \bfseries QM force-component MAE / (eV/Å) & \bfseries Environment force-component MAE / (eV/Å) \\
\midrule
\endhead
\midrule
\multicolumn{6}{r}{\footnotesize Continued on next page}\\
\endfoot
\bottomrule
\endlastfoot
Original held-out domain & \added{417} & \added{0.079242} & \added{0.100816} & \added{0.029978} & \added{0.024046} \\
Additional negative-coordinate held-out window centered at \ensuremath{-}2.4 Å & \added{104} & \added{0.081962} & \added{0.104369} & \added{0.033248} & \added{0.023784} \\
\end{longtable}
\endgroup

\hypertarget{fs-doxcny9vz4WtVNyJ3uGU3miRvsh}{}
The two domains contain 521 held-out configurations in total.

\hypertarget{fs-doxcnrVIgz8kqalFGJUI4V305ec}{}
The 6,560 frames are partitioned by complete umbrella windows into training, validation, and test sets containing 5,519, 520, and 521 frames, respectively. The negative-coordinate window subset contains 1,560 frames from earlier bare GFN2-xTB QM/MM sampling, including 104 frames in the held-out test set. Training excludes frames from production trajectories generated by the retrained DPRc model.

\hypertarget{fs-doxcn5WEo0ZmGzNqFEgBoglmxob}{}
All 6,560 configurations are labeled with PBE0/6-31G* QM/MM minus GFN2-xTB QM/MM energy and force differences. Model MAE is evaluated on the 521-frame held-out set.

\FloatBarrier
\Needspace{9\baselineskip}
\hypertarget{fs-doxcndbpvlDNQfwCI6gzQZTaDEg}{}
\section*{S9. WHAM free-energy analysis and statistical diagnostics}
\addcontentsline{toc}{section}{S9. WHAM free-energy analysis and statistical diagnostics}

\FloatBarrier
\Needspace{8\baselineskip}
\hypertarget{fs-doxcnB8bGsnizgLAFGNAEAehaDh}{}
\subsection*{S9.1 Statistical treatment of production trajectories}
\addcontentsline{toc}{subsection}{S9.1 Statistical treatment of production trajectories}

\hypertarget{fs-doxcnIRMCWjkAt25t1fgukqBB0g}{}
WHAM combines histogram counts from all windows and production trials~\citep{Kumar_JComputChem_1992_v13_p1011}. The binning range is [\ensuremath{-}3.600,2.100] Å, the bin width is 0.020 Å, the analysis temperature is 298 K, and the dimensionless convergence tolerance is 10\textsuperscript{-10}. WHAM removes the reaction-coordinate umbrella bias while retaining the common attack-angle restraint.

\hypertarget{fs-doxcnF43AtOVHm5o9RiRqtY2Pnd}{}
For each trial\textendash{}window series, statistical inefficiencies are estimated separately for the reaction coordinate and attack angle. The larger value characterizes the temporal correlation of that series~\citep{Chodera_JChemTheoryComput_2016_v12_p1799,Flyvbjerg_JChemPhys_1989_v91_p461,Geyer_StatSci_1992_v7_p473}. Let Ntw be the number of samples in trial t and window w. The pooled effective sample size is

\hypertarget{fs-doxcnN3NCAbxlglrr82GdWH1hog}{}
\begin{equation}
g_{tw}=\max\!\left(g_{\xi,tw},g_{\alpha,tw}\right),\qquad N_{\mathrm{eff},w}=\sum_t\frac{N_{tw}}{g_{tw}}.
\end{equation}

\hypertarget{fs-doxcnna79GcCMwzYm9qfdeZnJ4f}{}
Block resampling is performed independently for each trial\textendash{}window series, with a block length of ceil(g) samples. Resampled counts are pooled and WHAM is reconstructed in each of 500 rounds. Every resampled curve is shifted to its global observed minimum. The 2.500\% and 97.500\% quantiles of finite PMF values in each bin define pointwise 95\% intervals. PMF and barrier central values are reported in kcal mol\textsuperscript{-1}. These pointwise 95\% intervals are descriptive. Their interpretation is restricted to individual PMF bins and excludes joint confidence claims for peak--valley or barrier differences.

\hypertarget{fs-doxcnVZOfcrzkcKW1E7pRAbQule}{}
Each window is equilibrated for 100 ps, and production statistics begin after this prescribed segment. No additional burn-in is removed from the production segments, and automatic equilibration detection is not used to replace the predefined equilibration segment. Raw and retained frame counts, statistical inefficiency, effective sample size, failures, and bootstrap coverage are recorded for each trial\textendash{}window series using the definitions in this section.

\FloatBarrier
\Needspace{8\baselineskip}
\hypertarget{fs-doxcncoLCjVKqLFIftLuqRfhcGg}{}
\subsection*{S9.2 Diagnostic definitions in the analysis implementation}
\addcontentsline{toc}{subsection}{S9.2 Diagnostic definitions in the analysis implementation}

\hypertarget{fs-doxcn5UL7VI4nTkHRTrjMBqjpWe}{}
The analysis implementation defines adjacent-window overlap, resampling coverage, and between-trial consistency. Adjacent windows use the same bins. Their normalized histograms give the Bhattacharyya coefficient

\hypertarget{fs-doxcn6f1tDTfMsDboWcuB2icNOf}{}
\begin{equation}
O_{w,w+1}=\sum_b\sqrt{p_{w,b}p_{w+1,b}},\qquad \sum_b p_{w,b}=1.
\end{equation}

\hypertarget{fs-doxcnLys1ZP6nh04MFY6CK5DW6d}{}
The implementation estimates statistical inefficiency using Geyer's initial monotone positive sequence. Non-overlapping circular blocks wrap around at the end of the series. Resampling draws complete blocks and any required short prefix to retain the sample count of each trial\textendash{}window series. The original coverage diagnostic considers every observed bin in the full analysis range and requires a finite PMF in at least 0.900 of the 500 resampling rounds for each bin. Coverage of the 235 points within the window-center range is reported separately in Section S9.3. For between-trial consistency, curves are first aligned at the lowest pooled-PMF bin supported by every trial. Comparisons then use the core region where the pooled PMF is at most 8 kcal/mol.

\begingroup
\fontsize{10}{13.5}\selectfont
\setlength{\tabcolsep}{3pt}
\renewcommand{\arraystretch}{1.18}
\setlength{\LTcapwidth}{\linewidth}
\hypertarget{fs-doxcn6pQ5HZxPI9gOt1LjIgX6eb}{}
\begin{longtable}{@{}>{\RaggedRight\arraybackslash}p{\dimexpr 0.300000\linewidth-0.600000\tabcolsep\relax}>{\RaggedRight\arraybackslash}p{\dimexpr 0.700000\linewidth-1.400000\tabcolsep\relax}@{}}
\caption*{\hypertarget{fs-doxcnENQIQRLacBkO8atoUM6wlt}{}Table S9a. WHAM and block-resampling parameters.}\\
\toprule
\bfseries Parameter & \bfseries Value or definition \\
\midrule
\endfirsthead
\toprule
\bfseries Parameter & \bfseries Value or definition \\
\midrule
\endhead
\midrule
\multicolumn{2}{r}{\footnotesize Continued on next page}\\
\endfoot
\bottomrule
\endlastfoot
Analysis temperature & 298 K \\
Binning range and bin width & [\ensuremath{-}3.600,2.100] Å/0.020 Å \\
WHAM convergence tolerance & Dimensionless tolerance of 10\textsuperscript{-10} \\
Effective sample size & Sum of N/max(g\ensuremath{\xi},g\ensuremath{\alpha}) over trials \\
Block length and number of resampling rounds & ceil(max(g\ensuremath{\xi},g\ensuremath{\alpha}))/500 \\
Interval and zero reference & Pointwise 95\% intervals, with each curve shifted to its global observed minimum \\
\end{longtable}
\endgroup

\begingroup
\fontsize{10}{13.5}\selectfont
\setlength{\tabcolsep}{3pt}
\renewcommand{\arraystretch}{1.18}
\setlength{\LTcapwidth}{\linewidth}
\hypertarget{fs-doxcnDVGaZMdj629xLylTRT2Ckg}{}
\begin{longtable}{@{}>{\RaggedRight\arraybackslash}p{\dimexpr 0.300000\linewidth-0.600000\tabcolsep\relax}>{\RaggedRight\arraybackslash}p{\dimexpr 0.700000\linewidth-1.400000\tabcolsep\relax}@{}}
\caption*{\hypertarget{fs-doxcnM547gjeOjJElh1rqgHRB8e}{}Table S9b. Analysis parameters and diagnostic definitions.}\\
\toprule
\bfseries Parameter & \bfseries Value or definition \\
\midrule
\endfirsthead
\toprule
\bfseries Parameter & \bfseries Value or definition \\
\midrule
\endhead
\midrule
\multicolumn{2}{r}{\footnotesize Continued on next page}\\
\endfoot
\bottomrule
\endlastfoot
Maximum WHAM iterations & 100,000 \\
Statistical-inefficiency estimator & Geyer initial monotone positive sequence \\
Resampling random seed & 982451653 \\
Minimum and target adjacent-window overlap & 0.050/0.200 \\
Minimum pooled effective sample size per window & 50 \\
Per-bin resampling coverage & Every observed bin in the original analysis range should have a finite-PMF resampling fraction of at least 0.900. The minimum fraction over the original observed bins is 0.602. In a separate check, all 235 points within the window-center range have a fraction of 1.000. Section S9.3 specifies that range. \\
Between-trial consistency & Each trial supports the region with pooled PMF\ensuremath{\leq}8 kcal/mol, and the maximum pairwise difference is\ensuremath{\leq}2 kcal/mol \\
\end{longtable}
\endgroup

\FloatBarrier
\Needspace{8\baselineskip}
\hypertarget{fs-doxcnlhRAmo70tBHRGb0J2SKYgk}{}
\subsection*{S9.3 Literature comparison and DPRc statistical results}
\addcontentsline{toc}{subsection}{S9.3 Literature comparison and DPRc statistical results}

\hypertarget{fs-doxcnOF6dKEXOcQ6FGFDrJgU7se}{}
The reference comprises 705 digitized points from the red gwTP-PBE0 curve in \href{https://doi.org/10.1021/acs.jpca.2c06201}{Fig. 3a} of Giese, Zeng, and York (2022). Each simulated curve and the reference curve is shifted by its own A(0), obtained by linear interpolation. MAE and RMSE are evaluated with equal weights at 233 shared grid centers from \ensuremath{-}3.070 to 1.570 Å with spacing 0.020 Å. The horizontal coordinate is retained without fitting a shift. The reference-curve source, pixel calibration, pointwise aligned values, and residuals follow the calculation definitions in this section.

\hypertarget{fs-doxcnlw8iIX2z01aExyDUhJrLWb}{}
The red reference curve is a PBE0/6-31G* free-energy estimate obtained by gwTP reweighting of sampling from four MNDO/d QM/MM+DPRc reference potentials~\citep{Giese_JPhysChemA_2022_v126_p8519}. The original high-level QM/MM calculations used SANDER/HFDF with ambient-potential composite Ewald, 8 Å cutoffs for both real-space electrostatics and Lennard\textendash{}Jones interactions, and an LJ tail correction. Labels in the present study use AmberTools/QUICK PBE0/6-31G* with a 9 Å cutoff. The resulting MAE and RMSE therefore include differences in Hamiltonian, computational protocol, and sampling. Both sampling sets in this work retain the 180\ensuremath{^{\circ}} attack-angle restraint.

\hypertarget{fs-doxcngexdQHb7DoC0sdsIqv833c}{}
The comparison uses the 705 digitized points, pixel calibration, and pointwise residuals according to the definitions in this section.

\hypertarget{fs-doxcn4WoUG8zMwa9BrmU7DdlKce}{}
For the retrained DPRc model, the minimum adjacent-window overlap coefficient is 0.567 and the minimum pooled effective sample size per window is 105.013. Both meet the original thresholds. The minimum finite-bootstrap fraction over observed bins in the original analysis range is 0.602, below 0.900, so the original coverage and between-trial support criteria are incompletely satisfied. Between-trial support is missing from +1.870 to +1.910 Å. Within the window-center range from \ensuremath{-}3.100 to 1.600 Å, all 235 analysis grid points (\ensuremath{\xi}\textsubscript{i}=\ensuremath{-}3.09+0.02i, i=0,\ldots{},234) have finite results in 500/500 bootstrap rounds.

\hypertarget{fs-doxcnPNUWG6mVCB3ocfgXzz3MEd}{}
Free-energy differences among the principal peaks, reactant, and intermediate are descriptive central values. The digitization covers the literature central curve and leaves its error bars unquantified. The original PMF and its pointwise 95\% intervals use the global observed minimum of each curve as the zero reference. Alignment at A(0) is used to compare central curves. The RMSE values of the bare QM/MM and DPRc curves are 6.61 and 0.866 kcal mol\textsuperscript{-1}, respectively. The free-energy differences of the two principal DPRc peaks relative to the reactant are 9.21 and 16.45 kcal mol\textsuperscript{-1}. Their differences from the reference central values are +0.17 and \ensuremath{-}0.33 kcal mol\textsuperscript{-1}, respectively.

\FloatBarrier
\Needspace{9\baselineskip}
\hypertarget{fs-doxcnWWPnLPJfSFV7Z5LZr4q5tb}{}
\section*{S10. Data sources and licenses}
\addcontentsline{toc}{section}{S10. Data sources and licenses}

\hypertarget{fs-doxcnwRiVtPUXYJ0kP3NmMQUqdd}{}
Use, citation, and redistribution of the original QM9 and OMol25 data follow their respective upstream licenses. Those terms continue to govern the upstream data when used in the summaries, figures, and analyses generated in this study.

\hypertarget{fs-doxcnDAhg6IxWRfffisNng5F9Uf}{}
The xTBloom repository (\nolinkurl{https://github.com/jinzhezenggroup/xtbloom}) provides source code, build instructions, interfaces, and benchmark protocols. The LAMMPS-DPRc repository (\nolinkurl{https://github.com/jinzhezenggroup/lammps-dprc}) provides the periodic QM/MM plugin and coordinator.

\FloatBarrier
\nocite{Bannwarth_WiresComputMolSci_2021_v11_pe1493,Darden_JChemPhys_1993_v98_p10089,Essmann_JChemPhys_1995_v103_p8577,Nam_JChemTheoryComput_2005_v1_p2,Giese_JChemTheoryComput_2016_v12_p2611}
\bibliography{zenglib}